\documentclass[11pt]{article}

\usepackage{svg}
\usepackage{authblk}
\usepackage{tablefootnote}
\usepackage{array}

\usepackage[left=2cm,top=2cm,bottom=1.5cm,right=2cm]{geometry}

\usepackage[colorlinks=true]{hyperref}
\hypersetup{allcolors=blue, breaklinks=true}

\usepackage[sort&compress,numbers]{natbib}
\usepackage{doi}

\usepackage[T1]{fontenc}
\makeatletter
\providecommand{\tabularnewline}{\\}
\makeatother
\usepackage{babel}
\usepackage{array}
\usepackage{varwidth}
\usepackage{stmaryrd}
\usepackage{graphicx} 	
\usepackage{amsmath}

\usepackage{enumitem}
\setitemize{noitemsep,topsep=0pt,parsep=0pt,partopsep=0pt,leftmargin=*}
\usepackage{amssymb}

\usepackage{nopageno}
\usepackage{enumitem}

\usepackage{fancyhdr}
\usepackage{tabularx}
\usepackage{ragged2e} 

\newcommand{\ome}{\mathsf{EVE}}
\newcommand{\omeco}{\ome[I]}
\newcommand{\omeeco}{\ome_{\epsilon}[I]}

\newcommand{\evecob}{\ome_{\epsilon}}

\newcommand{\cA}{\mathcal{A}}
\newcommand{\cC}{\mathcal{C}}
\newcommand{\cQ}{\mathcal{Q}}

\newcommand{\bbE}{\mathbb{E}}
\newcommand{\abs}[1]{| {#1} |}
\newcommand{\qcost}{\cQ}
\newcommand{\ccost}{\cC}
\newcommand{\poly}{\mathrm{poly}}

\newcommand{\nocontentsline}[3]{}
\let\origcontentsline\addcontentsline
\newcommand\stoptoc{\let\addcontentsline\nocontentsline}
\newcommand\resumetoc{\let\addcontentsline\origcontentsline}

\newtheorem{corollary}{Corollary}
\newtheorem{definition}{Definition}[section]

\newtheorem{remark}{Remark}

\usepackage{xcolor}
\usepackage{enumitem}
\newcommand{\ignore}[1]{}

\providecommand{\tabularnewline}{\\}

\makeatother

\begin{document}
\addtocounter{page}{-1}

\title{On the Importance of Error Mitigation for Quantum Computation}

\author[1,2]{Dorit Aharonov\footnote{\texttt{dorit.aharonov@qedma.com}}}
\author[1]{Ori Alberton} 
\author[1,3]{Itai Arad}
\author[1]{Yosi Atia}
\author[1]{Eyal Bairey}
\author[1,4]{Zvika Brakerski}
\author[1]{\\ Itsik Cohen}
\author[1]{Omri Golan\footnote{\texttt{omri.golan@qedma.com}}}
\author[1]{Ilya Gurwich}
\author[1,5]{Oded Kenneth}
\author[1]{Eyal Leviatan}
\author[1,5]{Netanel H. Lindner}
\author[1]{Ron Aharon Melcer}
\author[1]{Adiel Meyer}
\author[1]{Gili Schul}
\author[1]{Maor Shutman}

\affil[1]{\textit{Qedma Quantum Computing, Tel Aviv, Israel}}
\affil[2]{\textit{The Benin School of Computer Science and Engineering, Hebrew University, Jerusalem, Israel}}
\affil[3]{\textit{Centre for Quantum Technologies, National University of Singapore, Singapore}}
\affil[4]{\textit{Faculty of Mathematics and Computer Science,
Weizmann Institute of Science, Israel}}
\affil[5]{\textit{Department of Physics, Technion, Haifa, Israel}}

\date{}

\maketitle

\begin{abstract}
Quantum error mitigation (EM) is a family of hybrid quantum-classical 
methods for eliminating or reducing the effect of noise and decoherence on quantum algorithms run on quantum hardware, 
without applying quantum error correction (EC).
While EM has many benefits compared to EC, specifically that it requires no (or little) qubit overhead, this benefit comes with a painful price: EM seems to necessitate an overhead in quantum run time which grows as a (mild) exponent \cite{takagi2022fundamental, Takagi2023universal, Tsubouchi2023universal, Quek_Eisert}. Accordingly, recent results show that EM alone cannot enable exponential quantum advantages (QAs), for an average variant of the expectation value estimation problem \cite{Schuster_Yao}. These works raised concerns regarding the role of EM in the road map towards QAs. 

We aim to demystify the discussion, and provide a clear picture of the role of EM in achieving QAs, both in the near and long term. We first propose a clear distinction between \textit{finite} QA and \textit{asymptotic} QA, which is crucial to the understanding of the question, and present the notion of \textit{circuit volume boost}, which we claim is an adequate way to quantify the benefits of EM. Using these notions we can argue straightforwardly that EM is expected to have a significant role in achieving QAs. Specifically, that EM is likely to be the first error reduction method for useful finite QAs, before EC;
that the first such QAs are expected to be achieved using EM in the very near future; and that EM is expected to maintain its important role in quantum computation even when EC will be routinely used
-- for as long as high-quality qubits remain a scarce resource.
\end{abstract}

\setcounter{tocdepth}{2}
\tableofcontents

\section{Introduction}
\subsection{Error correction (EC)}

The necessity of protecting quantum computations from noise had been realized already close to three decades ago, soon
after the remarkable theoretical discoveries  
indicating a potential for exponential algorithmic speedups (or \textit{quantum advantages}, QAs) over
classical computing devices \cite{bernstein_vazirani_1993,
simon_1997}, including Shor's algorithm \cite{shor_1999}. The no
less remarkable discovery of  \textit{quantum error correction} (EC)
\cite{steane_1996, shor_1995} and the follow-up proofs of
\textit{threshold theorems} showing that EC  can be
made \textit{fault-tolerant} \cite{aharonov_1997, kitaev_1997,
knill_laflamme_1996,preskillaliferis}, laid the foundations for the
approach taken today to handle noise: large scale quantum computers (or \textit{quantum processing units}, QPUs)
are to be protected from the detrimental effect of noise by redundantly encoding computational {\it logical} qubits into physical qubits, and performing
fault-tolerant EC, periodically measuring and
correcting errors. Fault-tolerant EC has since become a
lively active field on its own, and most quantum hardware companies
have adopted this path as the main approach in their roadmap, to handle noise in future
large scale QPUs \cite{google_quantum_ai,
ibm_quantum_technology, quera_qec_2024, quantinuum_roadmap_2023,
ionq_roadmap_2024, iqm_roadmap_2024}.  

However, the EC
approach suffers from several severe practical drawbacks.  First, it
requires very large overheads in qubit numbers, due to the redundant encoding of physical qubits into logical qubits, as well additional qubits needed to periodically measure error syndromes. 

The second drawback is of
course the requirement for extremely high quality qubits; fault
tolerance requires that the gates applied on the qubits be of
fidelity better than a certain threshold, the ``fault tolerance
threshold''. EC is only beneficial if the infidelity is
below that threshold.  Moreover, the closer to the threshold the
actual infidelity is, the larger the qubit overhead required in
order to achieve the same effective (logical) error rate. This
importantly implies that even when the infidelity is below the
quantum fault tolerance threshold, the noise still limits the size
of the quantum computation.  

By far the most important leading candidate for a quantum error
correcting code to be used on fault tolerant quantum devices, had
been, for a long time, the surface code \cite{fowler2012surface},
due to its high threshold value \cite{wang2011surface}. However, due
to the large qubit overhead it requires, recently the interest of
the industry had shifted to other codes, and in particular to
quantum low-density parity check (qLDPC) codes \cite{xu2024constant,
bravyi_2024, scruby2024high}, as well as concatenated codes
\cite{yamasaki2024time, goto2024high}.  The past year has seen
tremendous progress on the experimental side of various fault
tolerance components \cite{1-bluvstein2024logical,
2-mayer2024benchmarking,3-da2404demonstration,
4-reichardt2024demonstration,5-acharya2024quantum,
putterman2024hardware}.  However, it is still unclear when quantum
computers with sufficiently many high quality logical qubits will be
available to enable the first useful QAs using EC alone. By {\it useful} QA we mean any computation performed by a QPU faster or better than possible by classical computers, and whose output is of interest outside of the question of whether it was achieved by a QPU or by some other computational
means.\footnote{QAs have
already been argued in several experiments  \cite{googlesupremacy1, wu2021strong, QAphotons, zhu2022quantum, Morvan2024, 1-bluvstein2024logical, decross2024computational}, including
random circuit sampling (RCS) experiments; we will mention
those in Section \ref{Sec: QuAlA}. As remarkable as these
achievements are, they do not belong to this class of {\it useful}
QAs, since their output is in general not useful, for any purpose other than demonstrating QA (though Ref.~\cite{scottcertifiable} suggested using RCS to generate certifiable randomness).} 
 According to the public roadmaps of both IBM
\cite{ibm_quantum_technology} and Quantinuum 
\cite{quantinuum_roadmap_2023}, for example, achieving  the first useful
QAs with EC is a matter of at least five years.

\subsection{Error mitigation (EM)} 
In light of these hurdles, and the long anticipated time until useful
QA with EC, side by side with
the immense effort and many successes in improving fault tolerant
constructions and reducing their requirements, another approach had
been developed - a family of protocols which goes by the name of
\textit{quantum error mitigation} (EM) \cite{emreview}. Such EM
protocols replace the execution of a given target circuit with
multiple noisy circuit executions; the results of which are
post-processed to provide an estimate for the error-free output of
the target circuit. The first EM protocols, based on
quasi-probability (QP) distributions and zero-noise extrapolation
(ZNE) were developed in \cite{bravyitemmegambetta, li2017efficient,
PhysRevX.8.031027} and followed by a plethora of additional
protocols and extensions
\cite{emreview,PhysRevA.104.052607,cai2021multi,
filippov2024scalability}. 

The advantages of EM methods over EC are threefold. 
\begin{enumerate} 
    \item Unlike EC, the EM approach does not require adaptive operations.\footnote{Namely, mid-circuit measurements, followed by real-time classical processing of measurement results, which are in turn followed by unitary gates conditioned on the classical processing output.} Thus the requirements on the hardware are significantly relaxed.  This lack of adaptive operations is the theoretically defining feature of EM \cite{takagi2022fundamental, Takagi2023universal, Tsubouchi2023universal, Quek_Eisert}. 
  \item EM methods entail no (or little) overhead in qubit 
    number. This is  critical, given the fact that increasing the
    number of high quality qubits has proven to be a steady but
    unfortunately slow process, at least in the currently leading publicly available 
    hardware platforms of superconducting qubits and trapped ions  \cite{ibm_quantum_technology,
    quantinuum_roadmap_2023}.  Qubit numbers may be larger, though still limited, in upcoming neutral atom platforms \cite{quera_qec_2024}. 

\item Last but not least, EM methods do not require the error rate 
  to be below any threshold! They work with any infidelity. 
\end{enumerate} 
These advantages make EM applicable on current quantum processors. 
Indeed, EM protocols are now being routinely used in essentially all large
scale quantum computing experiments, 
showcasing impressive achievements and demonstrating the ability of
those methods to bring current quantum processors much closer to
QA \cite{kim2023evidence, shinjo2024unveiling,
yu2023simulating, mi2024stable, gyawali2024observation,
iqbal2024qutrit, greene2024measuring, chen2024benchmarking}. 

Importantly, already from the first EM protocols, it was clear that
these protocols come with a painful price: they exhibit a sampling
overhead that grows like a mild exponent with the size of the
computation.

This exponential price of EM raises important questions: Is EM worth
the effort and the time overhead? Is it a viable approach for
addressing quantum noise on the way to achieving useful QAs? And very importantly, what would its role be when
EC becomes available for end-users? In
particular, concerns were recently raised following several
papers \cite{takagi2022fundamental, Takagi2023universal, Tsubouchi2023universal, Quek_Eisert} that highlighted 
exponential lower bounds on the sampling overhead of EM. Importantly, a recent strong result by \cite{Schuster_Yao} managed to rule out the possibility that EM can provide exponential QA for a certain computational problem
(we discuss the results of those papers and their implications in 
Sec. \ref{Sec: EM alone cannot}). In light of all this, what is
the role of EM in the future of quantum computation? Perhaps
it should be considered as a temporary method, interesting for
investigations but to be abandoned very soon when EC
takes over?

\subsection{This paper, in a nutshell} 
The question regarding the role of EM in the evolution of quantum computation is crucial for the clarification of the path that the
quantum eco-system will take as it moves forward towards algorithmic
QAs. In this manuscript, we aim to remove some of the misconceptions in the public discussion regarding this question, and by that, clarify the role of EM in this roadmap towards algorithmic QAs. 

The main message of this paper can be summarized as follows: There are two different types of QA which should not be confused: \textit{asymptotic}  QA, which is a mathematical notion, and \textit{finite} QA, which is a practical notion. While asymptotic QA is probably indeed not achievable by EM alone, EM is expected to provide dramatic finite QA in the near future, and in fact be the first error reduction approach to provide useful QA, before this will be achieved by EC. Furthermore, 
even when fault-tolerant EC becomes available, it is a misconception that this will completely resolve the problem of noise in quantum computation. 

Since quantum  computers are expected to have a limited number of high-fidelity qubits in the foreseeable future, we expect severe limitations on the quantum circuit volumes available even when fault tolerant EC becomes available.  At that time, EM combined with EC would provide a much more effective error reduction solution, leading to circuit volumes that are orders of magnitude larger than those available by EC alone. Thus, EM is here to stay, and is expected to play a crucial role in all important milestones of the quantum ecosystem in the near and far future.

The paper follows the above sequence of claims and provides details behind each of the key points. 
Section \ref{sec:overview} provides a more detailed overview of the flow of the paper. We attempt to keep the discussion simple, and gloss over some
details, referring the reader to other resources where these details appear.  

Table \ref{Tab: tab0} provides an overview of
different methods for handling errors in quantum computation, along
with the main take-away messages we propose for each method. EM, EC,
and logical error mitigation (LEM) are discussed throughout the
paper, as detailed below. We do not discuss error suppression (ES) except in this table, and assume
throughout the paper that `bare', or `noisy', quantum circuit execution
already includes the best available ES methods.

\renewcommand{\arraystretch}{1.2}
\begin{table}[h!]
\fontsize{8}{10}\selectfont
\begin{centering}
\caption{Overview of different approaches for dealing with errors in quantum
computation. ES is not further discussed in this paper, and we assume throughout that the best available ES methods have been incorporated into the `bare', or `noisy' circuit execution. EM is described in Sec.~\ref{Sec: EM alone cannot}-\ref{Sec: Error Mitigation Provides}, while EC and LEM are discussed in Sec.~\ref{Sec: EMEC}.   
\vspace{5pt}
\label{Tab: tab0}
}

\begin{tabular}{V{\linewidth}V{\linewidth}V{\linewidth}V{\linewidth}V{\linewidth}}
\hline 
\hline
 & \textbf{Error Suppression (ES)} & \textbf{Error Mitigation (EM)} & \textbf{Error Correction (EC)} & \textbf{Logical EM (LEM)}\tabularnewline
\hline 
Relation & Always done. & On top of ES, 

instead of EC. & On top of ES, 

instead of EM. & On top of ES and EC.\tabularnewline
\hline 
Main 

idea & Cancel coherent errors

(directly, or between 

layers or shots).\tablefootnote{Realistically, dissipative dynamics is due to coherent interaction with an environment. The distinction we make between `coherent' and `dissipative' errors then depends on the dynamical time scales of the environment relative to the time scales for implementing quantum gates: interaction with a `slow' (`fast') environment manifests as  `coherent' (`dissipative'). \label{tabfoot3}}

& Run multiple noisy 

computations and 

post-process.  & Encode qubits 

redundantly, 

measure and invert 

errors in real time.  & Perform EC and 

mitigate remaining 

logical errors. \tabularnewline
\hline 
Main 

resource & Essentially free. & QPU time.\tablefootnote{Beyond the main resources stated, we note that EM can actually benefit from extra qubits via e.g.,
parallelization and coherent operations on multiple copies of the noisy computation, and EC does actually require a QPU time overhead, since logical operations are composed of multiple physical gate layers.  \label{tabfoot1}} & Qubits.$^\text{\ref{tabfoot1}}$ & Trade QPU time for 

qubits.\tabularnewline
\hline 
Examples & Gate calibration, 

Pauli twirling, 

dynamical decoupling.  & Zero Noise Extrapolation 

(ZNE), Quasi Probability

(QP), Clifford Data 

Regression (CDR). & Steane, Surface, quantum 

Low Density Parity Check 

(qLDPC), concatenation.  & External LEM, 

Error correction with 

post-selection (EC+PS), 

Syndrome-aware LEM.\tabularnewline
\hline 
Maximal 

circuit 

volume & $\boldsymbol{V_{bare}:=V_{ES}\sim\epsilon/\gamma}$,
\medskip{}

for allowed inaccuracy

$\epsilon$ and (un-suppressed) 

gate infidelity $\gamma$.  & $\boldsymbol{V_{EM}\sim(1-10)/\gamma}$,
\medskip{}

independent of $\epsilon$.\tablefootnote{For EM and LEM, we assume a negligible mitigation bias $b\ll\epsilon$, and the range $1-10$ depends the details of the EM method used, and on the allowed shot \textit{overhead} (fixing the allowed shot \textit{number} leads to a logarithmic dependence on $\epsilon$, see Sec. \ref{Sec:CVB}).  \label{tabfoot2}}  & $\boldsymbol{V_{EC}\sim\epsilon/\gamma'}$,
\medskip{}

for allowed inaccuracy $\epsilon$ and 

logical gate infidelity $\gamma'$.  & $\boldsymbol{V_{LEM}\sim(1-10)/\gamma'}$, 
\medskip{}

independent of $\epsilon$.$^\text{\ref{tabfoot2}}$\tabularnewline
\hline 
Take-away 

message  & Should be done as much

as possible, but 

fundamentally cannot 

handle dissipation,$^\text{\ref{tabfoot3}}$ and 

therefore limited to small 

volume.  & Volume boosted by a 

factor $\sim(1-10)/\epsilon$ 

over ES. & If physical qubit number 

is limited, tension between 

logical accuracy (or depth) 

and width:

\medskip{}

Higher accuracy requires 

smaller $\gamma'$, which requires 

more physical qubits per 

logical qubit, leaving less 

logical qubits.  & Break EM volume 

barrier with EC. 

Ease accuracy-width

tension in EC.

\medskip{}

Volume boosted by a

factor $\sim(1-10)/\epsilon$ 

over EC.\tabularnewline
\hline 
Volumes 

anticipated 

based 

on Hardware 

roadmaps & Not indicated. & $V\sim10^{3}$ within 2025

\cite{ibm_quantum_technology, quantinuum_roadmap_2023}.

$V\sim10^{4}$ by 2027-2028

\cite{ibm_quantum_technology, quantinuum_roadmap_2023}.

Larger volumes not 

anticipated with EM

\cite{ibm_quantum_technology, quantinuum_roadmap_2023}. & $V\sim10^{3}-10^{4}$ within 2026 

\cite{quera_qec_2024}.

$V\sim10^{4}-10^{9}$ within 2029 

\cite{quantinuum_roadmap_2023}.

$V\sim10^{8}$ within 2029 \cite{ibm_quantum_technology}. & Not indicated.\tabularnewline
\hline 
\hline
\end{tabular}
\par\end{centering}
\end{table}

\subsection{Overview of key points}
\label{sec:overview}

\begin{itemize} 
\item {\bf Asymptotic vs. finite Quantum advantages} Our starting point (Sec.~\ref{Sec: QuAlA}) is a distinction, that we believe is too often overlooked, between two notions of QA. One is an \textit{asymptotic} notion, where the most famous example, perhaps, is Shor's factoring algorithm that, as far as we know, performs exponentially better than any classical factoring algorithm, when considering the asymptotic cost. The other is \textit{finite} QA, which refers to a concrete experiment where a QPU computes ``much faster'' than a classical computer. Ideally, the experiment can be repeated with larger and larger settings to establish a \textit{scalable} behavior, where the QA is expected to strengthen as the problem size grows. We discuss some of the subtleties involved in defining finite QA, such as the proper way to define what is meant by ``much faster'', i.e., what is the proper way to quantify  the cost of the computation.
\item {\bf The expectation value estimation problem, and definition of active volume}
We next present the computational problem on which we focus our discussion of the above notions of QA: \textit{expectation value estimation} ($\ome$), in which for a given circuit $C$ and an observable $O$, one is required to estimate the expectation value of $O$ in the state obtained by applying $C$ to the all zero input state, with a prescribed accuracy. $\ome$ is one of the most basic problems in physics and has a variety of applications within the field of quantum algorithms. Furthermore, $\ome$ is particularly appropriate for the discussion of EM.  We define $\ome$ in Sec.~\ref{sec:eve}, together with the notion of \textit{active volume} -- the relevant notion of quantum circuit volume within the context of $\ome$.  

\item {\bf EM is likely not capable of providing exponential asymptotic QA}
Equipped with the definitions of $\ome$ and active volume, we next proceed to discuss the question of whether and in what context EM is useful for QA. We first consider the question of asymptotic QA. Section \ref{Sec: EM alone cannot} describes the barriers towards the usability of EM for such QA. We start by describing examples and general results, providing strong evidence that EM is expected to require a number of shots that grows exponentially with $\gamma V$, in order to ensure a required accuracy. Here $\gamma$ is the gate infidelity and $V$ the active volume of a given instance of $\ome$. In cases where this exponential behavior holds,  we show that if EM solves $\ome$ for a given circuit in time $T_{EM}$, the same task can be accomplished classically in a time which is bounded (roughly) as $T_c=O(T_{EM}^{1/\gamma})$, i.e., $T_c$ is at most polynomial, with degree $1/\gamma$, in $T_{EM}$. This rules out exponential (and in fact any non-polynomial) asymptotic QA.  We explain that it is still an open question whether the above exponential scaling as a function of $\gamma V$ is a generic behavior for all EM protocols and inputs to the $\ome$ problem. Nevertheless, recent work  significantly strengthened the case against exponential asymptotic QA with EM \cite{Schuster_Yao}. By deriving a classical polynomial time simulation of {\it any} noisy quantum circuit (albeit when averaged over the inputs,  and with the polynomial degree scaling as $1/\gamma^2$), the authors deduced that for the $\ome$ problem in which the required accuracy is defined on average over inputs to the circuit, exponential asymptotic QA is ruled out. Interestingly,  this leaves room for the possibility of exponential asymptotic QA with EM in non-generic cases; however, the main message of  Sec.~\ref{Sec: EM alone cannot} is that EM cannot generically provide  exponential asymptotic QA. 

\item {\bf The polynomial bound on the advantage of EM leaves ample room for finite QA}

Naively, the above statement could be interpreted to mean that the execution of quantum circuits using EM can be replaced by a classical simulation, so there is no point in using EM on a QPU at all. However, we observe that the
large power $1/\gamma$ (e.g., a thousand!) in the above-mentioned polynomial relation between $T_c$ and $T_{EM}$ quantifies a large slowdown of classical computation compared to
EM. This implies that even though EM probably 
cannot exhibit exponential asymptotic QA,
there is ample room for EM to enable dramatic {\it finite} QAs. In Sec.~\ref{Sec: Error Mitigation Provides} we indeed argue why EM is expected to enable such finite QA in the very near future, for the useful $\ome$ problem.   
    
\item {\bf Circuit Volume Boosts can quantify the contribution of EM} We therefore put forth a quantity that properly quantifies the contribution of EM to the execution of quantum circuits. 
Indeed, we claim that using EM, it is possible to use noisy hardware to execute quantum circuits of \emph{higher volume} than is possible with the `bare' execution, without EM. Namely, we fix the \textit{allowed shot overhead}  and the desired output accuracy, and consider the maximal active volume that can be obtained by naive circuit execution, as opposed to the maximal active volume possible with EM. In Sec.~\ref{Sec:CVB} we define the ratio of these maximal volumes to be the \emph{circuit volume boost} (CVB) of EM in the particular setting; Fig.~\ref{fig: CVB} shows the expected CVBs as a function of the required output accuracy. For example, EM is expected to enable circuit volume boosts of over a $100$ (!) for required accuracy of  $99\%$ with  a mild allowed shot overhead of $10$. Furthermore, the better the required accuracy (and allowed shot overhead), the larger the CVBs that will be provided, namely the more beneficial it is to apply EM. We argue that the CVB is the proper way to quantify the contribution of EM to quantum circuit execution on noisy hardware. 

\item{\bf Dramatic finite QAs expected using EM, way before EC} 
 In Sec.~\ref{Sec: predictions} we discuss the expected impact 
    of EM in the near future. We argue that EM is likely to be the
    first error reduction method to enable useful finite QA. We predict that this will be achieved for 
    the fundamental task of $\ome$, namely,
    extracting accurate expectation value estimations from 
    ideal (noise free) generic
    quantum circuits. We expect this milestone to be achieved as soon
    as hardware platforms reach $50$ qubits with an average two-qubit
    gate fidelity of $99.9\%$. We note that these hardware
    requirements will certainly be achieved before quantum
    computations on $50$ logical qubits will be available using error
    correction alone (see Tab.~\ref{Tab: tab}); which means that finite QAs
    will be achieved first, and by a large margin, by EM before it
    is achieved by EC! 
    Our predictions are given in Fig.~\ref{fig: QESEM runtime}.   We additionally discuss the relation between the circuit geometry (shape of the active volume) and the relative run time of EM compared to classical simulation, providing geometrical guidelines for the construction of early finite QA demonstrations. 
\item {\bf EM, combined with EC, will continue to provide large circuit volume boosts even in the fault tolerant EC era}
  In Sec.~\ref{Sec: EMEC} we turn to the important discussion 
    of the role of quantum EM when fault-tolerant EC
    becomes available. It is a common misconception that 
    once EC is available to end users, there will not be any need 
    in other error reduction methods. We explain why this is indeed a misconception: even when fault-tolerant
    EC is available to end users, the fact that the number of high
    quality qubits is likely to remain limited in the foreseeable
    future, means that the logical error cannot be reduced to an
    arbitrary small value, and thus, fault tolerant quantum
    computations will also be limited in their volumes. We then
    explain how EM can provide large circuit volume boosts over
    those available by EC alone, making it extremely
    useful also when EC is available. This
    will be achieved by combining EC with EM, and will dramatically push the limits of circuit
    volumes provided by EC alone, multiplying
    available circuit volumes by orders of magnitude.
    The expected CVBs provided by this combination, which we call logical error mitigation, are given in Fig.~\ref{fig: EMEC}. This figure is based on simulations of  
    the prototypical Steane EC code combined with EM, and we provide in Tab.~\ref{Tab: tab} order of magnitude estimates for state of the art EC codes. Thus, we argue that EM is not just a ``transition technology'' for the NISQ era that bridges the gap until fault tolerant EC is achieved and becomes available to end-users, but rather an important resource that is applicable whenever the number of high-quality qubits cannot be scaled without a significant cost. Specifically, we believe that it is the combination of EM and EC that will first enable  reaching the milestone of
     MegaQuOp (one million logical gates)  \cite{riverlane2024qec, preskill2024megaquop}.
     
\item {\bf Conclusions regarding the central role of EM and the quantum advantage timeline}
We conclude our manuscript (Sec.~\ref{sec: summary}) with our future prospects, expressed as a QA timeline (Tab.~\ref{tab: time line}).
 This Table draws the implications of the main points in the paper, and highlights the central role that EM is expected to play in the future of 
the field.
\end{itemize}

\section{Asymptotic vs.\ finite quantum advantage (QA)}
\label{Sec: QuAlA}

The notion of {\it quantum algorithmic advantage} or in short {\it
quantum advantage} (QA) aims to capture the ability to solve
problems with a real quantum computer (QPU) using much fewer resources
than without it.
 
This notion is of major
interest to both industry and academia; however, it is not well defined in the literature and there are substantial
misconceptions about its meaning.  In this section we point out an
important distinction between different flavors of this notion: the
mathematical concept of {\it asymptotic QA}; the concept of finite
QA achieved in a {\it finite-size} real system - and the concept of
{\it scalable} QA which is in fact a finite sequence of finite QAs
whose behavior fits a certain scaling. 
In the current public discourse about QA, all these notions (and
perhaps others as well) are often used interchangeably, as if they are one and
the same. 
However, for the discussion regarding the usefulness of EM, the
question of which QA we are considering is crucial.  We devote this
section to clarifying these notions and the differences between
them. The definitions are not fully formal, and there are also many
possible variants; however the important points that distinguish
between the various flavors is what we care about here. 

\subsection{Asymptotic QA}
Let us first recall the well known mathematical notion of QA, which
we call \emph{asymptotic} QA. It considers the relation between the
computational complexity (complexity, in short) of solving a certain
problem using a classical computer versus a QPU (the
QPU may also interact with a classical computer in a hybrid
manner). The classical and quantum complexities are denoted $S_C(n)$
and $S_Q(n)$, respectively where $n$ is the size of the input to the
problem (or some parameter closely related to the input size that
characterizes the scaling of the problem).  These complexities are
often measured in the quantum computer science literature by the
number of elementary gates used in the computation; alternatively,
those can refer to the depth of the circuits used, which is an
abstraction of the run time that the computation requires.  

\begin{definition}[Asymptotic QA]\label{def:asympqa}
Asymptotic QA on a given computational problem is achieved if
$\lim_{n \to \infty}\frac{S_Q(n)}{S_C(n)}=0$. If in addition
$S_C(n) \le \poly(S_Q(n),n)$ we say that the QA is (at most)
\emph{polynomial}. If $S_Q(n) \le \poly(\log(S_C(n)),n)$ we say
  that the asymptotic QA is (at least) \emph{exponential}. 
\end{definition}

It is widely agreed by the scientific
community that asymptotic QA holds for a variety of problems \cite{quantum_algorithm_zoo}, in
particular, the factoring problem (due to Shor's quantum algorithm
\cite{shor_1999}) and the problem of simulating the dynamics of quantum systems \cite{lloyd1996universal}.\footnote{In computer science, providing a rigorous proof that quantum computation cannot be simulated in polynomial time by classical computers, would imply a resolution of the long standing seminal open problem of whether $P=PSPACE$. However it is widely conjectured that exponential asymptotic QA holds, and restricted proofs exist in the black box model for several oracle problems, e.g., Simon's problem \cite{simon_1997}.}

We note that Definition \ref{def:asympqa} can be instantiated where
the underlying quantum model is either an ideal QPU or a noisy
QPU, due to the quantum fault tolerance theorem
\cite{aharonov_1997, kitaev_1997,
knill_laflamme_1996,preskillaliferis}. Therefore, the asymptotic
notion is robust to the noise in the model, as long as the noise rate is below the fault tolerance threshold,  and the number of physical qubits with this noise rate is unlimited.  

\subsection{Finite QA} 
A major question that concerns the quantum community is whether and
when we will see experimental demonstrations of QAs for a variety of
useful problems.  We thus need to formalize notions of QA that are
closer to experimental realizations. In particular, this requires
quantifying the physical resources that are used by real physical
quantum and classical computers. Indeed, the complexity measures
mentioned above do not seem to directly correspond to a physically
measurable quantity of interest in the real world. 

 Before we discuss the appropriate complexity
measures that are more adequate to experimental settings, we first
clarify the context in which such experiments might take place. We
consider two competitors attempting to solve a computational
problem: one with access only to a classical computer, and the other
with access to a quantum and a classical computer combined in a
hybrid manner. To compare the physical resources
required by the two competitors for solving a certain computational
problem, we need to put their resources on the same scale. This is a
challenging task since the resources used by quantum and classical
computational devices are very different in character, and strongly
depend on the particulars of the hardware.  We will return to this
question at the end of this section, but for now presume that we can
quantify the {\it cost} for solving a given computational problem
using the best available quantum and classical hardware and software
as $\qcost = \qcost(n)$, ($n$ being the aforementioned scaling
parameter of the problem) and its cost using the best available
classical hardware and software as $\ccost=\ccost(n)$. Importantly,
these definitions consider the best {\it available} hardware and
software, rather than the best {\it possible} as in the asymptotic
definition. This is because when we discuss experimental
demonstrations of QA, we attempt to capture a snapshot of the added
value of quantum computers over classical at a specific point in
time, rather than provide a prediction for idealized future
hardware. The latter question is in the regime of computational
complexity theory, while here we are attempting to formally address
the question of achieving QA on existing physical devices, at a
given point in time.  

We first define QA in a non-asymptotic setting, i.e.\ \emph{finite}
QA which considers a fixed value of $n$. It is somewhat harder to
make a precise definition in this context, but we can consider the
rather informal one below.
\begin{definition}[Finite QA]\label{def:finqa}
For a given computational problem $\cal L$, let $I_n$ be the set of inputs to $\cal L$ of size $n$.  Finite QA for $\cal L$ on inputs of size $n$ is
  achieved if $\qcost(n) \ll \ccost(n)$, 
  where $\qcost(n)$ is the expected  cost of the quantum algorithm over the inputs in 
  $I_n$, and similarly $\ccost(n)$ is the expected cost of the classical algorithm over the inputs in $I_n$.  
\end{definition}
The reason for requiring that $\qcost(n) \ll \ccost(n)$ and not just
$\qcost(n) < \ccost(n)$ is that the evaluation of the costs for the
best currently available computers involves a large amount of
arbitrariness, and may change from one day to the next. Indeed a
robust definition will require a large concrete gap between the two
that will be agreed upon universally. 

We note that Definition~\ref{def:finqa} does not specify a procedure for the jury in a QA competition to \emph{verify} that finite QA had been achieved; this may be a challenging task on its own.
One may imagine a competition attempting to demonstrate finite QA for some problem, for some given fixed $n$. How would that work? The competition may  
involve asking the competitors to run their respective algorithms on inputs chosen from some distribution over $I_n$ for the fixed $n$, and the judge would then need to determine first whether the algorithms indeed output correct results, and secondly, the judge would also need to estimate the average costs and determine whether indeed $\qcost(n)\ll \ccost(n)$.

\begin{remark}[{The size of $I_n$}] \label{re:single1} 
We note that it is desirable to choose
$I_n$ to be a large set, rather than a single instance or just a few. The reason is that in practice, if the competition is defined to be on a single instance, both players will fine-tune their apparatus and algorithm to work optimally on that specific instance. Demonstrating an advantage for that particular instance will then not be indicative regarding performance on any other instance. Presumably, the goal of the QA competition is to provide a proof of principle that the QPU can (or cannot) achieve the advantage not just for the particular instance on which it was tested, but also more generally, for some interesting class of instances.\footnote{
In the theory of computer science this question is related to the question of whether or not the resources invested in preprocessing are limited or not; more formally, this can be viewed as the question of whether we are working in the uniform or non-uniform circuit model.}
\end{remark}

While there are still debates about particular experiments, it is generally agreed in the community that finite QA had already been demonstrated experimentally. Experiments claiming such finite QA include RCS and Boson Sampling, as well as other examples \cite{googlesupremacy1, wu2021strong, QAphotons, zhu2022quantum, Morvan2024, 1-bluvstein2024logical, decross2024computational}.\footnote{As an example, let us see how to view the RCS experiments of \cite{googlesupremacy1, Morvan2024} as instances of  Def.~\ref{def:finqa}. In this case  $I_n$ is the ensemble of $n$-qubits circuits from which the random circuit is drawn. The sampling task is to output a sequence of $k$ $n$-bit strings $x_i$ such that the cross entropy benchmark fidelity, $F_{\text{XEB}}=\frac{2^n}{k}\sum_{i=1}^k |\langle 0^n|C|x_i\rangle|^2-1$,
is above a certain threshold.}
 As mentioned in the introduction, {\it useful} finite QA is yet to be demonstrated, and we discuss in Sec.~\ref{sec:eve} a leading candidate problem for this goal, namely, expectation value estimation.
\subsection{Scalable QA} 
The experimental demonstration of finite QA is a remarkable achievement of the quantum community, but it is still not close enough to the fantastic notion of exponential asymptotic QAs promised by the theoretical discoveries of quantum algorithms. One may consider a notion which is closer to asymptotic QA, in which  one could demonstrate a {\it sequence of finite QAs, over some range of $n$'s}, such that those finite QAs fit a certain scaling (polynomial or exponential) as a function of $n$.    Given that this is a real world notion, it is impossible to ``prove'' that this scaling is maintained for all $n$, as one would hope for in order to validate the mathematical asymptotic notion of QA -- however one may consider it a pretty convincing evidence if the  scaling seems to behave like an exponent or a polynomial in some large regime of $n$'s. 
One can refer to this type of QA as 
{\it scalable finite QA}. Achieving   experimental demonstration of scalable QA is a major goal of the quantum community; we do not consider this notion of QA further in this exposition (for recent progress on scalable QA see, e.g., \cite{lidarscalable}).

\subsection{Asymptotic QA vs. Finite QA}

Despite the evident difference between finite and asymptotic QA, there is often a confusion in the literature regarding which of these notions of QA is appropriate to use in  different contexts. We make several important remarks regarding the relations between these two notions. 

\begin{remark}[Does Asymptotic QA imply finite QA, or vice versa?]\label{re:notimply} We note that  asymptotic QA and finite QA are in general incomparable, and it is possible that one holds but not the other, for a given problem. For example consider a problem for which we believe that asymptotic QA holds (e.g., factoring). Mathematically, for any reasonable choice of definition for $\qcost(n)$ and $\ccost(n)$ (and their relation to 
$S_Q(n)$ and $S_C(n)$ of Def.~\ref{def:asympqa}), respectively, such asymptotic QA implies that for a large enough $n$ we will have $\qcost(n)\ll\ccost(n)$. Based on this, and using quantum fault tolerance, in principle one should be able to experimentally demonstrate finite QA, for some large enough finite $n$. However it might be that this requires working with numbers of qubits which are too large to reach due to some technological barriers, and hence finite QA would not be achievable in our physical world, at least  for this particular computational problem. On the other hand, it is conceivable that finite QA is achievable in our real world for some problem, for some fixed $n$, but on the other hand this problem does not exhibit asymptotic QA at all, 
and the finite QA is simply an artifact of some finite size phenomenon that does not scale. 
\end{remark}

Which of the two notions, finite QAs or asymptotic QAs, is more interesting or relevant for quantum computation? 
This is a difficult and possibly confusing question. Much of the interest in quantum computation is practical, and comes from the industry and applications side; it relies on the strong belief that QAs will be realizable in the {\it real} world, and in particular, that for certain sizes of inputs for problems of practical interest, the actual relevant measure of costs on a QPU, say the wall-clock run time, will be shorter than that required to achieve this task on a classical computer. Achieving finite QAs is thus a leading goal.
 However, since such finite QAs may be due to a ``finite size effect'' that may become irrelevant as the system sizes grow. It is theoretical asymptotic algorithmic results such as Shor's algorithm, namely {\it asymptotic}  QAs, that provide the scientific basis and motivation to seek for the experimental realization of finite QAs in the first place. 
As explained in Remark~\ref{re:notimply},  asymptotic QA implies that finite QAs for this problem are to be expected starting from some ``crossover point''. Indeed, humanity's current endeavour towards experimental demonstrations of QA implicitly assumes that the crossover point is within our reach, and that the required $n$ is not too large for our finite world. Whereas theoretical study does not guarantee that this is the case, we may draw inspiration from the technological ability to scale classical computing systems quite rapidly. As we argue in this manuscript, the important role of EM is to enable higher circuit volumes for a given hardware, thus reaching the crossover point earlier (see Sec.~\ref{Sec: Error Mitigation Provides}). 

This discussion reveals an inherent tension between the practical, finite world and the asymptotic approach of computational complexity; it is indeed rightfully confusing, and has been a constant source of misunderstandings in the literature and study of QAs.  
We believe that maintaining the distinction between asymptotic and finite QAs may significantly help avoid these misunderstandings.

\subsection{Quantifying  resources}\label{sec:quantifyingresources}  Before we move on, we remark on the question which we have allowed ourselves to evade so far in this section - how to quantify the realistic costs $\qcost$ and $\ccost$. 
We discuss several possibilities in broad strokes. 
One possibility is to adapt the notion of circuit size (the number of quantum/classical gates used in an execution) to a metric which is adequate for finite QA. Quantum and classical gates are not comparable. To make $\qcost, \ccost$ measure resources on the same scale, we can decide on a conversion factor between them. Such a conversion factor does not matter from an asymptotic point of view, so does not concern us when we discuss asymptotic QA. 
However there seem to be severe drawbacks to this approach. Firstly, this metric seems to be much more cumbersome to use in the context of finite QA, where the constants are crucial - especially when hybrid quantum computation making use of  classical computers is involved. 
More importantly, the circuit complexity metric of number of gates does not distinguish between time and space (i.e.\ number of bits/qubits being used), since it considers, essentially, only their product. However space and time might be very different in their importance for resource counting; and their importance within quantum and classical computation is also very different, a difference which is hard to assign a convergence rate to. 

One can define instead $\qcost, \ccost$ using wall-clock time (which roughly corresponds to another metric commonly used in computational complexity: the {\it depth} of the circuit, namely the number of its layers). 
This choice is in line with contemporary QA demonstrations, where the goal is to beat the best available classical super-computer in terms of run time, regardless of its size, energy consumption and other considerations.  

 Both above metrics ignore many resources, such as cost in dollars, energy consumption, economic costs, and more. For example, see  Ref.~\cite{markov2018quantum} for an estimation of the classical cost for simulating RCS in $\$$s, and  Ref.~\cite{googlesupremacy1} for energy consumption estimations.  
The number of qubits is another important resource, of course; see, e.g., Ref.~\cite{finegrained1} for a detailed exposition of qubit-number estimates. 
We note that no matter what metric is chosen, it can be mapped to a volume crossover point above which finite QA is reached w.r.t. this metric. 

We further note that in all these metrics we did not account for the resources that go into preprocessing; it does not seem realistic nor practical to take into account the resources invested by the different competitors in preprocessing, as it seems extremely difficult to put a measure on such efforts.\footnote{In the language of theoretical computer science, this means that we are working in the non-uniform model that allows infinite preprocessing.} However, to a certain extent, the requirement that $I_n$ is large (Remark~\ref{re:single1}) is a practical way to significantly reduce the importance of preprocessing performed by the competitors, and prevent it from making the competition meaningless. Once $I_n$ is large enough, the competitors would have to put an enormous amount of effort to fine tune their systems so that the algorithm will work for all (or most of) these inputs.

\section{Expectation value estimation\label{sec:eve}}

To further discuss finite and asymptotic QA, we focus on a concrete and ubiquitous problem, which is of great importance in quantum physics and chemistry, as well as in a variety of quantum algorithms, namely, the \emph{expectation value estimation} ($\ome$) problem.  This is the problem of estimating the expected value of some observable in a pure quantum state that is generated in some prescribed manner.

Achieving finite QA for $\ome$ is fundamentally important, as it would imply that an existing QPU is better at approximating ideal (noise-free) quantum computation than classical simulation methods run on HPC hardware.\footnote{This should be contrasted with finite QA for RCS \cite{googlesupremacy1}, which may be viewed as implying that an existing QPU is better at approximating \textit{noisy} quantum computation than classical HPC. Indeed, in RCS it is sufficient to sample from a distribution whose distance from the output distribution of an ideal quantum circuit can be as large as $1-F$, where $F$ is very small - it is the experimental \textit{circuit} fidelity (e.g., $F\sim 10^{-3}-10^{-2}$ \cite{googlesupremacy1, Morvan2024}). The large error bound $1-F\lesssim1$ should be contrasted with the small additive error $\epsilon\ll1$ in $\ome$. \label{Foot: RCS 2}} 
Practically, we expect that a QPU capable of demonstrating such finite QA will be useful for two purposes. First, such a machine will allow for the otherwise-impossible simulation of important problems in quantum many-body physics, and in particular condensed matter physics, e.g., Floquet and Hamiltonian time evolution \cite{Ayral2023}. 
Secondly, this may have an important impact on researchers working on {\it designing} quantum algorithms and adapting them to specific applications. For lack of an alternative, researchers currently working on quantum algorithm design make extensive use of classical HPC to simulate their quantum algorithm's performance \cite{Intel2020}, and are thus very limited in the sizes of the instances they can study. The ability to solve the $\ome$ problem for larger and more general input quantum circuits would allow those researchers much more effective exploration of the behavior of their algorithms.  

The important $\ome$ problem lends itself naturally to EM methods, and thus it will be very convenient to focus on it in the remainder of this paper.    
We remark that one may consider other computational problems as a basis for finite QA, such as deterministic problems like Factoring \cite{shor_1999}, sampling problems like RCS \cite{googlesupremacy1},  Boson-Sampling and IQP  \cite{samplingreview} and even more generally interactive protocols for computational tasks such as certifiable randomness \cite{BCMVV18}.  While EM methods are commonly referred to as appropriate mainly for $\ome$, there is an ongoing effort towards the application of EM to sampling problems (e.g., \cite{liu2025quantum}).

\subsection{Problem statement}

We define the problem of expectation value estimation in the realistic setting where we are interested in circuits and observables taken from some restricted  prespecified family. Hence we consider a fixed family of sets $I=\{I_n\}_n$ where for each $n$, $I_n$ is a set of pairs $(C,O)$  with $C$ a circuit acting on $n$ qubits and $O$ a tensor product observable acting on its output. We require that there is an a-priori polynomial bound on the  size of the circuits, as a function of $n$ (formally, that there is a polynomial $p$, so that all circuits $C$ in $I_n$ have size at most $p(n)$.)

\begin{definition}[Expectation Value Estimation ($\ome$)]\label{def:ome}
The input to the expectation value estimation problem $\omeco$ (we omit the parameter $I$ when it is clear from the context) is a pair $(C,O)\in I_n$ for some $n$, comprising of a description of a quantum circuit $C$ acting on $n$ qubits and a description of a tensor product observable $O$ over the output domain of $C$. The output $\ome(C,O)$ is an estimation of the expected value of $O$ over the state $C|0^n\rangle=|C(0^n)\rangle$ derived by  applying $C$ to the all $0$ input state, namely, $\langle O \rangle_{C}:= \langle C(0^n)|O|C(0^n)\rangle$. 
We say that an algorithm $\cA$ $\epsilon$-approximates $\ome(C,O)$ if the input pair $(C,O)$
the mean-square error in the  approximation
(averaged over the randomness in the algorithm) is bounded by $\epsilon$:
\begin{align}
    \sqrt{\bbE_{\substack{\cA}}\big[\abs{\cA(C,O)
    -\langle O \rangle_{C}}^2\big]} \le \epsilon=\epsilon(n)~.
\end{align}
With the convention that $O$ is normalized in operator norm, $\|O\|_{op}= 1$, we consider a small $\epsilon\ll1$, which may be a function of $n$. Note that the value $\langle O \rangle_{C}$ is a fixed number given the pair $(C,O)$, and  the expectation $\mathbb{E}_{\cA}$ is taken over the randomness in the algorithm $\cA$, due to classical coin flips, quantum measurements, or both. 
The problem of $\epsilon$-approximating $\omeco$ is denoted $\omeeco$. 
\end{definition}

\begin{remark} \textit{(Variations of $\ome$)}
The above definition considers  \emph{worst-case} accuracy, requiring that $\cA$ performs well on the entire set $I_n$. We may also consider average case versions, where $\epsilon$ corresponds to the mean-squared-error over $I_n$.
We also note that in the above definition, we allow only the all-zero state as an input to the circuit. Indeed, this is how circuits are executed in the lab in current implementations. In a variety of settings it is natural to consider all $2^n$ inputs to the circuit; this is equivalent to considering $2^n$ circuits with the all zero input, where the circuits are derived from the original one by adding a layer of single qubit gates at the beginning, which change the all-zero input to some other input string. 
\end{remark}

\subsection{Active circuit volume \label{Sec: active volume}}

In the remainder of this paper, in particular Sec.~\ref{Sec: EM alone cannot} and \ref{Sec: Error Mitigation Provides}, we will study the potential for both asymptotic and finite QA in $\evecob$, by comparing the run time of EM protocols with that of classical simulation methods. Each instance of $\evecob$ is defined by the circuit and observable $(C,O)$, and one may naively expect the hardness of each instance to be controlled by the \textit{total} number of qubits on which $C$ is defined, and by the \textit{total} volume of $C$, measured as e.g., the total number of two-qubit gates in the circuit. These, however, are naive metrics for the circuit size. A more refined convenient notion is the {\it circuit active volume}, which takes into account also the observable $O$ and the accuracy 
$\epsilon$:    
\begin{definition}[Active circuit volume]\label{def: active volume}
Given an instance $(C,O)$ of $\evecob$, its active circuit volume $V$ is given by the size of the minimal subset of 2-qubit gates in $C$, such that arbitrary modifications to the gates in the complement of the subset lead to a negligible change of $\left<O\right>_{C}$ relative to $\epsilon$. That is,
\begin{align}
    V=\min\left\{|S|:\ S\subset C,\ \forall \tilde{S^c}\    |\left<O\right>_{(C\setminus S^c)\cup \tilde{S^c}}-\left<O\right>_{C} |\ll \epsilon\right\},\label{Eq: active volume}
\end{align}
where $S^c$ is the complement of a subset $S\subset C$, and $\tilde{S^c}$ is any modification to the complement. 
\end{definition}

\begin{remark} {\it (Allowed modifications)}
We intentionally leave vague the precise details of the allowed `arbitrary modification' in Definition \ref{def: active volume}, as different conventions can be appropriate for different applications. For the purpose of this paper, it will suffice to consider the replacement of each $k$-qubit unitary gate in the complement by an arbitrary $k$-qubit unitary.\footnote{The unitary replacement  captures both the {\it removal} of gates outside of $V$, as well as the addition of errors outside of $V$, with commonly used error models (e.g., local Pauli channels), corresponding to independent probability distributions over $k$-qubit unitaries occurring after each $k$-qubit gate.} 
\end{remark}

By a small abuse of notation, we sometime refer to $V$ as a subset of 2-qubit gates, and not just as the size of a subset. In the remainder of this paper, whenever we use the shorthand `active volume', `circuit volume', or simply `volume', we always refer to active circuit volume, unless stated otherwise.

The notion of active circuit volume captures, in some sense, the ``relevant'' volume for the problem; intuitively, one 
doesn't care about gates outside of it. 
Classical algorithms for $\evecob$ may benefit in run time from the removal of gates outside of $V$.  Quantum-classical algorithms (such as EM), may alternatively ignore errors occurring outside of $V$. Note that algorithms that require the active volume to be a-priori identified in order to benefit from it must incorporate an \text{active volume identification} subroutine \cite{QESEMpaper}, and their performance will be affected by the performance of this subroutine. 

Finally, we also restrict attention to the ``relevant'' subset of qubits: 

\begin{remark} {\it (Active qubits)}\label{re:activequbits} 
The active qubits are those qubits acted upon by any of the gates in the active volume. 
In the remainder of the paper we make the assumption that all $n$ qubits in the circuit are indeed active. Otherwise some qubits are not acted upon by any 2-qubit gate (after the removal of all gates outside of $V$) and in principle can be removed from the circuit. 
The number $n$ thus represents the `number of active qubits' in the circuit $C$, with respect to the observable $O$ and accuracy $\epsilon$, and in particular, satisfies $n\leq V$.  
\end{remark}

\section{Evidence against asymptotic QA with EM}
\label{Sec: EM alone cannot}

It is long known that in order to handle the effect of the most disruptive type of errors in quantum computers, namely non-unitary or dissipative errors, one must constantly remove the entropy accumulated in the quantum system (e.g., \cite{aharonov_1997,preskillFT, quantumrefrigerator}). 
Quantum error correction (EC) does this by employing a large overhead in qubit numbers due to (i) the encoding of logical qubits onto physical qubits, and (ii) the addition of measurement qubits, which are periodically reset or replaced between EC cycles.
In contrast to EC, error mitigation (EM) is a collective name for a family of methods to reduce the effect of errors in quantum computation without introducing fresh qubits during the computation, and importantly, while avoiding the large qubit overheads of EC. To do this, EM replaces
the execution of a given target circuit by multiple noisy circuit executions; the results of which are post-processed to provide an estimate for the error-free output of the target circuit. The fact that EM has little or no qubit overhead is a great benefit of EM over EC; however, this benefit comes at a painful price, which is the subject of this section. 

\vspace{5pt}

\subsection{Evidence for exponential sampling overhead of EM \label{Sec: evidence for exp}} From its inception, it was understood that EM probably cannot lead to exponential QA. The reason is that, starting with the development of the first EM protocols \cite{bravyitemmegambetta, li2017efficient, PhysRevX.8.031027}, and continuing in subsequent extensions and protocols \cite{emreview,PhysRevA.104.052607,cai2021multi, filippov2024scalability}, it was observed that, in order to produce an accurate output, the proposed protocols generically require a number of shots $M$ which is exponential in the total infidelity affecting the measured observable: 
\begin{equation}
\label{eq:emscaling}
M\ge \frac{1}{\epsilon^2}e^{\lambda \gamma V},
\end{equation}
where $\epsilon$ is the required statistical error, $\gamma$ is the 2-qubit (average) gate infidelity, $V$ is the active circuit volume (Definition \ref{def: active volume}). The product $\gamma V$ is the total infidelity within the active volume. The parameter $\lambda$ appearing in the exponent may be referred to as the `blow-up rate' of the specific EM method being used, and in the 
above described cases, it was found to be a constant (independent of $n$). 
For example, in EM based on QP distributions \cite{bravyitemmegambetta}, it has been argued that $\lambda\geq 4$, with the lower bound achieved in e.g., commonly-used constructions for Pauli channels \cite{xiong2020sampling, ferracin2024efficiently, van2023probabilistic}. 
As another example, in exponential ZNE (i.e.\ ZNE with an exponential extrapolation function), it has been argued that $\lambda \geq 2$, with the lower bound achieved by the two-point scheme described in \cite{filippov2024scalability}.\footnote{There are a few subtleties involved in accurately determining the shot overhead of EM protocols. First, there are generically $O(\gamma^2 V)$ corrections in the exponent. Additionally, there may be multiplicative factors which are polynomial in $\gamma V$. Finally, we work with the active circuit volume as it is an intrinsic property of instances of the problem $\evecob$. However, the shot overhead of some EM protocols (e.g., exponential ZNE) depends on the \textit{effective circuit volume}, $V_{eff}=\gamma^{-1}\log(\left<O\right>_{ideal}/\left<O\right>_{noisy})$, which corresponds not to a subset of gates in the circuit, but to the decay of the expectation value due to noise. The latter is generically proportional to the active volume (with an $n$-independent proportionality constant), but the two can scale differently with $n$ in some cases (see e.g., Ref.~\cite{granet2024dilution}). 
As another example, for EM based on QP distributions, mitigation is usually performed locally, and the relevant volume corresponds to the gates (or gate errors) one chooses to mitigate to insure a negligible bias, see \cite{tran2023locality, eddins2024lightcone, QESEMpaper}. This is closely related, but not identical, to the active volume. Since Eq.~\eqref{eq:emscaling} is expressed in terms of the active volume, in methods in which other volumes are relevant, the difference is swallowed into $\lambda$.
\label{foot: 1}
}

The above case by case study seems to suggest that a lower bound in the form 
of Eq.~\eqref{eq:emscaling} may hold 
more generally; the following toy example may give some intuition for why that might be true. 
Consider the global $n$-qubit depolarizing error model, where 
a channel $D_q:\rho\mapsto (1-q)\rho+q\text{tr}(\rho)I/2^{n}$, with depolarizing probability $q\in[0,1]$, acts after each of the $V$ ideal unitary gates in the circuit. Since $D_q$ commutes with all unitary gates, all $V$ copies of $D_q$ can be pulled to the end of the circuit, where they contract a trace-less observable $O$ as $O\mapsto (1-q)^V O$. To mitigate this effect, we may consider the following toy EM protocol. The protocol estimates the noisy expectation value $\left<O\right>_{noisy}$ by running $M$ shots on a QPU leading to a statistical error $\leq 1/\sqrt{M}$. Assuming perfect knowledge of $q$, the protocol then needs to "stretch" the result by a factor $(1-q)^{-V}$, producing an unbiased estimate of $\left<O\right>_{ideal}$. However, unfortunately, the statistical error is also stretched, to $(1-q)^{-V}/\sqrt{M}$. Thus, $M\geq \epsilon^{-2}(1-q)^{-2V}$ shots must be used to ensure a statistical estimation error at most  $\epsilon$. Using the infidelity $\gamma=(1-4^{-n})q$  of the depolarizing channel,\footnote{We work with the `entanglement infidelity', as opposed to the `average gate infidelity' \cite{PRXQuantum.2.010201}.} this can be written as $M\geq \epsilon^{-2}e^{2\gamma V/(1-4^{-n})+O(\gamma^2 V)}$, which demonstrates Eq.~\eqref{eq:emscaling}, with blowup rate $\lambda=2/(1-4^{-n})\in(2,8/3]$.
Though this toy model is over-simplified,
it seems to capture an essential point - the need to effectively stretch the output by an exponentially large factor, in order to combat the effect of exponential decay due to noise. This stretch increases the variance of the output exponentially, implying a similar increase in the number of shots, which is needed in order to meet the required statistical error with such a large variance.

Another toy EM protocol that one may consider is given by an idealized post-selection that manages to reject all erred shots \cite{emreview}. 
In this case only a fraction $(1-q)^V$ of shots is maintained, leading to a sampling overhead $(1-q)^{-V}$, which again reproduces Eq.~\eqref{eq:emscaling}, now with $\lambda=1/(1-4^{-n})$. 

The above arguments have led researchers to seek for a lower bound along the lines of Eq.~\eqref{eq:emscaling} that may hold for {\it any} EM method. 
We survey those below. 

\subsection{Towards general exponential lower bounds on the sampling overhead}\label{sec:rigLB} 

There are a number of results providing lower bounds on the required shot number that hold for \textit{general}, rather than specific, EM methods \cite{takagi2022fundamental, Takagi2023universal, Tsubouchi2023universal, Quek_Eisert}. 

Before we survey these results, we note that known EM protocols \cite{emreview}, as well as the lower bounds of
Ref.~\cite{takagi2022fundamental, Takagi2023universal, Tsubouchi2023universal, Quek_Eisert}, show that one may avoid the exponential scaling of Eq.~\eqref{eq:emscaling} only at the expense of a significant {\it bias}, that is, a large estimation error that persists even with infinitely many shots.\footnote{This is the behavior of e.g., polynomial ZNE (with a low-degree polynomial) which requires a small sampling overhead, but can lead to significant biases unless $\gamma V\ll 1$ \cite{bravyitemmegambetta}.} Such biases are notoriously difficult to bound, often leading to EM protocols with no accuracy guarantee; a situation which  is especially problematic in the context of computational problems that are not efficiently verifiable. In the remainder of this paper we restrict attention to EM protocols that produce a negligible bias $b\ll\epsilon $ for the circuits in question.\footnote{There's no problem allowing for a small, but non-negligible, bias, say $b<\epsilon/2$. This effectively replaces $\epsilon$ by $\epsilon/2$ in Eq.~\eqref{eq:emscaling}, as well as in   Eq.~\eqref{Eq: total error EM} below. In fact, trading bias for error bar in a controlled manner is an important aspect of designing practical EM protocols.} This includes (i) unbiased EM protocols, where the bias is rigorously guaranteed to be negligible for any circuit and observable; and (ii) heuristic EM protocols, where no such guarantee exists, but nevertheless the bias is still negligible in particular cases.\footnote{An example for the first type, namely unbiased protocols, is EM protocols based on QP distributions that rely on a good-enough characterization of error channel. An example for heuristic EM protocols in which the bias is negligible for specific cases, is exponential ZNE, where the bias vanishes  for Clifford circuits with Pauli errors, but not generally.} 
 
We now survey the results of Ref.~\cite{takagi2022fundamental, Takagi2023universal, Tsubouchi2023universal, Quek_Eisert}, for EM protocols with negligible bias. These results are derived under a general definition of EM protocols provided in Ref.~\cite{takagi2022fundamental}, the essence of which is the exclusion of adaptive quantum  operations. In a nutshell, 
a lower bound along the lines of Eq.~\eqref{eq:emscaling}  (with $\lambda$ a constant independent of $n$) is not known to hold for all circuits and observables; in fact, recent work suggests the existence of fine-tuned counter examples. Nevertheless, Eq.~\eqref{eq:emscaling} does seem to hold very generally. 

To the best of our knowledge, the strongest known lower bounds on the number of shots $M$ that hold for any circuit, observable and EM protocol, are exponential in the circuit depth $D$ as opposed to the volume $V$ \cite{Takagi2023universal,Tsubouchi2023universal, Quek_Eisert}. In particular, assuming (i) a local-depolarizing error model, (ii) a trace-less observable $O$, and (iii) a large-enough ideal expectation value $|\left<O\right>_{ideal}|>2\epsilon$; the number of shots $M$ required by any EM protocol (with negligible bias) to reach a statistical error at most $\epsilon$ in estimating the expectation value $\langle O \rangle_C$ for any observable $O$ and circuit $C$, satisfies  
\begin{equation}
\label{eq:emscalingD}
M\geq \frac{1}{\epsilon^2}e^{\lambda \gamma D}. 
\end{equation}
This result is based on the known decay of the output state towards the maximally mixed state, which, assuming a local depolarizing model, decreases exponentially with $\gamma D$ \cite{ABIN96, GaoDuan2018, Deshpande2022, 10.1063/1.4804995}. 

The works \cite{Tsubouchi2023universal, Quek_Eisert} further demonstrated the existence of circuits for which the number of shots $M$ for any EM protocol must be exponential in the total circuit volume $nD/2\geq V$ and not only in $D$.  By limiting their circuit construction to the connectivity of a $d$-dimensional lattice, Reference \cite{Quek_Eisert} further showed compelling evidence that, for the above circuits, it is the active volume (or `light cone') that controls the shot overhead of EM, as opposed to the total circuit volume. We describe these results, which are closely related to the above global depolarizing toy model, in more detail in Appendix \ref{app:lowerbounds}. 
 
Despite the above results,  
compelling evidence that  Eq.~\eqref{eq:emscaling} is violated for a different family of circuits, for restricted input states and observables, 
was recently presented in  Ref.~\cite{granet2024dilution}. 
In the examples discussed in~\cite{granet2024dilution} 
the behavior of the shot overhead is exponential in $D$ and {\it not} in $V$. 
The authors consider certain 2d Trotter circuits, where $n\sim D^2$ and $V\sim nD\sim D^3$, and observe that for very particular initial states and observables, the decay of the observable due to errors can scale with depth, as $e^{-\Omega(\gamma D)}$, rather than with volume.  Accordingly, EM methods with a sampling overhead of only $e^{O(\gamma D)}$ may in principle be used in these particular cases. 

We conclude that while Eq.~\eqref{eq:emscaling} seems to hold very broadly, it fails to 
hold in all cases with $V$ in the exponent; it does hold generally with $D$ in place of $V$. This is a strong lower bound only for circuits whose depth $D$ grows super-logarithmically with $n$.  We now proceed to  
deduce the negative implications of such exponential lower bounds on the prospects for \textit{asymptotic} QA with EM. 

{~}

\subsection{EM with exponential overheads cannot enable exponential asymptotic QA}  It is easy to see that in cases where the EM shot overhead scales according to Eq.~\eqref{eq:emscaling},  exponential asymptotic QA is ruled out. Furthermore, for {\it any}  negligibly-biased EM protocol, applied to any circuit and observable with $D>n$,
exponential asymptotic QA is ruled out due to the general lower bound in Eq.~\eqref{eq:emscalingD}. 

Let $T_c$ be the time required to classically solve the expectation value estimation problem $\evecob$ using the best available algorithm, and $T_{EM}$ the time to  solve it on a quantum processor using an EM protocol; we claim

\begin{corollary}\label{cor}{\bf (Exponentially scaling EM cannot provide exponential asymptotic QA)} Consider an EM protocol, and a family of circuits and observables for which one of the following two conditions hold: (i) The exponential lower bound in $V$, Eq.~\eqref{eq:emscaling}, is satisfied  with $\lambda \gamma\geq \overline{\gamma}$ for some $n$-independent bound $\overline{\gamma}\in(0,1]$, or (ii) the depths of the circuits satisfy $D\geq n$, 
and we also have that the conditions for the exponential lower bound in $D$, Eq.~\eqref{eq:emscalingD}, hold; namely, the circuit is affected by a local depolarizing model, the observable $O$ is trace-less, and its ideal expectation value is non-negligible,  $|\left<O\right>_{ideal}|>2\epsilon$. In either case we have that the classical run time is polynomially bounded by the EM run time:   
\begin{equation}
T_c=O(T_{EM}^{1/\overline{\gamma}}). \label{Eq: T_c <> T_EM}
\end{equation}
Since the classical run time $T_c$  is at most {\it polynomial} (with degree $1/\overline{\gamma}$) in the EM run time $T_{EM}$, any asymptotic QA with EM is at most polynomial. In particular, there is no exponential asymptotic QA of EM over classical computation. 
\end{corollary}

{\it Proof:} 
We prove the first case, the second one is similar. 
First, we translate  
the shot number in
Eq.~\eqref{eq:emscaling} to a lower bound on the run time of EM for a
quantum circuit of depth $D$,
\begin{equation}
  T_{EM}=\Omega(DM)=\Omega \left(\frac{1}{\epsilon^2}D 
    e^{\lambda \gamma V}\right).\label{Eq: T_EM}
\end{equation}
Note that we use here the total circuit depth over all quantum circuits in the EM protocol to bound the run time,\footnote{We comment here on the relation of this simple bound  to realistic experimental settings. In experiments, one can bound $T_{EM}\geq t_s M$, where $t_s$ is the shot time, which can itself be bounded as $t_s\geq t_l D$, where $t_l$ is the time per circuit layer. The lower
bound for the shot time assumes that the relevant QPU can
parallelize all gates appearing in each circuit layer, and
accounts only for the time to implement these layers. Limited
parallelization, or additional delays due to e.g., state preparation
and measurement, will increase the shot time. Similarly, the
bound $T_{EM}\geq t_s M$ accounts only for the physical time to
implement the quantum operations in each shot. In practice, current
QPUs suffer significant additional delays (`circuit time'), due to the loading of
circuit instructions onto classical control electronics.
\label{foot: 2}} without including the classical run time which will of course only increase the lower bound.

We can now compare this run time to that of the
best available classical simulation. The run time of that would be at most that of the simplest classical simulation of quantum circuits, namely, state-vector simulation, which would require 
\begin{equation}
  T_c= O(V2^n),\label{Eq: T_c}
\end{equation}
floating-point operations, for an ideal $n$-qubit quantum circuit including $V$  2-qubit gates; see details in Sec.~\ref{Sec: predictions}.
Further note that $V$ is bounded by the maximal number of 2-qubit gates in the circuit, $V\leq nD/2$, where $D$ is the circuit depth and $n/2$ is the maximal number of 2-qubit gates per layer.
Equations \eqref{Eq: T_EM} and \eqref{Eq: T_c} then together imply the desired relation: 
Using
$\epsilon<1$, $V\geq n$ (see Remark~\ref{re:activequbits}) and $e^{n}>n2^n$ (for, say, $n>10$),
the corollary follows from $\epsilon^{-2}De^{\lambda \gamma V} \geq
De^{\overline{\gamma} V}\geq De^{\overline{\gamma} n}> D(n2^n)^{ \overline{\gamma}} >
(Dn2^n)^{\overline{\gamma}}>(V2^n)^{\overline{\gamma}}$. $\square$

\begin{remark}
\textit{(Role of the lower bound $\overline{\gamma}$)} The requirement that $\lambda \gamma\geq \overline{\gamma}$ may be refined to separate $n$-independent lower bounds on $\gamma$ and $\lambda$. The first lower bound is a very realistic assumption, since there is no experimental indication for error rates decreasing with the number of qubits. In fact, error rates tend to increase with the number of qubits \cite{mckay2023benchmarking}, since maintaining high quality at scale is a more challenging engineering task.  The second lower bound, on $\lambda$, is  meant to exclude the known examples in which the sampling overhead is exponential in $D$ instead of $V$ (see Subsection \ref{sec:rigLB}). 
\end{remark}

As we expect Eq.~\eqref{eq:emscaling} to hold generically (with $\lambda$ bounded from below by a constant independent of $n$), we expect that, \textit{generically}, EM cannot provide exponential asymptotic QA.

\begin{remark}(What about polynomial QA?) 
Corollary \ref{cor} does not exclude polynomial QA using EM. However, we note that classical simulation becomes more efficient compared to EM (namely, the upper bound of Eq.~\eqref{Eq: T_c <> T_EM} becomes less tight), the larger $V$ is relative to $n$, such that polynomial asymptotic QA is only possible when $V=O(n)$, namely for constant depth circuits. 
In the noiseless setting, constant depth circuits are conjectured to provide exponential asymptotic QA for sampling problems, e.g., in the model of Instantaneous Quantum Polytime (IQP) \cite{bremnerIQP1, bremnerIQP2,bremnerIQP3, bremnerIQP4,IQP5}. We pose here as an open problem whether 
EM for expectation values of heavy weight tensor product observables measured at the end of constant depth quantum circuits may lead to polynomial asymptotic  QAs.
\end{remark} 

The results in this section still do not rule out exponential asymptotic QA using EM in special cases. We next describe a recent result which proves a no-go theorem in a more general setting.

\subsection{No-go for exponential
asymptotic QA} 

The assumptions stated in Corollary \ref{cor} (namely, that Eq.~\eqref{eq:emscaling} holds or that $D\geq n$) were needed because of the current limitations of rigorous lower bounds on the shot overhead of EM, as described in Sec.~\ref{sec:rigLB}. In this section we describe recent work that excludes exponential asymptotic QA without relying on such assumptions, and which holds for every EM protocol, circuit and observable, but only on average over input states \cite{Schuster_Yao}. The main result of Ref.~\cite{Schuster_Yao} is a classical algorithm for approximating expectation values in any noisy quantum circuit (up to error $\epsilon'$). The algorithm's run time is $O((1/\epsilon')^{\tilde{O}(1/\gamma^2)})$, which is polynomial in the required accuracy $\epsilon'$, with degree $\tilde{O}(1/\gamma^2)$; albeit a large degree, this is a polynomial algorithm. Notably, this performance guarantee  holds for {\it any} fixed circuit and observable, improving over previous work \cite{Aharonov2022} which provided a similar result but only on average over random circuits. Importantly, the accuracy guarantee of Ref.~\cite{Schuster_Yao} does not hold for any input to the quantum circuit, but only on average over computational basis input states. 

Ref.~\cite{Schuster_Yao} use their algorithm to provide limitations on any EM protocol  for the problem of $\evecob$, based on the following idea.  They use their classical algorithm for simulating noisy circuits, in order to simulate each of the $M$ shots of the EM protocol. In order to arrive at total error $\epsilon$ in the final estimation of the expectation value, the authors run their classical simulation algorithm for each of the $M$ shots with required accuracy $\epsilon'=O(\epsilon/M)$. This leads to a total classical run time 
\begin{align}
T_c=O((M/\epsilon)^{\tilde{O}(1/\gamma^2)})=O((T_{EM}/\epsilon)^{\tilde{O}(1/\gamma^2)}).  \label{Eq: Yao} 
\end{align}
Notably, a lower bound which is significantly weaker than  Eq.~\eqref{Eq: T_c <> T_EM}, but still a polynomial. 
Thus, whenever the number of shots $M$ in the EM protocol for $\evecob$ is polynomial, Reference \cite{Schuster_Yao} constructs a simulation of  $\evecob$ for the ideal quantum circuit which takes polynomial time. We stress that the guarantee for the error to be smaller than  $\epsilon$ does not hold for any input state, but only on average over all inputs. 

Thus, Reference \cite{Schuster_Yao} shows that EM alone cannot lead to exponential asymptotic QA (or in fact to any non-polynomial asymptotic QA), for the variant of the problem $\evecob$ in which $\epsilon$ is the root-mean-squared error over all $2^n$ input states (see Sec.~\ref{sec:eve}).\footnote{One may notice that we can now turn back to the first question discussed in this section, and ask whether the results of Ref.~\cite{Schuster_Yao} can be used to deduce the lower bound of Eq.~\eqref{Eq: T_EM} for generic EM protocols.
As mentioned above, there are specific counter examples  in which this bound does not hold \cite{granet2024dilution}; however, one might expect the lower bound to still hold generically. In  Appendix \ref{app:samplingLB}
we show that such a line of reasoning can indeed be applied, to yield a lower bound which in some cases is stronger than the exponential in $D$ lower bound, however it is weaker than  Eq.~\eqref{Eq: T_EM}.} 
We note that this no-go result still
does not completely rule out exponential QA using EM, due to this average over all inputs to the circuits involved. This leaves a small crack in the door for exponential asymptotic QA to be achievable using EM when restricted to a special small set of input states, that will not significantly affect the average over inputs; in line with the recent examples of Ref.~\cite{granet2024dilution}, suggesting a violation of Eq.~\eqref{eq:emscaling} with fine-tuned input states. We do not concern ourselves further with this 
highly specific case; the main message is that, generically, EM alone cannot enable exponential asymptotic QA. 

\vspace{5pt}

To summarize, we see that EM cannot on its own  enable generic exponential asymptotic QAs. In light of this, and the immense importance of exponential asymptotic QAs in motivating the whole field of quantum computation, one must ask -- why bother with EM at all?

\section{Useful finite QAs expected in the near future using EM \label{Sec: Error Mitigation Provides}}

We now explain how, despite the lessons from Sec.~\ref{Sec: EM alone cannot}, EM does in fact provide a viable path to QA. The point is exactly the distinction between exponential asymptotic QA, which Sec.~\ref{Sec: EM alone cannot} argues cannot be achieved with EM alone, and finite QA. We argue here that EM is an extremely promising path to finite QAs, and is in fact expected to enable such advantages in the very near future -- much earlier than EC. Furthermore, based on the same reasoning which we will outline in this section, we will see in Sec.~\ref{Sec: EMEC} that though EM cannot provide exponential QA on its own, it can in fact crucially help in bringing it much earlier than would be possible without EM, namely with EC alone.    

The starting point is a very important observation regarding the exponent in the lower bound on the sampling overhead of EM, Eq.~\eqref{eq:emscaling}. The power of  the exponent in this lower bound contains a very small constant - the infidelity per gate $\gamma$. Thus, though the sampling overhead indeed grows exponentially, it is a {\it very} mild exponential growth. Current 
values of $\gamma$ are in the range $10^{-3}-10^{-2}$, and may reach $10^{-4}$ within the next few years \cite{quantinuum_roadmap_2023, loschnauer2024scalable}. Closely related, the degree $1/\gamma$ of the polynomial connecting the complexity of classically simulating quantum circuits and the complexity of EM protocols (Corollary \ref{cor}), is very large, currently in the range $10^2-10^3$, and possibly $10^4$ in the near term. 
Thus, while the classical run time is indeed only polynomial in the quantum run time as is derived in Corollary \ref{cor}, this polynomial classical computation could still be tremendously expensive compared to the quantum one. The upshot is that though EM is not expected to 
lead to {\it exponential} QA (and in fact not even polynomial, except possibly for very special cases of constant depth circuits, as discussed at the end of  Sec.~\ref{Sec: EM alone cannot}) it does offer a very challenging fight, for shallow circuits, to the classical competition, due to this large power of $1/\gamma$. This is what leaves a slack for EM to provide very significant {\it finite} QAs.

In this section we clarify the resulting potential benefits of using EM in real world quantum computations, in two ways. First, in Sec.~\ref{Sec:CVB} we introduce an important notion, which we term ``circuit
volume boost'' (CVB), that can be used to quantify the benefit of using EM over not using it. In Sec.~\ref{Sec: predictions}, we
provide estimation of the finite QAs that can
be achieved using EM, for different fidelity values of the hardware. 

\subsection{Circuit Volume Boosts \label{Sec:CVB}}

We introduce a useful definition for comparing the performance of quantum computation with and without EM, which we call \textit{circuit volume boost} (or CVB in short). The CVB captures the factor by which one can increase the volume of quantum circuits being run using EM, compared to the circuit volumes available without EM. 
In fact, the CVB can be used for comparing the performance of any two methods for executing quantum circuits. 

Recall that the number of shots required to ensure accuracy $1-\epsilon$ in the ideal (noiseless) computation is $\epsilon^{-2}$, since in this case the error is entirely  statistical. In EM, we allow a larger shot budget. We find it convenient to write the allowed shot budget as a multiple of the number of shots required by the ideal computation for the same accuracy,  
\begin{equation}\label{eq:MandR}
M=R/\epsilon^2,
\end{equation} 
with $R>1$. We thus introduce the \textit{allowed shot overhead} $R:=M\epsilon^2$.
In the discussion below, we will study the behavior of 
EM for fixed values of the allowed shot overhead $R$. 
With this convention, we now define the CVB.

\begin{definition}{\bf Circuit volume boost} (CVB) 
Let $V_A(\epsilon,R)$ denote the maximal circuit volume that can be achieved with circuit execution method $A$, given an allowed shot overhead $R\geq1$, and while meeting a required output accuracy $1-\epsilon$.  Let  $V_B(\epsilon,R)$ denote the maximal volume possible with circuit execution method $B$, under the same constraints. The circuit volume boost of method $A$ over method $B$, given an allowed shot overhead $R$ and required accuracy $1-\epsilon$, is then given by 
\begin{equation} 
    CVB(\epsilon,R)=\frac{V_A(\epsilon,R)}{V_B(\epsilon,R)}.
\end{equation} 

\end{definition}

This gives a very clear quantification of the benefit of one method over another, in terms of the ratio between the active volumes that the two methods can achieve.  
We would now like to estimate the circuit volume boost provided by executing circuits using EM, compared to `bare' circuit execution, without EM.  
We first estimate the circuit volume $V_{bare}$ which can be run  without EM, then estimate $V_{EM}$, and then compare.  

{~}

\subsubsection{Estimating $V_{bare}$}
 We do this under the same global-depolarizing toy-model used in Sec.~\ref{Sec: EM alone cannot}, where $\left<O\right>_{noisy}=e^{-\gamma V}\left<O\right>_{ideal}$,\footnote{We suppress here the small differences between $(1-q)^V=e^{-\gamma V/(1-4^{-n})+O(\gamma^2 V)}$ and $e^{-\gamma V}$.}  and $\|O\|_{op}=1$. The bias in the bare computation is given by $b=|\left<O\right>_{ideal}-\left<O\right>_{noisy}|\leq 1-e^{-\gamma V}$, and the statistical error is $\leq 1/\sqrt{M}$. Using these bounds, we can ensure that the total error,  given by the sum of the statistical error and the bias,\footnote{One may similarly work with the mean squared error instead of the sum.} is at most $\epsilon$, if
\begin{align}
    \epsilon = (1-e^{-\gamma V_{bare}}) + 1/\sqrt{M}.
\end{align}
Solving for $V_{bare}$ and using Eq.~\eqref{eq:MandR} then gives 
\begin{align}\label{eq:bare}
    V_{bare}&=-\frac{1}{\gamma}\log\left(1-\epsilon+\frac{1}{\sqrt{M}}\right)\\
    &=-\frac{1}{\gamma}\log\left(1-\epsilon+\frac{\epsilon}{\sqrt{R}}\right)\nonumber\\
    &\sim \frac{\epsilon}{\gamma}\left(1-\frac{1}{\sqrt{R}}\right).\nonumber
\end{align}
Note that the bare execution requires a non-trivial allowed shot overhead $R>1$ to support a volume $V_{bare}>0$. The reason is that the systematic error takes up part of the total allowed error budget $\epsilon$, leaving `less room' for the statistical error. On the other hand, due to the bias, even if the number of shots $M$ is taken to infinity, there is a  hard upper bound on the available volume: \begin{equation}\label{eq:vbareest} V_{bare}\sim \epsilon/\gamma.\end{equation} This upper bound is particularly strict in the high accuracy (small $\epsilon$) regime.  Equation \eqref{eq:vbareest} is a good rule of thumb for the maximal circuit volume possible without EM. 

{~}

\subsubsection{Estimating $V_{EM}$}
Let us now compute the maximal active circuit volume $V_{EM}$ possible with EM. 
We assume here that there's no systematic error - this is the case, e.g., when performing EM based on QP distributions and assuming an exact characterization.  However, the statistical error is the bottle neck - following Eq.~\eqref{eq:emscaling}, the statistical error grows exponentially with the circuit volume, but with a small pre-factor $\gamma$, the infidelity per gate:  
\begin{equation}
\epsilon=0+\sqrt{e^{\lambda \gamma V_{EM}}/M}.\label{Eq: total error EM}
\end{equation}
Solving for $V_{EM}$  
we get,  
\begin{equation}
V_{EM}=\frac{1}{\gamma}\frac{\log(\epsilon^2M)}{\lambda}=\frac{1}{\gamma}\frac{\log R}{\lambda}.\label{Eq:V_EM}
\end{equation}
As opposed to $V_{bare}$, this is independent of $\epsilon$ but is only a function of $R$.  

{~}

\subsubsection{Estimating the circuit volume boost}
\begin{figure}[h]\label{fig:boost}
\begin{centering}
\includegraphics[width=0.5\columnwidth]{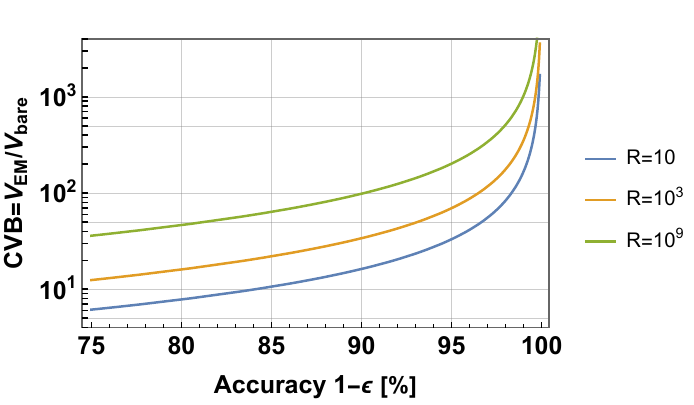}
\par\end{centering}
\caption{Circuit volume boost (CVB) due to unbiased error mitigation with blow up rate $\lambda=2$. As the allowed inaccuracy $\epsilon\rightarrow 0$, the maximal volume possible without EM vanishes, while the maximal volume possible with unbiased EM  is independent of $\epsilon$, given the same allowed shot overhead $R$ for both circuit execution methods. \label{fig: CVB}
}
\end{figure}

Given the above two estimates, Eq.~\eqref{eq:bare} and Eq.~\eqref{Eq:V_EM}, the CVB is given by 
\begin{align}
    CVB(\epsilon,R)=\frac{V_{EM}}{V_{bare}}\sim\frac{1}{\epsilon}\frac{\log R}{\lambda(1-1/\sqrt{R})}.\label{Eq: CVB}
\end{align}
Note that the CVB is independent of the infidelity $\gamma$! The function $\log(R)
/(1-1/\sqrt{R})$ depends only weakly on $R$, for example for allowed shot overhead $R$ in the range $[2,10^9]$ it takes values in the range $\approx[2,20]$.\footnote{Shot overheads as high as $10^9$ are reasonable in superconducting-qubit platforms, where gate times are of the order of tens of nanoseconds.}  As described in Sec.~\ref{Sec: EM alone cannot}, the blowup rate $\lambda$ is a constant of order $1$. Thus, an order of magnitude estimate for the CVB due to EM is 
\begin{align}\label{eq:cvbapprox}
    CVB(\epsilon,R)\sim \frac{1\text{--}10}{\epsilon}. 
\end{align}
 The important feature is that the $CVB$  scales like $\sim1/\epsilon$ (see Fig.~\ref{fig: CVB}) meaning that the higher the required accuracy is (smaller $\epsilon$), the larger the $CVB$. In particular, chemistry applications usually require reaching `chemical accuracy', which translates to $\epsilon\sim10^{-3}$ \cite{McArdle2020}. As an example, consider  e.g., $\lambda=2$ and a mild overhead $R=10$. This gives $CVB>10$ for $85\%$ accuracy; a $CVB> 100$ is obtained for $99\%$ accuracy; and a $CVB>1000$ is obtained for $99.9\%$ accuracy.  

\subsection{Predictions for finite QAs enabled 
using EM alone \label{Sec: predictions}} 
 
Given the impressive expected circuit volume boosts described in
Sec.~\ref{Sec:CVB}, what can be said about the expected actual
values of $V$, the volumes of the quantum circuits, to be enabled by
EM in the near future? 

In this subsection, we will argue that EM acting on devices with
$\gamma=10^{-3}$ will already enable achieving finite QA within
reasonable QPU time. This is the origin of our assertion in the
introduction that finite QA in $\ome$ (as defined in Sec.~\ref{Sec:
QuAlA}) is to be expected already in the very near future.  This
prediction relies on two claims: first, we provide estimates of the
QPU time for EM as a function of the active volumes with such
fidelities. According to these estimates, active volumes of over a
thousand (natively available on a given QPU) 2-qubit gates are
achievable in reasonable time with $\gamma=10^{-3}$. Secondly, we
claim that achieving active volume of over a thousand gates for
general quantum circuits suffices for finite QA in $\ome$. This is
done by showing that for generic circuits, such active volumes seem
to be well-beyond the reach of state of the art HPC. Most of this
subsection is devoted to explaining the analysis that supports this
latter claim. Figure \ref{fig: QESEM runtime} summarizes the
analysis of both the quantum and the classical run time as a
function of volume, and indicates where finite QA is expected for
two types of quantum hardware platforms: superconducting qubits and
trapped ions.\footnote{In drawing Fig.~\ref{fig: QESEM runtime}, we assume that both EM and classical simulation have access to the same active volume identification subroutine, and so $n$ and $V$ have the same meaning in the analysis of both.} Note that we consider here the ability of EM to execute generic quantum circuits of a given active volume
and geometry faster than the performance of classical HPC for the same active volume and geometry, as
marking the point of finite QA.\footnote{\label{foot:finiteQA} We stress, however, that designing finite QA demonstrations that can be considered as rigorously justified requires further work. In particular, to meet the formal requirements of Definition \ref{def:finqa}, one must choose a specific circuit family $I_n$ and demonstrate finite QA with respect to this family. The natural choice of defining $I_n$ to be the set of all quantum circuits of a given size consisting of native gates (for a given hardware), accompanied with some fixed observable, will likely not suffice, since random (or unstructured) circuit families tend to exhibit exponentially small expectation values with overwhelming probability, making the computational task trivial; see e.g., Ref.~\cite{larocca2024review}. Instead, it is required to construct a family of circuits $I_n$ containing a large number of circuits, such that  the variance of the ideal expectation values is non-negligible (in particular, it needs to be larger than the accuracy $\epsilon$ that is possible to achieve with EM in reasonable time); this is beyond the scope of the current paper. 
}

{~}

\subsubsection{Estimates of active volume achievable using EM}

\begin{figure}[h!]
\begin{centering}
\includegraphics[width=1\columnwidth]{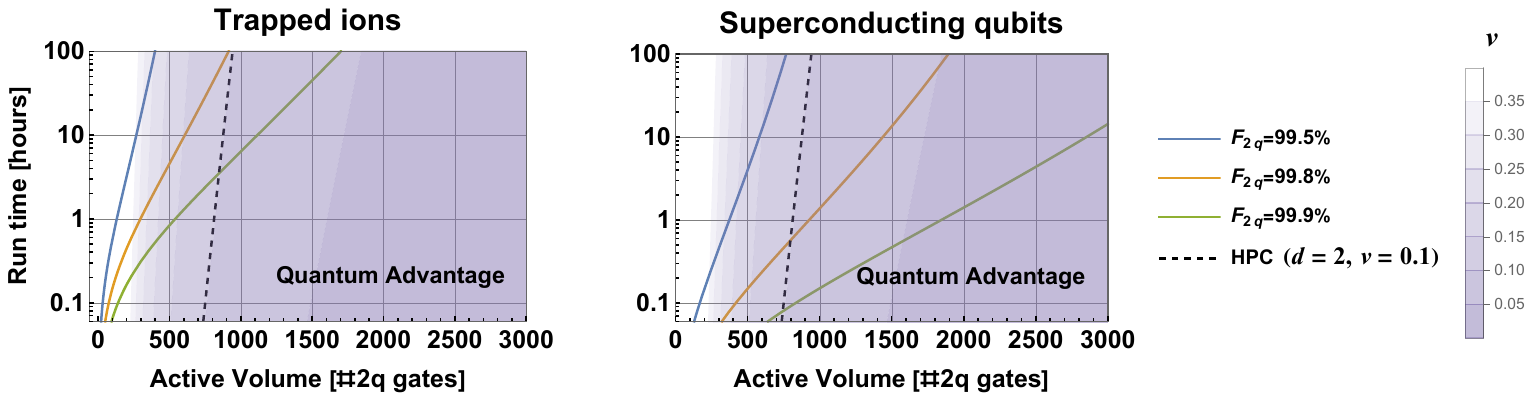}
\par\end{centering} \caption{ \label{fig: QESEM runtime} 
Predictions for finite QA with EM. Coloured lines indicate the QPU time as a function of active circuit volume, when using QESEM, an unbiased EM method introduced in Ref.~\cite{QESEMpaper}, on QPUs based on
 superconducting-qubits and trapped-ions,
 for different 2-qubit gate fidelities $F_{2q}=1-\gamma$, and requiring 95\% output accuracy. Differences between the left and right panels stem from the different time scales in the two types of QPUs. The dashed line estimates the run time of state-of-the-art classical simulation algorithms
 on HPC hardware, using  Eq.~\eqref{eq:Tc-of-V}, with dimension $d=2$ and operator spreading velocity $v=0.1$, as an example. Note that since we measure the
  active volume in native gates and work with $d=2$, the all-to-all
  connectivity advantage of trapped-ions is not explicitly
  indicated. Finite QA may be achieved when QESEM’s QPU time
  is significantly below the dashed line. The purple background shows
  the velocity $v$ (for $d=2$) as a function of the active volume
  $V$ and classical run time $T_c$, demonstrating how the dashed line would have looked like for with different choices of the velocity $v$. 
  }
\end{figure}

Similarly to the derivation of Eq.~\eqref{eq:cvbapprox}, we can use
Eq.~\eqref{Eq:V_EM} to obtain an order of magnitude estimate $V_{EM}
= \log R/\lambda\gamma \sim
(1\text{--}10)/\gamma$ for the volumes enabled by EM, as a function
of the infidelity $\gamma$, for a reasonable allowed shot overhead $R\in[2,10^9]$. For near-term
2-qubit gate infidelities $\gamma\sim10^{-3}$, this means 
\begin{align}
  V_{EM}\sim 10^{3}\text{--}10^{4}.\label{eq: near term V_EM}
\end{align}
Going beyond such order of magnitude estimates requires specifying a particular EM protocol and QPU, and accurately estimating the required number of shots and distinct circuits required by the protocol, as well as the shot and circuit times of the QPU (as already mentioned in Footnote \ref{foot: 2}). Figure \ref{fig: QESEM runtime} demonstrates the results of such an analysis, based on QESEM, an unbiased EM protocol introduced in Ref.~\cite{QESEMpaper}; and with characteristic time scales for current QPUs based on either superconducting-qubits or trapped-ions.

The solid colored lines in Fig.~\ref{fig: QESEM runtime} show estimates of the QPU time for QESEM as a function of the active volume, for a required output accuracy $1-\epsilon=95\%$.\footnote{We assume here, for simplicity, that the observable is a single Pauli string, with ideal expectation value $\left<O\right>_{ideal}=1$, and with an \textit{effective volume} $V_{eff}:=\gamma^{-1}\log(\left<O\right>_{ideal}/\left<O\right>_{noisy})$ identical to the active volume $V$. We refer to Ref.~\cite{QESEMpaper} for derivations, more general analytic QPU time estimates, improved estimates based on empirical data, and a comparison to experimental performance.}
We see that for both
superconducting-qubits and trapped-ions, a 2-qubit gate
infidelity $\gamma= 10^{-3}$ (green lines) suffices to achieve accurate
expectation values from circuits with an active volume of over a
thousand 2-qubit gates in reasonable time -- a few hours for trapped
ions, and under an hour for superconducting qubits.\footnote{ We consider here EM protocols whose run time is dominated by the required QPU time, with a negligible classical time for pre- and post-processing. It is possible to allow for significant HPC usage in EM, which can reduce the QPU time \cite{filippov2023scalable, lindner_yunoki_2024, fuller2025improved}. In such cases one must consider the total run time (or other appropriate resource), including both classical and quantum processing.}  
 
We devote the
remainder of this section to justify the claim that the ability of
EM to produce accurate output from circuits actively involving over
a thousand 2-qubit gates in reasonable time suffices for finite QA
in $\ome$, namely suffices to exceed state of the art quantum circuit simulation by classical HPC. We start in Sec.~\ref{Sec: HPC1} by considering state-vector simulation, and
restrict attention to circuits with a particular geometry to enable comparison to EM. In Sec.~\ref{Sec: TN contraction} we
argue that, for circuits with this geometry, state of the art simulation methods, based on tensor-network contraction, will not provide a significant advantage over the run
time of state-vector simulation. We comment on additional classical simulation methods, including approximate methods, in Sec.~\ref{Sec: TNS}.

{~}

\subsubsection{Comparison to classical HPC (1): State vector simulation \label{Sec: HPC1}}

We consider here the classical complexity of $\ome$, focusing on finite
problem sizes and time scales, as opposed to asymptotics. The
world's top supercomputers can now perform $\sim10^{18}$
floating-point operations per second (FLOPS) \cite{TOP500_Nov2024}.
A classical state-vector simulation of quantum circuits with volume
$V$ on $n$ qubits requires $4V 2^n$ complex floating-point operations,\footnote{For each
2-qubit gate, $2^{n-2}$ matrix-vector multiplications of complex
dimension 4. Conversion between
complex and real floating-point operations leads to an additional factor of 7.5 \cite{fang2022efficient}.} leading to a run time\footnote{We ignore here memory requirements, which only increase the run time, and revisit these in Sec.~\ref{Sec: TN contraction}.}
\begin{align}
  T_c\geq V 2^n 10^{-18}\text{sec}. 
  \label{eq:Tc}
  \end{align}

Equation~\eqref{eq:Tc} gives the run time of state vector
simulation as a function of both the active volume $V$ and the number of active qubits $n$. On the other hand, as discussed in Sec.~\ref{Sec: EM alone
cannot}, the run time of known EM methods is generically determined
solely by the active volume $V$, as in Eq.~\eqref{eq:emscaling}.
Therefore, in order to compare the two run times, we must fix a
functional relation between $n$ and $V$, by restricting attention
to a particular family of circuits, as we next describe. Once a functional relation $n(V)$ is fixed, it can be plugged into  
Eq.~\eqref{eq:Tc} to derive a classical HPC run time estimate as a function of $V$, as demonstrated in Fig.~\ref{fig: QESEM runtime} (dashed line).  

To demonstrate how a family of circuits with a fixed relation $n(V)$ may be defined, we consider a (hyper) cubic lattice of qubits in $d\geq1$ dimensions
on which we define a circuit of $D$ layers that is made of identical 2-qubit gates. The gates act on
the edges of the lattice, such that each qubit experiences a 2-qubit
gate in each of the $D$ circuit layers. In addition, we shall assume that at the end of the circuit single-qubit observable is measured. In such uniform circuits, it is natural to characterize the active volume
$V$ in terms of the \textit{operator spreading velocity} $v\in
[0,1]$ \cite{Haah1},
This velocity determines the number of active qubits in the support of the operator after evolving
through the $D$ circuit layers in the Heisenberg picture. When
$v=1$, 
the support of the operator grows by one qubit whenever the operator evolves through a 2-qubit gate that crosses its support. In practice, the spreading velocity can be much lower than its maximal value $1$, e.g., when the gates in the circuit have small rotation angles.

For the above circuit family, the active volume $V$ takes the shape of a $d$-dimensional pyramid with base area $n$ (the number of active qubits) and height $D$ (the circuit depth), with the velocity $v$ determining the opening angle of the pyramid. A simple geometrical calculation then gives the relation $n(V)=(4vV/c_d)^{c_d}$, with $c_d=d/(d+1)$, see Appendix \ref{app:velocity}. Substituting into Eq.~\eqref{eq:Tc}, we get the classical run time $T_c$ of state vector simulation, as a function of the active volume $V$, for the above circuit family: 
\begin{align}
\label{eq:Tc-of-V}
  T_c \sim V 2^{(4 v V/c_d)^{c_d}}. 
\end{align}
We use
Eq.~\eqref{eq:Tc-of-V} with the conservative values $d=2$ and
$v=0.1$ to obtain the dashed `HPC' line in Fig.~\ref{fig: QESEM
runtime}. Since the classical run time is very sensitive to the
velocity parameter $v$, we also include a colored background in
Fig.~\ref{fig: QESEM runtime}, where the color scale corresponds to
varying velocities $v$ for fixed $d=2$, showing how the graph $T_c(V)$ changes with different choices of the velocity $v$.  In the darker areas in Fig.~\ref{fig:
QESEM runtime}, finite
QA is less fine tuned, as it holds for broader circuit families of a given volume (namely, also for lower velocities). 

\begin{remark}(Circuit geometry and finite QA with EM) 
The above discussion provides insights regarding the choice of circuit geometry where it is best to look for early finite QAs with EM. Equation \eqref{eq:Tc-of-V} shows that $T_c$ grows exponentially with $V$ with an exponent that
becomes `worse' the larger $v$ or $d$ become. The dimension $d$ is essentially unlimited in QPUs
based on trapped ions and neutral atoms, but is restricted to $\leq
2$ in current superconducting qubit QPUs. For fixed $d$, lower  velocities correspond to more challenging circuits for EM relative to classical
simulation. Thus, for early finite QA using EM, we may therefore seek circuits with high velocity.  
Intuitively, we can
understand this directly from Eq.~\eqref{eq:Tc}, in which $T_c$
grows exponentially with $n$ and only linearly with $V$. It follows that, for fixed volume,
`balanced' circuits, where $D\sim n^{1/d}$, and so $n\sim V^{c_d}$, are harder to classically
simulate than circuits that are narrow and deep ($D\gg n^{1/d}$, or $n\ll V^{c_d}$), and therefore balanced circuits are more suited for early demonstrations of finite
QA.\footnote{To compensate for the `un-balanced' shape of the active volume in circuits with low velocity $v$, we may choose to measure observables supported on more than a single qubit, essentially chopping the top of the pyramid-shaped active volume. Such higher-weight observables will not, however, be advantageous for circuits with high velocity, where they lead to an imbalance in the opposite direction, $D< n^{1/d}$. As explained below, such circuits admit a reduced classical run time relative to Eq.~\eqref{eq:Tc}, via tensor-network contraction algorithms.} We expect such geometrical considerations to play an important role in near term demonstrations of finite QA with EM.
\end{remark}

\subsubsection{Comparison to classical HPC (2): Tensor network contraction\label{Sec: TN contraction}} 
So far, we explicitly considered the run time of state-vector
simulation, ignoring memory requirements and additional simulation
methods. In practice, state-vector simulation is limited by memory
requirements to under $50$ qubits, and the current leading method
for simulating generic quantum circuits seems to be \textit{tensor network contraction} \cite{kechedzhi2024effective, vallero2024state, PhysRevLett.128.030501}. The method is based on the representation of a given quantum circuit as
a single tensor network (TN), which is contracted via an optimized
ordering of tensors \cite{markov2008simulating, gray2021hyper}. Apart from potentially
reducing memory requirements relative to state-vector simulation (at
least for circuits with uniform connectivity, e.g., a
$d$-dimensional lattice) \cite{vallero2024state}, TN contraction can
significantly reduce the run time, if the circuit has a `tree-width'
which is significantly smaller than the number of qubits
\cite{markov2008simulating}. For circuits on a $d$-dimensional
lattice of $n$ qubits, this corresponds to the circuit depth $D$
being smaller than the lattice diameter $n^{1/d}$.

However, for $D>n^{1/d}$, the
expression in Eq.~\eqref{eq:Tc} is a good estimate for the run time
of TN contraction with state of the art classical algorithms and
hardware (see e.g. Sec.IV.A in Ref.~\cite{kechedzhi2024effective}).
In our example $D=(d/2v)n^{1/d}$ (see Eq.~\eqref{eq:nten} in Appendix \ref{app:velocity}), so for $d=2$, the above condition is met for  velocities $v$ which are not too
close to $1$. The color scale in Fig.~\ref{fig: QESEM
runtime} emphasizes low velocities, for which TN contraction cannot significantly
improve upon the run time of state-vector simulation.

\subsubsection{Comparison to classical HPC (3): Additional simulation methods\label{Sec: TNS}} 
Though TN contraction is arguably the leading method for simulating generic quantum circuits \cite{kechedzhi2024effective, vallero2024state, PhysRevLett.128.030501}, a convincing claim of finite QA would need to address additional classical simulation methods. Each known simulation method can evade the lower bound in Eq.~\eqref{eq:Tc} only for a very restricted family of circuits; and finite QA can only be achieved for circuit families that do not significantly overlap with any of these easy-to-simulate families.    

As a example, an important set of classical methods to consider is based on a TN
representation of the initial state (TNS), and the evolution of this representation through the circuit.
Such methods are advantageous for circuits that create limited
entanglement (see e.g., Ref.~\cite{PhysRevLett.91.147902}), so to
ensure that they do not admit a run time significantly lower than
Eq.~\eqref{eq:Tc}, we must consider circuits in
which all $n$ qubits are sufficiently entangled. 
It is indeed expected that in generic cases 
the $n$ active qubits in Eq.~\eqref{eq:Tc} are significantly entangled, in which case TNS methods would not be advantageous over state vector.\footnote{As an interesting recent example, the TNS approach of
\cite{PRXQuantum.5.010308} 
proved useful for simulating the naively
large volume experiment of \cite{kim2023evidence}. However, as shown
in \cite{kechedzhi2024effective}, the number of active qubits was
significantly smaller than the total number of qubits in the
circuit, such that efficient state-vector simulations were also
possible.}
We note that 
Ref.~\cite{Haah2} constructed examples for high dimensional qudits where this is not the case, and the number of active qubits is much larger than the number of qubits which are significantly entangled. For such circuits, the classical run time of TNS methods might be significantly better than Eq.~\eqref{eq:Tc}. 
However, while the \textit{entanglement growth velocity} 
is generally only upper bounded by the operator spreading velocity $v$ \cite{Haah1,Haah2}, we are not aware of examples where the two velocities differ significantly for generic, dense,
quantum circuits, defined on qubits.  

Two additional notable examples which would require significantly less run time for classical simulation than Eq.~\eqref{eq:Tc}, include circuits with mostly Clifford gates, which can be simulated efficiently based on the stabilizer formalism \cite{Bravyi2019simulationofquantum};
as well as circuits with mostly `matchgates', which can be simulated efficiently based on a mapping of qubits to free fermions \cite{ReardonSmith2024improvedsimulation}.

The above easy-to-simulate circuit families and corresponding simulation algorithms were discovered as part of extensive (and ongoing) research effort in quantum computational complexity. It is possible that additional studies will identify additional restricted circuit families that may be simulated efficiently. Nevertheless, we take
Eq.~\eqref{eq:Tc} as an estimate for the run time of
state-of-the-art classical simulation of \textit{generic} quantum circuits actively involving $n$ qubits and $V$ gates; and with the connectivity of a $d$-dimensional lattice and circuit depth $D>n^{1/d}$. This may be viewed as a finite-size complexity assumption, on which claims of finite QA may be based. 

As a final remark, we note that the run time for EM scales as $\epsilon^{-2}$ (as in Eq.~\eqref{eq:emscaling}) with the
allowed inaccuracy $\epsilon$, while the expression \eqref{eq:Tc} for the classical run time is naively independent of
$\epsilon$. As a result, finite QA with EM seems to be easier for larger
$\epsilon$, and more challenging with small $\epsilon$. Of course,
one may ask whether classical simulation methods can utilize the
allowed inaccuracy to save in run time. First, note that since Eq.~\eqref{eq:Tc} involves the active volume and active qubit number, the run time is already reduced based on the allowed inaccuracy $\epsilon$. Beyond this, despite results showing significant reductions in the run time for \textit{noisy} circuit simulation based on its large deviation from the ideal circuit \cite{markov2018quantum, PhysRevX.10.041038}, we are not aware of significant reductions in the run time of ideal circuit simulation given a \textit{small} allowed inaccuracy $\epsilon\ll1$.

\section{EM in the era of fault tolerant EC \label{Sec: EMEC}}

\subsection{The fault tolerance misconception: Why additional error reduction is needed even when fault tolerant EC is available}
A common misconception
in the quantum ecosystem is that fault tolerant EC is thought to handle errors in the best and most efficient way, and therefore, once EC abilities become available to end-users, no other error reduction methods will be needed. 
The origin of this misconception comes from the fact that the original fault tolerance results \cite{aharonov_1997, kitaev_1997, knill_laflamme_1996, preskillaliferis} assumed that the model of quantum computation can be described by the set of gates available, accompanied by a single parameter $\gamma$ indicating the fidelity of the individual qubits or gates in the system. The assumption in all those results is that once the desired below-threshold infidelity $\gamma$ is achieved for single gates, an unbounded number of qubits with this fixed error rate is available upon demand. 
This picture is also in line with the usual picture of condensed matter physics, in which the local properties of the physical system are fixed and then one takes the size of the system to infinity to arrive at the thermodynamic limit.  This idealistic picture, however, is not how things work in the reality of quantum computer implementations. 

Over the past decade, qubit numbers in existing QPUs have been increasing steadily but slowly. This had been the case in all leading hardware platforms, in particular those based on superconducting qubits and trapped ions, and it is expected to continue to be the situation in the foreseeable future, according to the road maps of all major quantum computing manufacturers, see e.g., \cite{ibm_quantum_technology, quantinuum_roadmap_2023}. The qubit numbers predicted for neutral atom based quantum processors are larger, but nevertheless restricted, see e.g., \cite{quera_qec_2024}. 
The situation is somewhat different in photon based QPUs, which we exclude from this discussion.  

We therefore consider here hardware platforms in which the increase in the number of high quality qubits is gradual; at any point in time in the foreseeable future, the number of qubits is upper bounded and limited by some fixed number, and the end-users at that point in time can only make use of QPUs of that size. This limit on the number of qubits has severe consequences. 

When the number of high quality qubits is limited, the user must choose between spending the qubits on EC, or on logical qubits for computation (namely, the number of qubits in the ideal circuit). Thus there is a trade-off between the number of logical qubits, and their quality, namely, the logical error. Say, for example, that we are looking a few years ahead into the future, when QPUs consisting of $10^4$ high quality (below threshold) qubits are available.  Suppose then that we want to run a quantum circuit on $10^3$ logical qubits; in this case we can only use EC with a rate of $1/10$ logical qubits to physical qubits. Such an EC code can only reduce the logical error rate by a limited amount; denote the resulting logical error by 
$\gamma'$. Using Eq.~\eqref{eq:bare}, this directly translates to an upper bound on the volume of the logical quantum circuit that can be run, $V\sim \epsilon/\gamma'$, for a required accuracy $\epsilon$. 
Perhaps the user would resort to running a circuit of just $100$ logical qubits? That would leave her the possibility of using an EC scheme with rate $1/100$, which if chosen correctly, can reduce the logical error by many orders of magnitudes compared to the physical one, and allow her to run very deep quantum circuits on those $100$ qubits. We note that the effectiveness of EC protocols  (namely, the value of the logical error rate $\gamma'$ that is achieved) depends very strongly (in fact, exponentially) on the \textit{distance} of the EC code used. However, increasing the code distance usually comes at the expense of reducing the code rate, entailing an increase in the required number of physical qubits.\footnote{The inverse relation between code distance and code rate holds in commonly used code families, such as Surface codes or code concatenations. An exception occurs in qLDPC codes of \textit{constant} rate -- codes in which the rate is bounded from below by a constant $>0$ independent of the code length that may grow unboundedly. However even for such codes, increasing the code distance means investing more physical qubits; see Tab.~\ref{Tab: tab} and surrounding text regarding the LP qLDPC codes.} This means that reducing the number of logical qubits in the above example from $10^3$ to a $100$, namely by a factor of ten, leads to a dramatic improvement in the logical error rate, which in turn allows an increase by a factor much larger than $10$ in the volume.

 We see that there is an important trade off: if we want to reduce the logical error rate significantly (in order to run logical circuits with larger volume), we will need to invest a lot of our qubit resources in EC, which will limit the logical circuits we can run to much fewer logical qubits.  
The upshot is that in the foreseeable future, and for as long as the number of high quality qubits is limited, circuit volumes and logical qubit numbers will be limited due to noise, even in the era of fault tolerant EC.  Hence, while volumes available by EM are severely limited due to limitations on available QPU time (cf. Sec.~\ref{Sec: EM alone cannot}), the volumes possible with fault tolerant EC are also limited -- due to restricted qubit numbers.

Here is an example demonstrating the circuit volumes expected to be available when just using EC in the near future, say with an optimistic fidelity $\gamma=5\times 10^{-4}$. 
In this case, the bare volume available for required accuracy $\epsilon=1\%$ is $V_{bare}\sim \epsilon/\gamma=20$ gates (following the rule of thumb of Eq.~\eqref{eq:vbareest}). How much can we improve on this when using EC? This, as explained above, depends on how much of our qubit resource we want to spend on EC. 
Suppose we have $10^3$ physical qubits at hand, with this fidelity. If we want to run a circuit on $100$ logical qubits, we can only afford using an EC code of rate roughly $1/10$; we could use Steane's code. 
We note that as in the bare case, we can estimate the available logical volume by the same rule of thumb, 
\begin{align}
    V_{EC}\sim\epsilon/\gamma',\label{Eq: V_EC}
\end{align}
where the physical infidelity $\gamma$ is replaced by the logical infidelity $\gamma'$.
We see that EC only increases the available volume compared to the bare case, by the same factor by which 
Steane's code improves the logical error rate compared to the physical error rate. 
With a  fidelity of $5\times 10^{-4}$ and taking into account that the threshold for Steane's code with state-of-the-art fault-tolerant constructions is $\sim 10^{-3}$,  
we get that the increase in volume would be less than a factor $2$ (see Fig.~\ref{fig: EMEC}, blue line). This enables reaching logical circuit volumes $<40$, certainly not sufficient for achieving  finite QA. 

Luckily, this situation can be remedied. While noise limits the circuit volumes that can be executed even when fault tolerant EC is used, the volume available can be significantly increased, in fact by several orders of magnitude, by using EM {\it in conjunction} with EC. We explain this combination now. 

\subsection{Logical EM: Combining EM with EC to provide further circuit volume boosts \label{Subsec: EMEC}}

To address the challenges of EC and EM, and go beyond their respective limitations, the combination of the two methods was recently proposed \cite{Suzuki2022, Piveteau2021, Lostaglio2021, Xiong2020, Tsubouchi2024}, and very recently demonstrated \cite{zhang2025demonstrating}. We refer to this general approach as \textit{logical error mitigation} (LEM). In terms of quantum resources,  LEM enables to optimize the use of both available QPU time \textit{and} physical qubit numbers, to maximize accessible circuit volume and output accuracy, with the given resources. 

\subsubsection{External LEM} 
\begin{figure}[h!]\label{fig:emec}
\begin{centering}
\includegraphics[width=0.6\columnwidth]{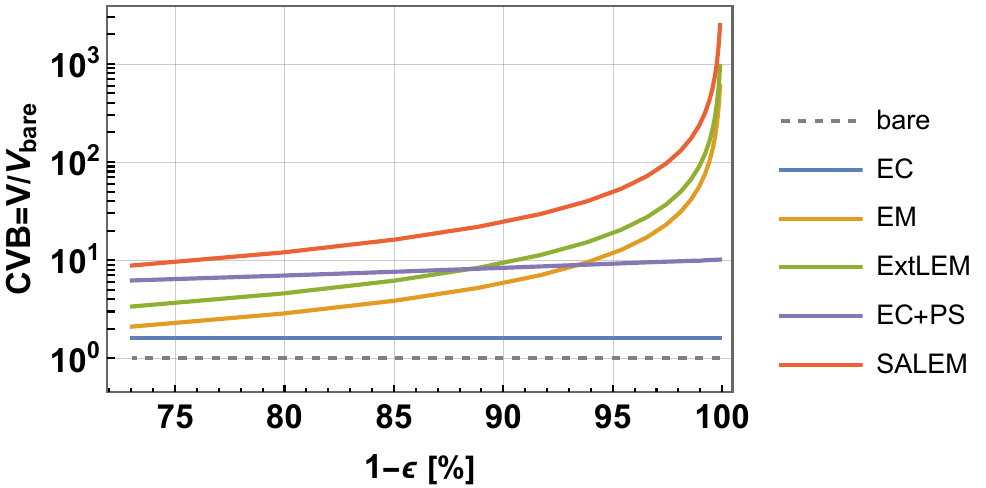}
\par\end{centering}
\caption{\label{fig: EMEC} An example for circuit volume boosts (CVBs) with different error reduction strategies, with and without EC. The different curves indicate the CVBs relative to bare circuit execution, due to EC, EM, an ‘external’ mitigation of logical errors (ExtLEM), the combination of EC and post-selection (EC+PS), and a syndrome-aware mitigation of logical errors (SALEM) \cite{SALEMpaper}. All methods are given a very mild allowed shot overhead $R=2$. The data presented was obtained from numerical simulations of repeated EC cycles (a logical memory circuit), with the well-known Steane code (7-qubit color code), based on the fault-tolerant construction of Ref.~\cite{Reichardt_2021}. The results hold qualitatively for any EC code, but will be quantitatively different, e.g. the blue line will be higher with better EC schemes (cf. Tab.~\ref{Tab: tab}). For ease of reference, focus on 95\% output accuracy. We compare the volume boosts for different methods.  For an infidelity of  $5\times 10^{-4}$, slightly below the Steane code's threshold, logical errors are slightly smaller than physical errors, and EC provides a small CVB (blue) compared to bare execution. This CVB does not improve with required accuracy, because EC suffers a significant bias due to logical errors (similarly to the bias in the bare computation, discussed in Sec.~\ref{Sec:CVB}). Unbiased EM (yellow), applied directly to physical qubits, is better than EC for the Steane code and chosen infidelity of  $5\times 10^{-4}$. ExtLEM (green), where EM is naively applied to logical gates, enjoys the benefits of both EC and EM, and provides a larger CVB than EM. The remaining methods, EC+PS and SALEM, are discussed in the main text. 
}
\end{figure}

The LEM methods described in the literature may be roughly summarized as follows: apply EC to generate logical operations with a reduced error rate relative to physical operations; then apply EM to these logical operations, as if they are physical operations. One can  view this is as follows. To create LEM, we start with some EM protocol, and replace each gate used by it by an error-corrected logical gate, including syndrome measurements, decoding and recovery operations. This results in the LEM protocol in which EM acts on the logical qubits. We refer to this approach as ‘external LEM’ (ExtLEM), since it does not make use of the wealth of internal syndrome data generated by logical operations.\footnote{We note that existing work does make use of several features of logical operations which are not present in physical operations, such as the ability to virtually apply Pauli operations by keeping track of a `Pauli frame' \cite{Suzuki2022}, as well as the expected lower logical error of Clifford gates relative to non-Clifford gates \cite{Piveteau2021, Lostaglio2021, Tsubouchi2024}.} Assuming that ExtLEM is based on an EM protocol which is either unbiased, or produces a bias which is negligible relative to the required accuracy $\epsilon$, we immediately obtain from Sec.~\ref{Sec:CVB} the order of magnitude estimate 
\begin{align}
    V_{LEM}\sim \frac{1\text{--}10}{\gamma'}\label{Eq: V_{LEM}}
\end{align}
for the logical circuit volume available with LEM. Accordingly, the circuit volume boosts (CVBs) of $\sim1\text{--}10/\epsilon$ shown in Fig.~\ref{fig: CVB} for EM over bare circuit execution, are maintained when comparing ExtLEM to EC. Thus, LEM enables $\sim1\text{--}10/\epsilon$  larger logical circuit volumes compared to those provides by EC alone, on the same hardware and with the same output accuracy, see blue and green curves in Fig. \ref{fig: EMEC}.

\subsubsection{Error correction combined with post-selection}

An alternative strategy to mitigate logical errors, which is routinely used in EC experiments, is based on post-selection (PS) \cite{smith2024mitigating, PrabhuReichardt2024}. Here, one employs decoders that reject certain syndromes, discarding the circuit repetitions (shots) in which they are observed. We refer to this strategy as EC+PS, to highlight the fact that the accepted set of syndromes is not only kept but also decoded and recovered from, as in standard EC.\footnote{We view PS without EC, that is, without decoding and recovering from accepted syndromes, as an EM method, as opposed to a LEM method. Note that PS can be performed even without EC codes, based on known symmetries of the ideal circuit \cite{emreview}.} Let us compare ExtLEM to the approach of EC+PS. In the latter method, one rejects a set of syndromes, which means that errors associated with these syndrome will not contribute to the output distribution, which is computed by conditioning on accepted (and corrected) syndromes only. Thus, in EC+PS only the logical errors in rejected syndromes are mitigated (by rejection of the corresponding shots), while leaving the logical errors in accepted syndromes intact.  This generically leads to a significant bias $b$. On the other hand, if the rejected set of syndromes is chosen appropriately, the above `rejected' logical errors are mitigated (by shot rejection) very efficiently, in the sense of a blowup rate $\lambda$ that can be as low as $1$ (cf. toy example in Sec.~\ref{Sec: EM alone cannot}). These features determine the behavior of the CVB due to EC+PS, which is generically better than that of ExtLEM when $b\ll\epsilon$, but significantly under-performs when $b\gg\epsilon$, see e.g., purple curve in Fig.~\ref{fig: EMEC}. 

\subsubsection{Syndrome Aware Logical EM (SALEM)}
 In fact, it turns out that both ExtLEM and EC+PS may be viewed as instances of a more general framework, that extends and improves upon both. We refer to this framework as \textit{syndrome-aware LEM} (SALEM), and provide its detailed technical description in Ref.~\cite{SALEMpaper}. In essence, SALEM allows to maintain the negligible bias possible with ExtLEM, while significantly improving upon its blowup rate. Figure ~\ref{fig: EMEC} (red curve) demonstrates these performance improvements of SALEM relative to ExtLEM and EC+PS.

\subsubsection{Logical EM with state of the art EC codes \label{Sec: LEM with SOTA codes}} 

Our discussion of LEM so far used the prototypical Steane code as an example. And while the simplicity and high rate of the Steane code made it a useful example for recent EC experiments, its low threshold value $\sim 10^{-3}$ makes it impractical for useful EC in the near term. This is reflected by the small CVB of EC relative to bare execution, even with an optimistic infidelity of $5\times 10^{-4}$ (blue curve in Fig.~\ref{fig: EMEC}).   We now provide examples for the benefits of LEM over mere EC. 
We consider three state of the art code families: Surface codes, BB qLDPC codes, and LP qLDPC codes \cite{xu2024constant,bravyi_2024}.  These are examples of codes expected to lead to significant performance improvements over bare circuit execution by the end of the decade, at least according to current hardware road maps \cite{google_quantum_ai, ibm_quantum_technology, quera_qec_2024, quantinuum_roadmap_2023}. 
In our analysis, we set aside the details of how LEM is performed, and use only the order of magnitude estimates in Eq.~\eqref{Eq: V_EC} and \eqref{Eq: V_{LEM}}. 
This is summarized in Tab.~\ref{Tab: tab}.  

Within each of the three code families, we consider two codes, the second one having a lower logical error
than the former. For the Surface and BB qLDPC code families, the
improvement in logical error rate comes at the cost of reducing the rate of the code, thus requiring more physical
qubits for a fixed number of logical qubits. The case of the LP qLDPC family is different, since the distance of these codes can increase  unboundedly  (leading to an exponential decrease in the logical error rate per qubit)
while keeping the rate of the code above a fixed constant $>0$. However, this requires working with codes in the family with whose block size ($n_{tot}$ in Tab.~\ref{Tab: tab0}) also grows unboundedly, eventually surpassing the total number of physical qubits in a given QPU. Thus, again, an increase in the number of physical qubits is generally required in order to reduce the logical error rate. In the examples given in Tab.~\ref{Tab: tab}, 
the improvement in logical error rate from the first LP qLDPC code to the second, is in fact accompanied by a slightly {\it higher} rate, but
comes at the price of a $1300$-qubit increase in $n_{tot}$ -- possibly
requiring more physical qubits than available on a near term QPU. The upshot is that in all three examples of code families, moving from the first code to the second code to improve the logical error rate requires increasing the required number of physical qubits for implementation. 

Assuming $\epsilon=0.1$, for each code family in Tab.~\ref{Tab: tab},
LEM of the first code enables to at least reproduce the available
volumes of the second code, thus trading run time for physical qubits.
For estimation error $\epsilon=0.01$, LEM of the first code provides
a significant improvement over the second code (without LEM). In particular, without
LEM, and for $\epsilon=0.01$, none of the listed codes can reach
volumes $\sim10^{6}$, the so-called MegaQuOp milestone \cite{preskill2024megaquop}, where one
may expect quantum applications with industrial value (see e.g., Ref.~\cite{clinton2024towards} for proposed quantum simulation algorithms of industrially-relevant materials at or below these circuit volumes). However, with
LEM, the second code in each family reaches MegaQuOp with a small allowed shot overhead $R$ (see Eq.~\eqref{eq:MandR}) while the first code can also reach MegaQuOp,
given a more significant allowed shot overhead  but with less physical qubits.

\renewcommand{\arraystretch}{1.2}
\begin{table}[h!]

\caption{Examples of state of the art EC codes, and corresponding logical circuit
volumes possible with EC alone, and with LEM. The total number of
physical qubits $n_{tot}$ includes the both code qubits and ancilla
qubits used for fault-tolerant syndrome measurements. The net rate
is given by $k/n_{tot}$. The logical error $\gamma'$ quoted here
is defined per syndrome measurement cycle and per logical qubit, and
at a physical infidelity $\gamma=10^{-3}$. Note that we consider
here logical memory circuits, without any
logical operations. Current proposals for logical operations for qLDPC
codes are based on teleportation to surface code patches, which 
degrades net rates \cite{xu2024constant,bravyi_2024}. The columns $V_{EC}$ and $V_{LEM}$ represent
the maximal logical circuit volumes possible with EC and LEM, based
on the order of magnitude estimates \eqref{Eq: V_EC} and \eqref{Eq: V_{LEM}}, respectively. 
\vspace{5pt}
\label{Tab: tab}}

\begin{centering}
\begin{tabular*}{1\linewidth}{@{\extracolsep{\fill}}ccccccc}
\hline 
\hline
Code family & $\left\llbracket n,k,d\right\rrbracket $ & $n_{tot}$ & $k/n_{tot}$ & $\gamma'$ & $V_{EC}$ & $V_{LEM}$\tabularnewline

\hline 
Surface, & $\left\llbracket 81,1,9\right\rrbracket $ & $161$ & $1/161$ & $9\times10^{-6}$ & $10^{5}\epsilon$ & $10^{5}-10^{6}$\tabularnewline
\cline{2-7}
Ref. \cite{bravyi_2024} & $\left\llbracket 121,1,11\right\rrbracket $ & $241$ & $1/241$ & $9\times10^{-7}$ & $10^{6}\epsilon$ & $10^{6}-10^{7}$\tabularnewline
\hline 
\hline 
BB qLDPC, & $\left\llbracket 72,12,6\right\rrbracket $  & $144$ & $1/12$ & $6\times10^{-6}$ & $2\times10^{5}\epsilon$ & $2\times(10^{5}-10^{6})$\tabularnewline
\cline{2-7}
Ref. \cite{bravyi_2024} & $\left\llbracket 144,12,12\right\rrbracket $  & $288$ & $1/24$ & $2\times10^{-8}$ & $6\times10^{7}\epsilon$ & $6\times(10^{7}-10^{8})$\tabularnewline
\hline 
\hline 
LP qLDPC, & $\left\llbracket 544,80,\leq12\right\rrbracket $  & $1367$ & $1/17$ & $10^{-6}$ & $8\times10^{5}\epsilon$ & $8\times(10^{5}-10^{6})$\tabularnewline
\cline{2-7}
Ref. \cite{xu2024constant} & $\left\llbracket 1428,184,\leq24\right\rrbracket $ & $2670$ & $1/15$ & $10^{-7}$ & $9\times10^{6}\epsilon$ & $9\times(10^{6}-10^{7})$\tabularnewline
\cline{2-7}
\hline
\hline
\end{tabular*}
\par\end{centering}

\end{table}

\subsection{Concluding remarks about logical EM} To summarize, given a restricted number of physical qubits, LEM provides access to significantly larger logical circuit volumes relative to just EC, at the expense of a tolerable overhead in QPU time. And while adding more physical qubits to a given QPU is generally impossible for end users (and a formidable challenge for hardware manufacturers), allowing for more QPU time is always a possibility, which  should be exploited based on its cost-effectiveness for the application at hand.
The implication is that at any given moment in time, when EC becomes available, the combination of EM and EC, namely logical EM, will provide significantly larger circuit volumes than with EC alone, and thus should be viewed as the way to optimally use the available quantum resources.

\section{Summary and forecast:  QA timeline \label{sec: summary}}

This paper refutes two common misconceptions

\begin{itemize} 
\item 
{\bf Misconception 1}: {\it Error mitigation cannot provide any quantum advantage due to its exponentially scaling shot overhead}. 

We explained in this paper why this is a misconception. 
The origin of the misconception is a confusion between asymptotic QA and finite QA; 
we clarify the difference between these two notions in Sec.~\ref{Sec: QuAlA}. 
We then show that exponential asymptotic QA
 is indeed generically impossible using EM alone (Sec. ~\ref{Sec: EM alone cannot}) and at the same time, dramatic finite QAs are expected very soon using EM (Sec.~\ref{Sec: predictions}).   

\item 
{\bf Misconception 2}: {\it Error mitigation will not be important once error correction becomes available.} 

In Sec.~\ref{Sec: EMEC} we explained why even when EC becomes available, this will not eliminate the problem of errors, as  residual logical errors will still limit the volumes of quantum circuits that can be executed. We described how circuit volume boosts of a few orders of magnitude are expected using \textit{logical} EM -- the combination of EM and EC -- compared to what can be achieved using EC alone on a given hardware platform. 
\end{itemize} 

{~}

We conclude that the expected role to be played by EM on the road to  the first finite QAs, and beyond, is crucial; EM is highly likely to be the first to provide the first useful finite QAs, and is expected to continue to have an important role also in the long run, when EC becomes a daily routine, by enabling significantly larger quantum circuits than what EC will enable on its own; thus enabling reaching a variety of finite QAs much earlier than would be possible with EC alone.

Based on the arguments presented here, Table \ref{tab: time line} provides predictions regarding the expected advances along the finite QA timeline, with and without EM. 
In particular, it highlights  
two milestones that  demonstrate the crucial role of EM on the road to QAs: 
\begin{enumerate} 
\item  The first milestone is to experimentally demonstrate finite QA for the expectation value estimation problem.  As argued in Sec.~\ref{Sec: predictions}, and based on the hardware road maps cited in Tab.~\ref{Tab: tab0}, this is expected to be achieved in the very near future, by demonstrating the ability to execute generic quantum circuits of over $1000$ two-qubit gates, using EM alone. This will constitute a demonstration of the {\bf first useful finite QAs}. 

\item  The second milestone is to reach (logical) circuit volumes $\sim 10^5-10^7$, where the \textbf{first industry-relevant applications} are expected, using \textit{logical} EM (LEM). As discussed in Sec.~\ref{Sec: LEM with SOTA codes}, LEM significantly reduces the physical qubit numbers required for reaching these (logical) circuit volumes, relative to EC alone, thus expediting the corresponding finite QAs. Based on the hardware road maps cited in Tab.~\ref{Tab: tab0}, we expect this milestone to be achieved within the next few years.   
\end{enumerate}

\begin{table}[h!]
    \centering
    \renewcommand{\arraystretch}{1.15} 
    
    \caption{Conservative road map for finite QAs with EM and logical EM (LEM). In the LEM column we consider the state of the art qLDPC codes in Tab.~\ref{Tab: tab}, but we allow for a larger net rate, up to $1/100$, as may be needed for logical operations (beyond memory). For both EM and LEM we assume a mild allowed shot overhead, leading to a circuit volume boost $\sim 1/\epsilon$, as a function of the required output accuracy $1-\epsilon$. Note that while a $99.9\%$ fidelity suffices for the third milestone with LEM, any improvement in fidelity translates to a significantly larger improvement in \textit{logical} fidelity, enabling much larger volumes. 
    \vspace{5pt} 
\label{tab: time line} }
    
    \begin{tabularx}{1\columnwidth}{>{\raggedright\arraybackslash}X 
                                   >{\raggedright\arraybackslash}X 
                                   >{\raggedright\arraybackslash}X 
                                   >{\raggedright\arraybackslash}X} 
        \hline 
        \hline
        Milestone & $V\sim10^{3}$ & $V\sim10^{4}$ & $V\sim10^{5}-10^{7}$ \\
        \hline
        Achieved by & EM & EM & LEM \\
        \hline 
        Hardware requirements & 
        50 physical qubits. 

        2-qubit gate fidelity $99.9\%$.

        Existing circuit execution rates. 

        & 50-100 physical qubits.

        Higher circuit execution rates, 
        and/or
        higher 2-qubit gate infidelity $99.95-99.99\%$.  
        & 2,000-10,000 physical qubits.

        100 logical qubits. 

        2-qubit gate fidelity $99.9\%$. \\
        \hline 
        Use cases & 
        Quantum many-body physics research 
        & + Quantum algorithm development 
        & + First industry relevant applications \\
        \hline 
        Volumes without 
        
        EM/LEM 
        & 
        $V\sim10^{3}\epsilon$ 
        
        & $V\sim10^{4}\epsilon$ 
        
        & $V\sim(10^{5}-10^{7})\epsilon$ \\
        \hline 
        \hline
    \end{tabularx}
    
\end{table}

This manuscript makes a case for the importance of EM both as the enabler of the first useful QAs, 
as well as an accelerator of  further finite QAs  when combined with EC, once the latter becomes available. We believe these insights place EM as a crucial component towards truly useful quantum computation.

\vspace{10pt}

\textbf{Acknowledgements} We thank Ryan Babbush, Sergio Boixo, Jay Gambetta, David Hayes, Abhinav Kandala, Chris Langer, Sabrina Maniscalco, Tom O'Brien, John Preskill, Ittai Rubinstein, Thomas Schuster, Sarah Sheldon and Norman Yao for useful discussions and helpful comments. 

\vspace{10pt}

\textbf{Note added} Towards finalization of this manuscript, Reference \cite{zimboras2025myths} appeared on the arXiv, which provides a  perspective on the role of EM in quantum computing. The views and arguments presented in this reference are generally in line with the present paper, and we were pleased to see a consensus with the authors as to the important role of EM. Relative to  Ref.~\cite{zimboras2025myths}, the present paper provides a more detailed and technical account to explain both the misconceptions and the justification of our predictions. We hope these expositions will be of use to the community.





\begin{thebibliography}{120}%
\makeatletter
\providecommand \@ifxundefined [1]{%
 \@ifx{#1\undefined}
}%
\providecommand \@ifnum [1]{%
 \ifnum #1\expandafter \@firstoftwo
 \else \expandafter \@secondoftwo
 \fi
}%
\providecommand \@ifx [1]{%
 \ifx #1\expandafter \@firstoftwo
 \else \expandafter \@secondoftwo
 \fi
}%
\providecommand \natexlab [1]{#1}%
\providecommand \enquote  [1]{``#1''}%
\providecommand \bibnamefont  [1]{#1}%
\providecommand \bibfnamefont [1]{#1}%
\providecommand \citenamefont [1]{#1}%
\providecommand \href@noop [0]{\@secondoftwo}%
\providecommand \href [0]{\begingroup \@sanitize@url \@href}%
\providecommand \@href[1]{\@@startlink{#1}\@@href}%
\providecommand \@@href[1]{\endgroup#1\@@endlink}%
\providecommand \@sanitize@url [0]{\catcode `\\12\catcode `\$12\catcode `\&12\catcode `\#12\catcode `\^12\catcode `\_12\catcode `\%12\relax}%
\providecommand \@@startlink[1]{}%
\providecommand \@@endlink[0]{}%
\providecommand \url  [0]{\begingroup\@sanitize@url \@url }%
\providecommand \@url [1]{\endgroup\@href {#1}{\urlprefix }}%
\providecommand \urlprefix  [0]{URL }%
\providecommand \Eprint [0]{\href }%
\providecommand \doibase [0]{http://dx.doi.org/}%
\providecommand \selectlanguage [0]{\@gobble}%
\providecommand \bibinfo  [0]{\@secondoftwo}%
\providecommand \bibfield  [0]{\@secondoftwo}%
\providecommand \translation [1]{[#1]}%
\providecommand \BibitemOpen [0]{}%
\providecommand \bibitemStop [0]{}%
\providecommand \bibitemNoStop [0]{.\EOS\space}%
\providecommand \EOS [0]{\spacefactor3000\relax}%
\providecommand \BibitemShut  [1]{\csname bibitem#1\endcsname}%
\let\auto@bib@innerbib\@empty
\bibitem [{\citenamefont {Takagi}\ \emph {et~al.}(2022)\citenamefont {Takagi}, \citenamefont {Endo}, \citenamefont {Minagawa},\ and\ \citenamefont {Gu}}]{takagi2022fundamental}%
  \BibitemOpen
  \bibfield  {author} {\bibinfo {author} {\bibfnamefont {R.}~\bibnamefont {Takagi}}, \bibinfo {author} {\bibfnamefont {S.}~\bibnamefont {Endo}}, \bibinfo {author} {\bibfnamefont {S.}~\bibnamefont {Minagawa}}, \ and\ \bibinfo {author} {\bibfnamefont {M.}~\bibnamefont {Gu}},\ }\href {https://www.nature.com/articles/s41534-022-00618-z} {\bibfield  {journal} {\bibinfo  {journal} {npj Quantum Information}\ }\textbf {\bibinfo {volume} {8}},\ \bibinfo {pages} {114} (\bibinfo {year} {2022})}\BibitemShut {NoStop}%
\bibitem [{\citenamefont {Takagi}\ \emph {et~al.}(2023)\citenamefont {Takagi}, \citenamefont {Tajima},\ and\ \citenamefont {Gu}}]{Takagi2023universal}%
  \BibitemOpen
  \bibfield  {author} {\bibinfo {author} {\bibfnamefont {R.}~\bibnamefont {Takagi}}, \bibinfo {author} {\bibfnamefont {H.}~\bibnamefont {Tajima}}, \ and\ \bibinfo {author} {\bibfnamefont {M.}~\bibnamefont {Gu}},\ }\href {\doibase 10.1103/PhysRevLett.131.210602} {\bibfield  {journal} {\bibinfo  {journal} {Phys. Rev. Lett.}\ }\textbf {\bibinfo {volume} {131}},\ \bibinfo {pages} {210602} (\bibinfo {year} {2023})}\BibitemShut {NoStop}%
\bibitem [{\citenamefont {Tsubouchi}\ \emph {et~al.}(2023)\citenamefont {Tsubouchi}, \citenamefont {Sagawa},\ and\ \citenamefont {Yoshioka}}]{Tsubouchi2023universal}%
  \BibitemOpen
  \bibfield  {author} {\bibinfo {author} {\bibfnamefont {K.}~\bibnamefont {Tsubouchi}}, \bibinfo {author} {\bibfnamefont {T.}~\bibnamefont {Sagawa}}, \ and\ \bibinfo {author} {\bibfnamefont {N.}~\bibnamefont {Yoshioka}},\ }\href {\doibase 10.1103/PhysRevLett.131.210601} {\bibfield  {journal} {\bibinfo  {journal} {Phys. Rev. Lett.}\ }\textbf {\bibinfo {volume} {131}},\ \bibinfo {pages} {210601} (\bibinfo {year} {2023})}\BibitemShut {NoStop}%
\bibitem [{\citenamefont {Quek}\ \emph {et~al.}(2024)\citenamefont {Quek}, \citenamefont {Stilck~Fran{\c{c}}a}, \citenamefont {Khatri}, \citenamefont {Meyer},\ and\ \citenamefont {Eisert}}]{Quek_Eisert}%
  \BibitemOpen
  \bibfield  {author} {\bibinfo {author} {\bibfnamefont {Y.}~\bibnamefont {Quek}}, \bibinfo {author} {\bibfnamefont {D.}~\bibnamefont {Stilck~Fran{\c{c}}a}}, \bibinfo {author} {\bibfnamefont {S.}~\bibnamefont {Khatri}}, \bibinfo {author} {\bibfnamefont {J.~J.}\ \bibnamefont {Meyer}}, \ and\ \bibinfo {author} {\bibfnamefont {J.}~\bibnamefont {Eisert}},\ }\href {https://www.nature.com/articles/s41567-024-02536-7} {\bibfield  {journal} {\bibinfo  {journal} {Nature Physics}\ }\textbf {\bibinfo {volume} {20}},\ \bibinfo {pages} {1648} (\bibinfo {year} {2024})}\BibitemShut {NoStop}%
\bibitem [{\citenamefont {Schuster}\ \emph {et~al.}(2024)\citenamefont {Schuster}, \citenamefont {Yin}, \citenamefont {Gao},\ and\ \citenamefont {Yao}}]{Schuster_Yao}%
  \BibitemOpen
  \bibfield  {author} {\bibinfo {author} {\bibfnamefont {T.}~\bibnamefont {Schuster}}, \bibinfo {author} {\bibfnamefont {C.}~\bibnamefont {Yin}}, \bibinfo {author} {\bibfnamefont {X.}~\bibnamefont {Gao}}, \ and\ \bibinfo {author} {\bibfnamefont {N.~Y.}\ \bibnamefont {Yao}},\ }\href {https://arxiv.org/abs/2407.12768} {\bibfield  {journal} {\bibinfo  {journal} {arXiv preprint arXiv:2407.12768}\ } (\bibinfo {year} {2024})}\BibitemShut {NoStop}%
\bibitem [{\citenamefont {Bernstein}\ and\ \citenamefont {Vazirani}(1997)}]{bernstein_vazirani_1993}%
  \BibitemOpen
  \bibfield  {author} {\bibinfo {author} {\bibfnamefont {E.}~\bibnamefont {Bernstein}}\ and\ \bibinfo {author} {\bibfnamefont {U.}~\bibnamefont {Vazirani}},\ }\href {https://doi.org/10.1137/S0097539796300921} {\bibfield  {journal} {\bibinfo  {journal} {SIAM Journal on Computing}\ }\textbf {\bibinfo {volume} {26}},\ \bibinfo {pages} {1411} (\bibinfo {year} {1997})}\BibitemShut {NoStop}%
\bibitem [{\citenamefont {Simon}(1997)}]{simon_1997}%
  \BibitemOpen
  \bibfield  {author} {\bibinfo {author} {\bibfnamefont {D.~R.}\ \bibnamefont {Simon}},\ }\href {https://doi.org/10.1137/S0097539796298637} {\bibfield  {journal} {\bibinfo  {journal} {SIAM Journal on Computing}\ }\textbf {\bibinfo {volume} {26}},\ \bibinfo {pages} {1474} (\bibinfo {year} {1997})}\BibitemShut {NoStop}%
\bibitem [{\citenamefont {Shor}(1994)}]{shor_1999}%
  \BibitemOpen
  \bibfield  {author} {\bibinfo {author} {\bibfnamefont {P.~W.}\ \bibnamefont {Shor}},\ }in\ \href {https://doi.org/10.1109/SFCS.1994.365700} {\emph {\bibinfo {booktitle} {Proceedings of the 35th Annual Symposium on Foundations of Computer Science}}}\ (\bibinfo  {publisher} {IEEE},\ \bibinfo {year} {1994})\ pp.\ \bibinfo {pages} {124--134}\BibitemShut {NoStop}%
\bibitem [{\citenamefont {Steane}(1996)}]{steane_1996}%
  \BibitemOpen
  \bibfield  {author} {\bibinfo {author} {\bibfnamefont {A.~M.}\ \bibnamefont {Steane}},\ }\href {https://doi.org/10.1103/PhysRevLett.77.793} {\bibfield  {journal} {\bibinfo  {journal} {Physical Review Letters}\ }\textbf {\bibinfo {volume} {77}},\ \bibinfo {pages} {793} (\bibinfo {year} {1996})}\BibitemShut {NoStop}%
\bibitem [{\citenamefont {Shor}(1995)}]{shor_1995}%
  \BibitemOpen
  \bibfield  {author} {\bibinfo {author} {\bibfnamefont {P.~W.}\ \bibnamefont {Shor}},\ }\href {https://doi.org/10.1103/PhysRevA.52.R2493} {\bibfield  {journal} {\bibinfo  {journal} {Physical Review A}\ }\textbf {\bibinfo {volume} {52}},\ \bibinfo {pages} {R2493} (\bibinfo {year} {1995})}\BibitemShut {NoStop}%
\bibitem [{\citenamefont {Aharonov}\ and\ \citenamefont {Ben-Or}(1997)}]{aharonov_1997}%
  \BibitemOpen
  \bibfield  {author} {\bibinfo {author} {\bibfnamefont {D.}~\bibnamefont {Aharonov}}\ and\ \bibinfo {author} {\bibfnamefont {M.}~\bibnamefont {Ben-Or}},\ }in\ \href {https://doi.org/10.1145/258533.258579} {\emph {\bibinfo {booktitle} {Proceedings of the 29th Annual ACM Symposium on Theory of Computing}}}\ (\bibinfo  {publisher} {ACM},\ \bibinfo {year} {1997})\ pp.\ \bibinfo {pages} {176--188}\BibitemShut {NoStop}%
\bibitem [{\citenamefont {Kitaev}(1997)}]{kitaev_1997}%
  \BibitemOpen
  \bibfield  {author} {\bibinfo {author} {\bibfnamefont {A.}~\bibnamefont {Kitaev}},\ }\href {https://doi.org/10.1070/RM1997v052n06ABEH002155} {\bibfield  {journal} {\bibinfo  {journal} {Russian Mathematical Surveys}\ }\textbf {\bibinfo {volume} {52}},\ \bibinfo {pages} {1191} (\bibinfo {year} {1997})}\BibitemShut {NoStop}%
\bibitem [{\citenamefont {Knill}\ \emph {et~al.}(1996)\citenamefont {Knill}, \citenamefont {Laflamme},\ and\ \citenamefont {Zurek}}]{knill_laflamme_1996}%
  \BibitemOpen
  \bibfield  {author} {\bibinfo {author} {\bibfnamefont {E.}~\bibnamefont {Knill}}, \bibinfo {author} {\bibfnamefont {R.}~\bibnamefont {Laflamme}}, \ and\ \bibinfo {author} {\bibfnamefont {W.~H.}\ \bibnamefont {Zurek}},\ }\href {https://doi.org/10.1103/PhysRevLett.77.411} {\bibfield  {journal} {\bibinfo  {journal} {Physical Review Letters}\ }\textbf {\bibinfo {volume} {77}},\ \bibinfo {pages} {411} (\bibinfo {year} {1996})}\BibitemShut {NoStop}%
\bibitem [{\citenamefont {Aliferis}\ \emph {et~al.}(2007)\citenamefont {Aliferis}, \citenamefont {Gottesman},\ and\ \citenamefont {Preskill}}]{preskillaliferis}%
  \BibitemOpen
  \bibfield  {author} {\bibinfo {author} {\bibfnamefont {P.}~\bibnamefont {Aliferis}}, \bibinfo {author} {\bibfnamefont {D.}~\bibnamefont {Gottesman}}, \ and\ \bibinfo {author} {\bibfnamefont {J.}~\bibnamefont {Preskill}},\ }\href {https://arxiv.org/abs/quant-ph/0703264} {\enquote {\bibinfo {title} {Accuracy threshold for postselected quantum computation},}\ } (\bibinfo {year} {2007}),\ \Eprint {http://arxiv.org/abs/quant-ph/0703264} {arXiv:quant-ph/0703264 [quant-ph]} \BibitemShut {NoStop}%
\bibitem [{\citenamefont {{Google Quantum AI}}(2024)}]{google_quantum_ai}%
  \BibitemOpen
  \bibfield  {author} {\bibinfo {author} {\bibnamefont {{Google Quantum AI}}},\ }\href {https://quantumai.google/} {\enquote {\bibinfo {title} {{Google Quantum AI}},}\ } (\bibinfo {year} {2024}),\ \bibinfo {note} {accessed: 2024-10-21}\BibitemShut {NoStop}%
\bibitem [{\citenamefont {{IBM Quantum}}(2024)}]{ibm_quantum_technology}%
  \BibitemOpen
  \bibfield  {author} {\bibinfo {author} {\bibnamefont {{IBM Quantum}}},\ }\href {https://www.ibm.com/quantum/technology} {\enquote {\bibinfo {title} {Ibm quantum technology},}\ } (\bibinfo {year} {2024}),\ \bibinfo {note} {accessed: 2024-10-21}\BibitemShut {NoStop}%
\bibitem [{\citenamefont {{QuEra Computing}}(2024)}]{quera_qec_2024}%
  \BibitemOpen
  \bibfield  {author} {\bibinfo {author} {\bibnamefont {{QuEra Computing}}},\ }\href {https://www.quera.com/qec} {\enquote {\bibinfo {title} {Quera quantum error correction (qec)},}\ } (\bibinfo {year} {2024}),\ \bibinfo {note} {accessed: 2024-10-21}\BibitemShut {NoStop}%
\bibitem [{\citenamefont {Quantinuum}(2023)}]{quantinuum_roadmap_2023}%
  \BibitemOpen
  \bibfield  {author} {\bibinfo {author} {\bibnamefont {Quantinuum}},\ }\href {https://www.quantinuum.com/press-releases/quantinuum-unveils-accelerated-roadmap-to-achieve-universal-fault-tolerant-quantum-computing-by-2030} {\enquote {\bibinfo {title} {{Quantinuum Unveils Accelerated Roadmap to Achieve Universal Fault-Tolerant Quantum Computing by 2030}},}\ } (\bibinfo {year} {2023}),\ \bibinfo {note} {accessed: 2024-10-21}\BibitemShut {NoStop}%
\bibitem [{\citenamefont {{IonQ}}(2024)}]{ionq_roadmap_2024}%
  \BibitemOpen
  \bibfield  {author} {\bibinfo {author} {\bibnamefont {{IonQ}}},\ }\href {https://ionq.com/news/ionq-unveils-accelerated-roadmap-and-new-technical-milestones-to-propel} {\enquote {\bibinfo {title} {Ionq unveils accelerated roadmap and new technical milestones to propel quantum computing},}\ } (\bibinfo {year} {2024}),\ \bibinfo {note} {accessed: 2024-10-21}\BibitemShut {NoStop}%
\bibitem [{\citenamefont {{IQM Quantum Computers}}(2024)}]{iqm_roadmap_2024}%
  \BibitemOpen
  \bibfield  {author} {\bibinfo {author} {\bibnamefont {{IQM Quantum Computers}}},\ }\href {https://www.meetiqm.com/technology/roadmap} {\enquote {\bibinfo {title} {Iqm quantum computers technology roadmap},}\ } (\bibinfo {year} {2024}),\ \bibinfo {note} {accessed: 2024-10-21}\BibitemShut {NoStop}%
\bibitem [{\citenamefont {Fowler}\ \emph {et~al.}(2012)\citenamefont {Fowler}, \citenamefont {Mariantoni}, \citenamefont {Martinis},\ and\ \citenamefont {Cleland}}]{fowler2012surface}%
  \BibitemOpen
  \bibfield  {author} {\bibinfo {author} {\bibfnamefont {A.~G.}\ \bibnamefont {Fowler}}, \bibinfo {author} {\bibfnamefont {M.}~\bibnamefont {Mariantoni}}, \bibinfo {author} {\bibfnamefont {J.~M.}\ \bibnamefont {Martinis}}, \ and\ \bibinfo {author} {\bibfnamefont {A.~N.}\ \bibnamefont {Cleland}},\ }\href {https://journals.aps.org/pra/abstract/10.1103/PhysRevA.86.032324} {\bibfield  {journal} {\bibinfo  {journal} {Physical Review A—Atomic, Molecular, and Optical Physics}\ }\textbf {\bibinfo {volume} {86}},\ \bibinfo {pages} {032324} (\bibinfo {year} {2012})}\BibitemShut {NoStop}%
\bibitem [{\citenamefont {Wang}\ \emph {et~al.}(2011)\citenamefont {Wang}, \citenamefont {Fowler},\ and\ \citenamefont {Hollenberg}}]{wang2011surface}%
  \BibitemOpen
  \bibfield  {author} {\bibinfo {author} {\bibfnamefont {D.~S.}\ \bibnamefont {Wang}}, \bibinfo {author} {\bibfnamefont {A.~G.}\ \bibnamefont {Fowler}}, \ and\ \bibinfo {author} {\bibfnamefont {L.~C.}\ \bibnamefont {Hollenberg}},\ }\href {https://journals.aps.org/pra/abstract/10.1103/PhysRevA.83.020302} {\bibfield  {journal} {\bibinfo  {journal} {Physical Review A—Atomic, Molecular, and Optical Physics}\ }\textbf {\bibinfo {volume} {83}},\ \bibinfo {pages} {020302} (\bibinfo {year} {2011})}\BibitemShut {NoStop}%
\bibitem [{\citenamefont {Xu}\ \emph {et~al.}(2024)\citenamefont {Xu}, \citenamefont {Bonilla~Ataides}, \citenamefont {Pattison}, \citenamefont {Raveendran}, \citenamefont {Bluvstein}, \citenamefont {Wurtz}, \citenamefont {Vasi{\'c}}, \citenamefont {Lukin}, \citenamefont {Jiang},\ and\ \citenamefont {Zhou}}]{xu2024constant}%
  \BibitemOpen
  \bibfield  {author} {\bibinfo {author} {\bibfnamefont {Q.}~\bibnamefont {Xu}}, \bibinfo {author} {\bibfnamefont {J.~P.}\ \bibnamefont {Bonilla~Ataides}}, \bibinfo {author} {\bibfnamefont {C.~A.}\ \bibnamefont {Pattison}}, \bibinfo {author} {\bibfnamefont {N.}~\bibnamefont {Raveendran}}, \bibinfo {author} {\bibfnamefont {D.}~\bibnamefont {Bluvstein}}, \bibinfo {author} {\bibfnamefont {J.}~\bibnamefont {Wurtz}}, \bibinfo {author} {\bibfnamefont {B.}~\bibnamefont {Vasi{\'c}}}, \bibinfo {author} {\bibfnamefont {M.~D.}\ \bibnamefont {Lukin}}, \bibinfo {author} {\bibfnamefont {L.}~\bibnamefont {Jiang}}, \ and\ \bibinfo {author} {\bibfnamefont {H.}~\bibnamefont {Zhou}},\ }\href {https://www.nature.com/articles/s41567-024-02479-z} {\bibfield  {journal} {\bibinfo  {journal} {Nature Physics}\ ,\ \bibinfo {pages} {1}} (\bibinfo {year} {2024})}\BibitemShut {NoStop}%
\bibitem [{\citenamefont {Bravyi}\ \emph {et~al.}(2024)\citenamefont {Bravyi}, \citenamefont {Cross}, \citenamefont {Gambetta}, \citenamefont {Maslov}, \citenamefont {Rall},\ and\ \citenamefont {Yoder}}]{bravyi_2024}%
  \BibitemOpen
  \bibfield  {author} {\bibinfo {author} {\bibfnamefont {S.}~\bibnamefont {Bravyi}}, \bibinfo {author} {\bibfnamefont {A.~W.}\ \bibnamefont {Cross}}, \bibinfo {author} {\bibfnamefont {J.~M.}\ \bibnamefont {Gambetta}}, \bibinfo {author} {\bibfnamefont {D.}~\bibnamefont {Maslov}}, \bibinfo {author} {\bibfnamefont {P.}~\bibnamefont {Rall}}, \ and\ \bibinfo {author} {\bibfnamefont {T.~J.}\ \bibnamefont {Yoder}},\ }\href {https://arxiv.org/abs/2308.07915} {\bibfield  {journal} {\bibinfo  {journal} {Nature}\ }\textbf {\bibinfo {volume} {627}},\ \bibinfo {pages} {78} (\bibinfo {year} {2024})},\ \bibinfo {note} {accessed: 2024-10-21}\BibitemShut {NoStop}%
\bibitem [{\citenamefont {Scruby}\ \emph {et~al.}(2024)\citenamefont {Scruby}, \citenamefont {Hillmann},\ and\ \citenamefont {Roffe}}]{scruby2024high}%
  \BibitemOpen
  \bibfield  {author} {\bibinfo {author} {\bibfnamefont {T.~R.}\ \bibnamefont {Scruby}}, \bibinfo {author} {\bibfnamefont {T.}~\bibnamefont {Hillmann}}, \ and\ \bibinfo {author} {\bibfnamefont {J.}~\bibnamefont {Roffe}},\ }\href {https://arxiv.org/abs/2406.14445} {\bibfield  {journal} {\bibinfo  {journal} {arXiv preprint arXiv:2406.14445}\ } (\bibinfo {year} {2024})}\BibitemShut {NoStop}%
\bibitem [{\citenamefont {Yamasaki}\ and\ \citenamefont {Koashi}(2024)}]{yamasaki2024time}%
  \BibitemOpen
  \bibfield  {author} {\bibinfo {author} {\bibfnamefont {H.}~\bibnamefont {Yamasaki}}\ and\ \bibinfo {author} {\bibfnamefont {M.}~\bibnamefont {Koashi}},\ }\href {https://www.nature.com/articles/s41567-023-02325-8} {\bibfield  {journal} {\bibinfo  {journal} {Nature Physics}\ }\textbf {\bibinfo {volume} {20}},\ \bibinfo {pages} {247} (\bibinfo {year} {2024})}\BibitemShut {NoStop}%
\bibitem [{\citenamefont {Goto}\ \emph {et~al.}(2024)\citenamefont {Goto} \emph {et~al.}}]{goto2024high}%
  \BibitemOpen
  \bibfield  {author} {\bibinfo {author} {\bibfnamefont {H.}~\bibnamefont {Goto}} \emph {et~al.},\ }\href {\doibase 10.1126/sciadv.adp6388} {\bibfield  {journal} {\bibinfo  {journal} {Science Advances}\ }\textbf {\bibinfo {volume} {10}},\ \bibinfo {pages} {eadp6388} (\bibinfo {year} {2024})}\BibitemShut {NoStop}%
\bibitem [{\citenamefont {Bluvstein}\ \emph {et~al.}(2024)\citenamefont {Bluvstein}, \citenamefont {Evered}, \citenamefont {Geim}, \citenamefont {Li}, \citenamefont {Zhou}, \citenamefont {Manovitz}, \citenamefont {Ebadi}, \citenamefont {Cain}, \citenamefont {Kalinowski}, \citenamefont {Hangleiter} \emph {et~al.}}]{1-bluvstein2024logical}%
  \BibitemOpen
  \bibfield  {author} {\bibinfo {author} {\bibfnamefont {D.}~\bibnamefont {Bluvstein}}, \bibinfo {author} {\bibfnamefont {S.~J.}\ \bibnamefont {Evered}}, \bibinfo {author} {\bibfnamefont {A.~A.}\ \bibnamefont {Geim}}, \bibinfo {author} {\bibfnamefont {S.~H.}\ \bibnamefont {Li}}, \bibinfo {author} {\bibfnamefont {H.}~\bibnamefont {Zhou}}, \bibinfo {author} {\bibfnamefont {T.}~\bibnamefont {Manovitz}}, \bibinfo {author} {\bibfnamefont {S.}~\bibnamefont {Ebadi}}, \bibinfo {author} {\bibfnamefont {M.}~\bibnamefont {Cain}}, \bibinfo {author} {\bibfnamefont {M.}~\bibnamefont {Kalinowski}}, \bibinfo {author} {\bibfnamefont {D.}~\bibnamefont {Hangleiter}},  \emph {et~al.},\ }\href {https://www.nature.com/articles/s41586-023-06927-3} {\bibfield  {journal} {\bibinfo  {journal} {Nature}\ }\textbf {\bibinfo {volume} {626}},\ \bibinfo {pages} {58} (\bibinfo {year} {2024})}\BibitemShut {NoStop}%
\bibitem [{\citenamefont {Mayer}\ \emph {et~al.}(2024)\citenamefont {Mayer}, \citenamefont {Ryan-Anderson}, \citenamefont {Brown}, \citenamefont {Durso-Sabina}, \citenamefont {Baldwin}, \citenamefont {Hayes}, \citenamefont {Dreiling}, \citenamefont {Foltz}, \citenamefont {Gaebler}, \citenamefont {Gatterman} \emph {et~al.}}]{2-mayer2024benchmarking}%
  \BibitemOpen
  \bibfield  {author} {\bibinfo {author} {\bibfnamefont {K.}~\bibnamefont {Mayer}}, \bibinfo {author} {\bibfnamefont {C.}~\bibnamefont {Ryan-Anderson}}, \bibinfo {author} {\bibfnamefont {N.}~\bibnamefont {Brown}}, \bibinfo {author} {\bibfnamefont {E.}~\bibnamefont {Durso-Sabina}}, \bibinfo {author} {\bibfnamefont {C.~H.}\ \bibnamefont {Baldwin}}, \bibinfo {author} {\bibfnamefont {D.}~\bibnamefont {Hayes}}, \bibinfo {author} {\bibfnamefont {J.~M.}\ \bibnamefont {Dreiling}}, \bibinfo {author} {\bibfnamefont {C.}~\bibnamefont {Foltz}}, \bibinfo {author} {\bibfnamefont {J.~P.}\ \bibnamefont {Gaebler}}, \bibinfo {author} {\bibfnamefont {T.~M.}\ \bibnamefont {Gatterman}},  \emph {et~al.},\ }\href {https://arxiv.org/abs/2404.08616} {\bibfield  {journal} {\bibinfo  {journal} {arXiv preprint arXiv:2404.08616}\ } (\bibinfo {year} {2024})}\BibitemShut {NoStop}%
\bibitem [{\citenamefont {da~Silva}\ \emph {et~al.}()\citenamefont {da~Silva}, \citenamefont {Ryan-Anderson}, \citenamefont {Bello-Rivas}, \citenamefont {Chernoguzov}, \citenamefont {Dreiling}, \citenamefont {Foltz}, \citenamefont {Frachon}, \citenamefont {Gaebler}, \citenamefont {Gatterman}, \citenamefont {Grans-Samuelsson} \emph {et~al.}}]{3-da2404demonstration}%
  \BibitemOpen
  \bibfield  {author} {\bibinfo {author} {\bibfnamefont {M.}~\bibnamefont {da~Silva}}, \bibinfo {author} {\bibfnamefont {C.}~\bibnamefont {Ryan-Anderson}}, \bibinfo {author} {\bibfnamefont {J.}~\bibnamefont {Bello-Rivas}}, \bibinfo {author} {\bibfnamefont {A.}~\bibnamefont {Chernoguzov}}, \bibinfo {author} {\bibfnamefont {J.}~\bibnamefont {Dreiling}}, \bibinfo {author} {\bibfnamefont {C.}~\bibnamefont {Foltz}}, \bibinfo {author} {\bibfnamefont {F.}~\bibnamefont {Frachon}}, \bibinfo {author} {\bibfnamefont {J.}~\bibnamefont {Gaebler}}, \bibinfo {author} {\bibfnamefont {T.}~\bibnamefont {Gatterman}}, \bibinfo {author} {\bibfnamefont {L.}~\bibnamefont {Grans-Samuelsson}},  \emph {et~al.},\ }\href {https://arxiv.org/abs/2404.02280} {\bibinfo  {journal} {arXiv preprint arXiv:2404.02280}\ }\BibitemShut {NoStop}%
\bibitem [{\citenamefont {Reichardt}\ \emph {et~al.}(2024)\citenamefont {Reichardt}, \citenamefont {Aasen}, \citenamefont {Chao}, \citenamefont {Chernoguzov}, \citenamefont {van Dam}, \citenamefont {Gaebler}, \citenamefont {Gresh}, \citenamefont {Lucchetti}, \citenamefont {Mills}, \citenamefont {Moses} \emph {et~al.}}]{4-reichardt2024demonstration}%
  \BibitemOpen
\bibfield  {journal} {  }\bibfield  {author} {\bibinfo {author} {\bibfnamefont {B.~W.}\ \bibnamefont {Reichardt}}, \bibinfo {author} {\bibfnamefont {D.}~\bibnamefont {Aasen}}, \bibinfo {author} {\bibfnamefont {R.}~\bibnamefont {Chao}}, \bibinfo {author} {\bibfnamefont {A.}~\bibnamefont {Chernoguzov}}, \bibinfo {author} {\bibfnamefont {W.}~\bibnamefont {van Dam}}, \bibinfo {author} {\bibfnamefont {J.~P.}\ \bibnamefont {Gaebler}}, \bibinfo {author} {\bibfnamefont {D.}~\bibnamefont {Gresh}}, \bibinfo {author} {\bibfnamefont {D.}~\bibnamefont {Lucchetti}}, \bibinfo {author} {\bibfnamefont {M.}~\bibnamefont {Mills}}, \bibinfo {author} {\bibfnamefont {S.~A.}\ \bibnamefont {Moses}},  \emph {et~al.},\ }\href {https://arxiv.org/abs/2409.04628} {\bibfield  {journal} {\bibinfo  {journal} {arXiv preprint arXiv:2409.04628}\ } (\bibinfo {year} {2024})}\BibitemShut {NoStop}%
\bibitem [{\citenamefont {Acharya}\ \emph {et~al.}(2024)\citenamefont {Acharya}, \citenamefont {Aghababaie-Beni}, \citenamefont {Aleiner}, \citenamefont {Andersen}, \citenamefont {Ansmann}, \citenamefont {Arute}, \citenamefont {Arya}, \citenamefont {Asfaw}, \citenamefont {Astrakhantsev}, \citenamefont {Atalaya} \emph {et~al.}}]{5-acharya2024quantum}%
  \BibitemOpen
  \bibfield  {author} {\bibinfo {author} {\bibfnamefont {R.}~\bibnamefont {Acharya}}, \bibinfo {author} {\bibfnamefont {L.}~\bibnamefont {Aghababaie-Beni}}, \bibinfo {author} {\bibfnamefont {I.}~\bibnamefont {Aleiner}}, \bibinfo {author} {\bibfnamefont {T.~I.}\ \bibnamefont {Andersen}}, \bibinfo {author} {\bibfnamefont {M.}~\bibnamefont {Ansmann}}, \bibinfo {author} {\bibfnamefont {F.}~\bibnamefont {Arute}}, \bibinfo {author} {\bibfnamefont {K.}~\bibnamefont {Arya}}, \bibinfo {author} {\bibfnamefont {A.}~\bibnamefont {Asfaw}}, \bibinfo {author} {\bibfnamefont {N.}~\bibnamefont {Astrakhantsev}}, \bibinfo {author} {\bibfnamefont {J.}~\bibnamefont {Atalaya}},  \emph {et~al.},\ }\href {https://arxiv.org/abs/2408.13687} {\bibfield  {journal} {\bibinfo  {journal} {arXiv preprint arXiv:2408.13687}\ } (\bibinfo {year} {2024})}\BibitemShut {NoStop}%
\bibitem [{\citenamefont {Putterman}\ \emph {et~al.}(2024)\citenamefont {Putterman}, \citenamefont {Noh}, \citenamefont {Hann}, \citenamefont {MacCabe}, \citenamefont {Aghaeimeibodi}, \citenamefont {Patel}, \citenamefont {Lee}, \citenamefont {Jones}, \citenamefont {Moradinejad}, \citenamefont {Rodriguez} \emph {et~al.}}]{putterman2024hardware}%
  \BibitemOpen
  \bibfield  {author} {\bibinfo {author} {\bibfnamefont {H.}~\bibnamefont {Putterman}}, \bibinfo {author} {\bibfnamefont {K.}~\bibnamefont {Noh}}, \bibinfo {author} {\bibfnamefont {C.~T.}\ \bibnamefont {Hann}}, \bibinfo {author} {\bibfnamefont {G.~S.}\ \bibnamefont {MacCabe}}, \bibinfo {author} {\bibfnamefont {S.}~\bibnamefont {Aghaeimeibodi}}, \bibinfo {author} {\bibfnamefont {R.~N.}\ \bibnamefont {Patel}}, \bibinfo {author} {\bibfnamefont {M.}~\bibnamefont {Lee}}, \bibinfo {author} {\bibfnamefont {W.~M.}\ \bibnamefont {Jones}}, \bibinfo {author} {\bibfnamefont {H.}~\bibnamefont {Moradinejad}}, \bibinfo {author} {\bibfnamefont {R.}~\bibnamefont {Rodriguez}},  \emph {et~al.},\ }\href@noop {} {\bibfield  {journal} {\bibinfo  {journal} {arXiv preprint arXiv:2409.13025}\ } (\bibinfo {year} {2024})}\BibitemShut {NoStop}%
\bibitem [{\citenamefont {Arute}\ \emph {et~al.}(2019)\citenamefont {Arute}, \citenamefont {Arya}, \citenamefont {Babbush} \emph {et~al.}}]{googlesupremacy1}%
  \BibitemOpen
  \bibfield  {author} {\bibinfo {author} {\bibfnamefont {F.}~\bibnamefont {Arute}}, \bibinfo {author} {\bibfnamefont {K.}~\bibnamefont {Arya}}, \bibinfo {author} {\bibfnamefont {R.}~\bibnamefont {Babbush}},  \emph {et~al.},\ }\href {\doibase 10.1038/s41586-019-1666-5} {\bibfield  {journal} {\bibinfo  {journal} {Nature}\ }\textbf {\bibinfo {volume} {574}},\ \bibinfo {pages} {505} (\bibinfo {year} {2019})}\BibitemShut {NoStop}%
\bibitem [{\citenamefont {Wu}\ \emph {et~al.}(2021)\citenamefont {Wu}, \citenamefont {Bao}, \citenamefont {Cao}, \citenamefont {Chen}, \citenamefont {Chen}, \citenamefont {Chen}, \citenamefont {Chung}, \citenamefont {Deng}, \citenamefont {Du}, \citenamefont {Fan} \emph {et~al.}}]{wu2021strong}%
  \BibitemOpen
  \bibfield  {author} {\bibinfo {author} {\bibfnamefont {Y.}~\bibnamefont {Wu}}, \bibinfo {author} {\bibfnamefont {W.-S.}\ \bibnamefont {Bao}}, \bibinfo {author} {\bibfnamefont {S.}~\bibnamefont {Cao}}, \bibinfo {author} {\bibfnamefont {F.}~\bibnamefont {Chen}}, \bibinfo {author} {\bibfnamefont {M.-C.}\ \bibnamefont {Chen}}, \bibinfo {author} {\bibfnamefont {X.}~\bibnamefont {Chen}}, \bibinfo {author} {\bibfnamefont {T.-H.}\ \bibnamefont {Chung}}, \bibinfo {author} {\bibfnamefont {H.}~\bibnamefont {Deng}}, \bibinfo {author} {\bibfnamefont {Y.}~\bibnamefont {Du}}, \bibinfo {author} {\bibfnamefont {D.}~\bibnamefont {Fan}},  \emph {et~al.},\ }\href@noop {} {\bibfield  {journal} {\bibinfo  {journal} {Physical review letters}\ }\textbf {\bibinfo {volume} {127}},\ \bibinfo {pages} {180501} (\bibinfo {year} {2021})}\BibitemShut {NoStop}%
\bibitem [{\citenamefont {Pan}\ \emph {et~al.}(2020)\citenamefont {Pan}, \citenamefont {Lu}, \citenamefont {He} \emph {et~al.}}]{QAphotons}%
  \BibitemOpen
  \bibfield  {author} {\bibinfo {author} {\bibfnamefont {J.-W.}\ \bibnamefont {Pan}}, \bibinfo {author} {\bibfnamefont {C.-Y.}\ \bibnamefont {Lu}}, \bibinfo {author} {\bibfnamefont {Y.}~\bibnamefont {He}},  \emph {et~al.},\ }\href {\doibase 10.1126/science.abe8770} {\bibfield  {journal} {\bibinfo  {journal} {Science}\ }\textbf {\bibinfo {volume} {370}},\ \bibinfo {pages} {1460} (\bibinfo {year} {2020})}\BibitemShut {NoStop}%
\bibitem [{\citenamefont {Zhu}\ \emph {et~al.}(2022)\citenamefont {Zhu}, \citenamefont {Cao}, \citenamefont {Chen}, \citenamefont {Chen}, \citenamefont {Chen}, \citenamefont {Chung}, \citenamefont {Deng}, \citenamefont {Du}, \citenamefont {Fan}, \citenamefont {Gong} \emph {et~al.}}]{zhu2022quantum}%
  \BibitemOpen
  \bibfield  {author} {\bibinfo {author} {\bibfnamefont {Q.}~\bibnamefont {Zhu}}, \bibinfo {author} {\bibfnamefont {S.}~\bibnamefont {Cao}}, \bibinfo {author} {\bibfnamefont {F.}~\bibnamefont {Chen}}, \bibinfo {author} {\bibfnamefont {M.-C.}\ \bibnamefont {Chen}}, \bibinfo {author} {\bibfnamefont {X.}~\bibnamefont {Chen}}, \bibinfo {author} {\bibfnamefont {T.-H.}\ \bibnamefont {Chung}}, \bibinfo {author} {\bibfnamefont {H.}~\bibnamefont {Deng}}, \bibinfo {author} {\bibfnamefont {Y.}~\bibnamefont {Du}}, \bibinfo {author} {\bibfnamefont {D.}~\bibnamefont {Fan}}, \bibinfo {author} {\bibfnamefont {M.}~\bibnamefont {Gong}},  \emph {et~al.},\ }\href {https://www.sciencedirect.com/science/article/pii/S2095927321006733?casa_token=D5nwo1qClcEAAAAA:Vhy0l_X2MyTc815hnyntbs_zafO9MZV2xnVDMkMrjlltn9J2XPm9lRnWoXiKvsR2UBBxaFf1uNE} {\bibfield  {journal} {\bibinfo  {journal} {Science bulletin}\ }\textbf {\bibinfo {volume} {67}},\ \bibinfo {pages} {240} (\bibinfo {year} {2022})}\BibitemShut {NoStop}%
\bibitem [{\citenamefont {Morvan}\ \emph {et~al.}(2024)\citenamefont {Morvan}, \citenamefont {Villalonga}, \citenamefont {Mi}, \citenamefont {Mandr\`a}, \citenamefont {Bengtsson}, \citenamefont {Klimov}, \citenamefont {Chen}, \citenamefont {Hong}, \citenamefont {Erickson}, \citenamefont {Drozdov} \emph {et~al.}}]{Morvan2024}%
  \BibitemOpen
  \bibfield  {author} {\bibinfo {author} {\bibfnamefont {A.}~\bibnamefont {Morvan}}, \bibinfo {author} {\bibfnamefont {B.}~\bibnamefont {Villalonga}}, \bibinfo {author} {\bibfnamefont {X.}~\bibnamefont {Mi}}, \bibinfo {author} {\bibfnamefont {S.}~\bibnamefont {Mandr\`a}}, \bibinfo {author} {\bibfnamefont {A.}~\bibnamefont {Bengtsson}}, \bibinfo {author} {\bibfnamefont {P.~V.}\ \bibnamefont {Klimov}}, \bibinfo {author} {\bibfnamefont {Z.}~\bibnamefont {Chen}}, \bibinfo {author} {\bibfnamefont {S.}~\bibnamefont {Hong}}, \bibinfo {author} {\bibfnamefont {C.}~\bibnamefont {Erickson}}, \bibinfo {author} {\bibfnamefont {I.~K.}\ \bibnamefont {Drozdov}},  \emph {et~al.},\ }\href {\doibase 10.1038/s41586-024-07998-6} {\bibfield  {journal} {\bibinfo  {journal} {Nature}\ }\textbf {\bibinfo {volume} {618}},\ \bibinfo {pages} {56} (\bibinfo {year} {2024})}\BibitemShut {NoStop}%
\bibitem [{\citenamefont {DeCross}\ \emph {et~al.}(2024)\citenamefont {DeCross}, \citenamefont {Haghshenas}, \citenamefont {Liu}, \citenamefont {Rinaldi}, \citenamefont {Gray}, \citenamefont {Alexeev}, \citenamefont {Baldwin}, \citenamefont {Bartolotta}, \citenamefont {Bohn}, \citenamefont {Chertkov} \emph {et~al.}}]{decross2024computational}%
  \BibitemOpen
  \bibfield  {author} {\bibinfo {author} {\bibfnamefont {M.}~\bibnamefont {DeCross}}, \bibinfo {author} {\bibfnamefont {R.}~\bibnamefont {Haghshenas}}, \bibinfo {author} {\bibfnamefont {M.}~\bibnamefont {Liu}}, \bibinfo {author} {\bibfnamefont {E.}~\bibnamefont {Rinaldi}}, \bibinfo {author} {\bibfnamefont {J.}~\bibnamefont {Gray}}, \bibinfo {author} {\bibfnamefont {Y.}~\bibnamefont {Alexeev}}, \bibinfo {author} {\bibfnamefont {C.~H.}\ \bibnamefont {Baldwin}}, \bibinfo {author} {\bibfnamefont {J.~P.}\ \bibnamefont {Bartolotta}}, \bibinfo {author} {\bibfnamefont {M.}~\bibnamefont {Bohn}}, \bibinfo {author} {\bibfnamefont {E.}~\bibnamefont {Chertkov}},  \emph {et~al.},\ }\href@noop {} {\bibfield  {journal} {\bibinfo  {journal} {arXiv preprint arXiv:2406.02501}\ } (\bibinfo {year} {2024})}\BibitemShut {NoStop}%
\bibitem [{\citenamefont {Aaronson}\ and\ \citenamefont {Hung}(2023)}]{scottcertifiable}%
  \BibitemOpen
  \bibfield  {author} {\bibinfo {author} {\bibfnamefont {S.}~\bibnamefont {Aaronson}}\ and\ \bibinfo {author} {\bibfnamefont {S.-H.}\ \bibnamefont {Hung}},\ }in\ \href {\doibase 10.1145/3564246.3585145} {\emph {\bibinfo {booktitle} {Proceedings of the 55th Annual ACM Symposium on Theory of Computing}}},\ \bibinfo {series and number} {STOC 2023}\ (\bibinfo  {publisher} {Association for Computing Machinery},\ \bibinfo {address} {New York, NY, USA},\ \bibinfo {year} {2023})\ p.\ \bibinfo {pages} {933–944}\BibitemShut {NoStop}%
\bibitem [{\citenamefont {Cai}\ \emph {et~al.}(2023)\citenamefont {Cai}, \citenamefont {Babbush}, \citenamefont {Benjamin}, \citenamefont {Endo}, \citenamefont {Huggins}, \citenamefont {Li}, \citenamefont {McClean},\ and\ \citenamefont {O'Brien}}]{emreview}%
  \BibitemOpen
  \bibfield  {author} {\bibinfo {author} {\bibfnamefont {Z.}~\bibnamefont {Cai}}, \bibinfo {author} {\bibfnamefont {R.}~\bibnamefont {Babbush}}, \bibinfo {author} {\bibfnamefont {S.~C.}\ \bibnamefont {Benjamin}}, \bibinfo {author} {\bibfnamefont {S.}~\bibnamefont {Endo}}, \bibinfo {author} {\bibfnamefont {W.~J.}\ \bibnamefont {Huggins}}, \bibinfo {author} {\bibfnamefont {Y.}~\bibnamefont {Li}}, \bibinfo {author} {\bibfnamefont {J.~R.}\ \bibnamefont {McClean}}, \ and\ \bibinfo {author} {\bibfnamefont {T.~E.}\ \bibnamefont {O'Brien}},\ }\href {\doibase 10.1103/RevModPhys.95.045005} {\bibfield  {journal} {\bibinfo  {journal} {Rev. Mod. Phys.}\ }\textbf {\bibinfo {volume} {95}},\ \bibinfo {pages} {045005} (\bibinfo {year} {2023})}\BibitemShut {NoStop}%
\bibitem [{\citenamefont {Temme}\ \emph {et~al.}(2017)\citenamefont {Temme}, \citenamefont {Bravyi},\ and\ \citenamefont {Gambetta}}]{bravyitemmegambetta}%
  \BibitemOpen
  \bibfield  {author} {\bibinfo {author} {\bibfnamefont {K.}~\bibnamefont {Temme}}, \bibinfo {author} {\bibfnamefont {S.}~\bibnamefont {Bravyi}}, \ and\ \bibinfo {author} {\bibfnamefont {J.~M.}\ \bibnamefont {Gambetta}},\ }\href {\doibase 10.1103/PhysRevLett.119.180509} {\bibfield  {journal} {\bibinfo  {journal} {Phys. Rev. Lett.}\ }\textbf {\bibinfo {volume} {119}},\ \bibinfo {pages} {180509} (\bibinfo {year} {2017})}\BibitemShut {NoStop}%
\bibitem [{\citenamefont {Li}\ and\ \citenamefont {Benjamin}(2017)}]{li2017efficient}%
  \BibitemOpen
  \bibfield  {author} {\bibinfo {author} {\bibfnamefont {Y.}~\bibnamefont {Li}}\ and\ \bibinfo {author} {\bibfnamefont {S.~C.}\ \bibnamefont {Benjamin}},\ }\href {https://journals.aps.org/prx/abstract/10.1103/PhysRevX.7.021050} {\bibfield  {journal} {\bibinfo  {journal} {Physical Review X}\ }\textbf {\bibinfo {volume} {7}},\ \bibinfo {pages} {021050} (\bibinfo {year} {2017})}\BibitemShut {NoStop}%
\bibitem [{\citenamefont {Endo}\ \emph {et~al.}(2018)\citenamefont {Endo}, \citenamefont {Benjamin},\ and\ \citenamefont {Li}}]{PhysRevX.8.031027}%
  \BibitemOpen
  \bibfield  {author} {\bibinfo {author} {\bibfnamefont {S.}~\bibnamefont {Endo}}, \bibinfo {author} {\bibfnamefont {S.~C.}\ \bibnamefont {Benjamin}}, \ and\ \bibinfo {author} {\bibfnamefont {Y.}~\bibnamefont {Li}},\ }\href {\doibase 10.1103/PhysRevX.8.031027} {\bibfield  {journal} {\bibinfo  {journal} {Phys. Rev. X}\ }\textbf {\bibinfo {volume} {8}},\ \bibinfo {pages} {031027} (\bibinfo {year} {2018})}\BibitemShut {NoStop}%
\bibitem [{\citenamefont {Mari}\ \emph {et~al.}(2021)\citenamefont {Mari}, \citenamefont {Shammah},\ and\ \citenamefont {Zeng}}]{PhysRevA.104.052607}%
  \BibitemOpen
  \bibfield  {author} {\bibinfo {author} {\bibfnamefont {A.}~\bibnamefont {Mari}}, \bibinfo {author} {\bibfnamefont {N.}~\bibnamefont {Shammah}}, \ and\ \bibinfo {author} {\bibfnamefont {W.~J.}\ \bibnamefont {Zeng}},\ }\href {\doibase 10.1103/PhysRevA.104.052607} {\bibfield  {journal} {\bibinfo  {journal} {Phys. Rev. A}\ }\textbf {\bibinfo {volume} {104}},\ \bibinfo {pages} {052607} (\bibinfo {year} {2021})}\BibitemShut {NoStop}%
\bibitem [{\citenamefont {Cai}(2021)}]{cai2021multi}%
  \BibitemOpen
  \bibfield  {author} {\bibinfo {author} {\bibfnamefont {Z.}~\bibnamefont {Cai}},\ }\href {https://www.nature.com/articles/s41534-021-00404-3} {\bibfield  {journal} {\bibinfo  {journal} {npj Quantum Information}\ }\textbf {\bibinfo {volume} {7}},\ \bibinfo {pages} {80} (\bibinfo {year} {2021})}\BibitemShut {NoStop}%
\bibitem [{\citenamefont {Filippov}\ \emph {et~al.}(2024)\citenamefont {Filippov}, \citenamefont {Maniscalco},\ and\ \citenamefont {Garc{\'\i}a-P{\'e}rez}}]{filippov2024scalability}%
  \BibitemOpen
  \bibfield  {author} {\bibinfo {author} {\bibfnamefont {S.~N.}\ \bibnamefont {Filippov}}, \bibinfo {author} {\bibfnamefont {S.}~\bibnamefont {Maniscalco}}, \ and\ \bibinfo {author} {\bibfnamefont {G.}~\bibnamefont {Garc{\'\i}a-P{\'e}rez}},\ }\href {https://arxiv.org/abs/2403.13542} {\bibfield  {journal} {\bibinfo  {journal} {arXiv preprint arXiv:2403.13542}\ } (\bibinfo {year} {2024})}\BibitemShut {NoStop}%
\bibitem [{\citenamefont {Kim}\ \emph {et~al.}(2023)\citenamefont {Kim}, \citenamefont {Eddins}, \citenamefont {Anand}, \citenamefont {Wei}, \citenamefont {Van Den~Berg}, \citenamefont {Rosenblatt}, \citenamefont {Nayfeh}, \citenamefont {Wu}, \citenamefont {Zaletel}, \citenamefont {Temme} \emph {et~al.}}]{kim2023evidence}%
  \BibitemOpen
  \bibfield  {author} {\bibinfo {author} {\bibfnamefont {Y.}~\bibnamefont {Kim}}, \bibinfo {author} {\bibfnamefont {A.}~\bibnamefont {Eddins}}, \bibinfo {author} {\bibfnamefont {S.}~\bibnamefont {Anand}}, \bibinfo {author} {\bibfnamefont {K.~X.}\ \bibnamefont {Wei}}, \bibinfo {author} {\bibfnamefont {E.}~\bibnamefont {Van Den~Berg}}, \bibinfo {author} {\bibfnamefont {S.}~\bibnamefont {Rosenblatt}}, \bibinfo {author} {\bibfnamefont {H.}~\bibnamefont {Nayfeh}}, \bibinfo {author} {\bibfnamefont {Y.}~\bibnamefont {Wu}}, \bibinfo {author} {\bibfnamefont {M.}~\bibnamefont {Zaletel}}, \bibinfo {author} {\bibfnamefont {K.}~\bibnamefont {Temme}},  \emph {et~al.},\ }\href {https://www.nature.com/articles/s41586-023-06096-3} {\bibfield  {journal} {\bibinfo  {journal} {Nature}\ }\textbf {\bibinfo {volume} {618}},\ \bibinfo {pages} {500} (\bibinfo {year} {2023})}\BibitemShut {NoStop}%
\bibitem [{\citenamefont {Shinjo}\ \emph {et~al.}(2024)\citenamefont {Shinjo}, \citenamefont {Seki}, \citenamefont {Shirakawa}, \citenamefont {Sun},\ and\ \citenamefont {Yunoki}}]{shinjo2024unveiling}%
  \BibitemOpen
  \bibfield  {author} {\bibinfo {author} {\bibfnamefont {K.}~\bibnamefont {Shinjo}}, \bibinfo {author} {\bibfnamefont {K.}~\bibnamefont {Seki}}, \bibinfo {author} {\bibfnamefont {T.}~\bibnamefont {Shirakawa}}, \bibinfo {author} {\bibfnamefont {R.-Y.}\ \bibnamefont {Sun}}, \ and\ \bibinfo {author} {\bibfnamefont {S.}~\bibnamefont {Yunoki}},\ }\href {https://arxiv.org/abs/2403.16718} {\bibfield  {journal} {\bibinfo  {journal} {arXiv preprint arXiv:2403.16718}\ } (\bibinfo {year} {2024})}\BibitemShut {NoStop}%
\bibitem [{\citenamefont {Yu}\ \emph {et~al.}(2023)\citenamefont {Yu}, \citenamefont {Zhao},\ and\ \citenamefont {Wei}}]{yu2023simulating}%
  \BibitemOpen
  \bibfield  {author} {\bibinfo {author} {\bibfnamefont {H.}~\bibnamefont {Yu}}, \bibinfo {author} {\bibfnamefont {Y.}~\bibnamefont {Zhao}}, \ and\ \bibinfo {author} {\bibfnamefont {T.-C.}\ \bibnamefont {Wei}},\ }\href {https://journals.aps.org/prresearch/abstract/10.1103/PhysRevResearch.5.013183} {\bibfield  {journal} {\bibinfo  {journal} {Physical Review Research}\ }\textbf {\bibinfo {volume} {5}},\ \bibinfo {pages} {013183} (\bibinfo {year} {2023})}\BibitemShut {NoStop}%
\bibitem [{\citenamefont {Mi}\ \emph {et~al.}(2024)\citenamefont {Mi}, \citenamefont {Michailidis}, \citenamefont {Shabani}, \citenamefont {Miao}, \citenamefont {Klimov}, \citenamefont {Lloyd}, \citenamefont {Rosenberg}, \citenamefont {Acharya}, \citenamefont {Aleiner}, \citenamefont {Andersen} \emph {et~al.}}]{mi2024stable}%
  \BibitemOpen
  \bibfield  {author} {\bibinfo {author} {\bibfnamefont {X.}~\bibnamefont {Mi}}, \bibinfo {author} {\bibfnamefont {A.}~\bibnamefont {Michailidis}}, \bibinfo {author} {\bibfnamefont {S.}~\bibnamefont {Shabani}}, \bibinfo {author} {\bibfnamefont {K.}~\bibnamefont {Miao}}, \bibinfo {author} {\bibfnamefont {P.}~\bibnamefont {Klimov}}, \bibinfo {author} {\bibfnamefont {J.}~\bibnamefont {Lloyd}}, \bibinfo {author} {\bibfnamefont {E.}~\bibnamefont {Rosenberg}}, \bibinfo {author} {\bibfnamefont {R.}~\bibnamefont {Acharya}}, \bibinfo {author} {\bibfnamefont {I.}~\bibnamefont {Aleiner}}, \bibinfo {author} {\bibfnamefont {T.}~\bibnamefont {Andersen}},  \emph {et~al.},\ }\href {https://www.science.org/doi/abs/10.1126/science.adh9932} {\bibfield  {journal} {\bibinfo  {journal} {Science}\ }\textbf {\bibinfo {volume} {383}},\ \bibinfo {pages} {1332} (\bibinfo {year} {2024})}\BibitemShut {NoStop}%
\bibitem [{\citenamefont {Gyawali}\ \emph {et~al.}(2024)\citenamefont {Gyawali}, \citenamefont {Cochran}, \citenamefont {Lensky}, \citenamefont {Rosenberg}, \citenamefont {Karamlou}, \citenamefont {Kechedzhi}, \citenamefont {Berndtsson}, \citenamefont {Westerhout}, \citenamefont {Asfaw}, \citenamefont {Abanin} \emph {et~al.}}]{gyawali2024observation}%
  \BibitemOpen
  \bibfield  {author} {\bibinfo {author} {\bibfnamefont {G.}~\bibnamefont {Gyawali}}, \bibinfo {author} {\bibfnamefont {T.}~\bibnamefont {Cochran}}, \bibinfo {author} {\bibfnamefont {Y.}~\bibnamefont {Lensky}}, \bibinfo {author} {\bibfnamefont {E.}~\bibnamefont {Rosenberg}}, \bibinfo {author} {\bibfnamefont {A.~H.}\ \bibnamefont {Karamlou}}, \bibinfo {author} {\bibfnamefont {K.}~\bibnamefont {Kechedzhi}}, \bibinfo {author} {\bibfnamefont {J.}~\bibnamefont {Berndtsson}}, \bibinfo {author} {\bibfnamefont {T.}~\bibnamefont {Westerhout}}, \bibinfo {author} {\bibfnamefont {A.}~\bibnamefont {Asfaw}}, \bibinfo {author} {\bibfnamefont {D.}~\bibnamefont {Abanin}},  \emph {et~al.},\ }\href {https://arxiv.org/abs/2410.06557} {\bibfield  {journal} {\bibinfo  {journal} {arXiv preprint arXiv:2410.06557}\ } (\bibinfo {year} {2024})}\BibitemShut {NoStop}%
\bibitem [{\citenamefont {Iqbal}\ \emph {et~al.}(2024)\citenamefont {Iqbal}, \citenamefont {Lyons}, \citenamefont {Lo}, \citenamefont {Tantivasadakarn}, \citenamefont {Dreiling}, \citenamefont {Foltz}, \citenamefont {Gatterman}, \citenamefont {Gresh}, \citenamefont {Hewitt}, \citenamefont {Holliman} \emph {et~al.}}]{iqbal2024qutrit}%
  \BibitemOpen
  \bibfield  {author} {\bibinfo {author} {\bibfnamefont {M.}~\bibnamefont {Iqbal}}, \bibinfo {author} {\bibfnamefont {A.}~\bibnamefont {Lyons}}, \bibinfo {author} {\bibfnamefont {C.~F.~B.}\ \bibnamefont {Lo}}, \bibinfo {author} {\bibfnamefont {N.}~\bibnamefont {Tantivasadakarn}}, \bibinfo {author} {\bibfnamefont {J.}~\bibnamefont {Dreiling}}, \bibinfo {author} {\bibfnamefont {C.}~\bibnamefont {Foltz}}, \bibinfo {author} {\bibfnamefont {T.~M.}\ \bibnamefont {Gatterman}}, \bibinfo {author} {\bibfnamefont {D.}~\bibnamefont {Gresh}}, \bibinfo {author} {\bibfnamefont {N.}~\bibnamefont {Hewitt}}, \bibinfo {author} {\bibfnamefont {C.~A.}\ \bibnamefont {Holliman}},  \emph {et~al.},\ }\href {https://arxiv.org/abs/2411.04185} {\bibfield  {journal} {\bibinfo  {journal} {arXiv preprint arXiv:2411.04185}\ } (\bibinfo {year} {2024})}\BibitemShut {NoStop}%
\bibitem [{\citenamefont {Greene-Diniz}\ \emph {et~al.}(2024)\citenamefont {Greene-Diniz}, \citenamefont {Self}, \citenamefont {Krompiec}, \citenamefont {Coopmans}, \citenamefont {Benedetti}, \citenamefont {Ramo},\ and\ \citenamefont {Rosenkranz}}]{greene2024measuring}%
  \BibitemOpen
  \bibfield  {author} {\bibinfo {author} {\bibfnamefont {G.}~\bibnamefont {Greene-Diniz}}, \bibinfo {author} {\bibfnamefont {C.~N.}\ \bibnamefont {Self}}, \bibinfo {author} {\bibfnamefont {M.}~\bibnamefont {Krompiec}}, \bibinfo {author} {\bibfnamefont {L.}~\bibnamefont {Coopmans}}, \bibinfo {author} {\bibfnamefont {M.}~\bibnamefont {Benedetti}}, \bibinfo {author} {\bibfnamefont {D.~M.}\ \bibnamefont {Ramo}}, \ and\ \bibinfo {author} {\bibfnamefont {M.}~\bibnamefont {Rosenkranz}},\ }\href {https://arxiv.org/abs/2409.15908} {\bibfield  {journal} {\bibinfo  {journal} {arXiv preprint arXiv:2409.15908}\ } (\bibinfo {year} {2024})}\BibitemShut {NoStop}%
\bibitem [{\citenamefont {Chen}\ \emph {et~al.}(2024)\citenamefont {Chen}, \citenamefont {Nielsen}, \citenamefont {Ebert}, \citenamefont {Inlek}, \citenamefont {Wright}, \citenamefont {Chaplin}, \citenamefont {Maksymov}, \citenamefont {P{\'a}ez}, \citenamefont {Poudel}, \citenamefont {Maunz} \emph {et~al.}}]{chen2024benchmarking}%
  \BibitemOpen
  \bibfield  {author} {\bibinfo {author} {\bibfnamefont {J.-S.}\ \bibnamefont {Chen}}, \bibinfo {author} {\bibfnamefont {E.}~\bibnamefont {Nielsen}}, \bibinfo {author} {\bibfnamefont {M.}~\bibnamefont {Ebert}}, \bibinfo {author} {\bibfnamefont {V.}~\bibnamefont {Inlek}}, \bibinfo {author} {\bibfnamefont {K.}~\bibnamefont {Wright}}, \bibinfo {author} {\bibfnamefont {V.}~\bibnamefont {Chaplin}}, \bibinfo {author} {\bibfnamefont {A.}~\bibnamefont {Maksymov}}, \bibinfo {author} {\bibfnamefont {E.}~\bibnamefont {P{\'a}ez}}, \bibinfo {author} {\bibfnamefont {A.}~\bibnamefont {Poudel}}, \bibinfo {author} {\bibfnamefont {P.}~\bibnamefont {Maunz}},  \emph {et~al.},\ }\href {https://quantum-journal.org/papers/q-2024-11-07-1516/} {\bibfield  {journal} {\bibinfo  {journal} {Quantum}\ }\textbf {\bibinfo {volume} {8}},\ \bibinfo {pages} {1516} (\bibinfo {year} {2024})}\BibitemShut {NoStop}%
\bibitem [{\citenamefont {{Riverlane}}(2024)}]{riverlane2024qec}%
  \BibitemOpen
  \bibfield  {author} {\bibinfo {author} {\bibnamefont {{Riverlane}}},\ }\href {https://www.riverlane.com/quantum-error-correction-report-2024} {\enquote {\bibinfo {title} {Quantum error correction report 2024},}\ } (\bibinfo {year} {2024}),\ \bibinfo {note} {accessed: 2025-03-10}\BibitemShut {NoStop}%
\bibitem [{\citenamefont {Preskill}(2024)}]{preskill2024megaquop}%
  \BibitemOpen
  \bibfield  {author} {\bibinfo {author} {\bibfnamefont {J.}~\bibnamefont {Preskill}},\ }\href {https://quantumfrontiers.com/2024/12/14/beyond-nisq-the-megaquop-machine/} {\enquote {\bibinfo {title} {Beyond nisq: The megaquop machine},}\ } (\bibinfo {year} {2024}),\ \bibinfo {note} {keynote address at the Q2B 2024 Conference, Silicon Valley}\BibitemShut {NoStop}%
\bibitem [{\citenamefont {Berry}\ \emph {et~al.}(2020)\citenamefont {Berry}, \citenamefont {Oliver}, \citenamefont {Ashley},\ and\ \citenamefont {Brown}}]{quantum_algorithm_zoo}%
  \BibitemOpen
  \bibfield  {author} {\bibinfo {author} {\bibfnamefont {D.~W.}\ \bibnamefont {Berry}}, \bibinfo {author} {\bibfnamefont {R.~S.~T.}\ \bibnamefont {Oliver}}, \bibinfo {author} {\bibfnamefont {L.~M.~D.}\ \bibnamefont {Ashley}}, \ and\ \bibinfo {author} {\bibfnamefont {D.~M. W.~P.}\ \bibnamefont {Brown}},\ }\href@noop {} {\enquote {\bibinfo {title} {Quantum algorithm zoo},}\ }\bibinfo {howpublished} {\url{https://quantumalgorithmzoo.org/}} (\bibinfo {year} {2020}),\ \bibinfo {note} {accessed: 2025-02-16}\BibitemShut {NoStop}%
\bibitem [{\citenamefont {Lloyd}(1996)}]{lloyd1996universal}%
  \BibitemOpen
  \bibfield  {author} {\bibinfo {author} {\bibfnamefont {S.}~\bibnamefont {Lloyd}},\ }\href {https://www.science.org/doi/abs/10.1126/science.273.5278.1073} {\bibfield  {journal} {\bibinfo  {journal} {Science}\ }\textbf {\bibinfo {volume} {273}},\ \bibinfo {pages} {1073} (\bibinfo {year} {1996})}\BibitemShut {NoStop}%
\bibitem [{\citenamefont {Singkanipa}\ \emph {et~al.}(2025)\citenamefont {Singkanipa}, \citenamefont {Kasatkin}, \citenamefont {Zhou}, \citenamefont {Quiroz},\ and\ \citenamefont {Lidar}}]{lidarscalable}%
  \BibitemOpen
  \bibfield  {author} {\bibinfo {author} {\bibfnamefont {P.}~\bibnamefont {Singkanipa}}, \bibinfo {author} {\bibfnamefont {V.}~\bibnamefont {Kasatkin}}, \bibinfo {author} {\bibfnamefont {Z.}~\bibnamefont {Zhou}}, \bibinfo {author} {\bibfnamefont {G.}~\bibnamefont {Quiroz}}, \ and\ \bibinfo {author} {\bibfnamefont {D.}~\bibnamefont {Lidar}},\ }\href {https://arxiv.org/abs/2401.07934} {\bibfield  {journal} {\bibinfo  {journal} {Quant-Ph}\ } (\bibinfo {year} {2025})}\BibitemShut {NoStop}%
\bibitem [{\citenamefont {Markov}\ \emph {et~al.}(2018)\citenamefont {Markov}, \citenamefont {Fatima}, \citenamefont {Isakov},\ and\ \citenamefont {Boixo}}]{markov2018quantum}%
  \BibitemOpen
  \bibfield  {author} {\bibinfo {author} {\bibfnamefont {I.~L.}\ \bibnamefont {Markov}}, \bibinfo {author} {\bibfnamefont {A.}~\bibnamefont {Fatima}}, \bibinfo {author} {\bibfnamefont {S.~V.}\ \bibnamefont {Isakov}}, \ and\ \bibinfo {author} {\bibfnamefont {S.}~\bibnamefont {Boixo}},\ }\href@noop {} {\bibfield  {journal} {\bibinfo  {journal} {arXiv preprint arXiv:1807.10749}\ } (\bibinfo {year} {2018})}\BibitemShut {NoStop}%
\bibitem [{\citenamefont {Dalzell}\ \emph {et~al.}(2020)\citenamefont {Dalzell}, \citenamefont {Harrow}, \citenamefont {Koh},\ and\ \citenamefont {La~Placa}}]{finegrained1}%
  \BibitemOpen
  \bibfield  {author} {\bibinfo {author} {\bibfnamefont {A.~M.}\ \bibnamefont {Dalzell}}, \bibinfo {author} {\bibfnamefont {A.~W.}\ \bibnamefont {Harrow}}, \bibinfo {author} {\bibfnamefont {D.~E.}\ \bibnamefont {Koh}}, \ and\ \bibinfo {author} {\bibfnamefont {R.~L.}\ \bibnamefont {La~Placa}},\ }\href {\doibase 10.22331/q-2020-05-11-264} {\bibfield  {journal} {\bibinfo  {journal} {Quantum}\ }\textbf {\bibinfo {volume} {4}},\ \bibinfo {pages} {264} (\bibinfo {year} {2020})}\BibitemShut {NoStop}%
\bibitem [{\citenamefont {Ayral}\ \emph {et~al.}(2023)\citenamefont {Ayral}, \citenamefont {Besserve}, \citenamefont {Lacroix},\ and\ \citenamefont {Guzman}}]{Ayral2023}%
  \BibitemOpen
  \bibfield  {author} {\bibinfo {author} {\bibfnamefont {T.}~\bibnamefont {Ayral}}, \bibinfo {author} {\bibfnamefont {P.}~\bibnamefont {Besserve}}, \bibinfo {author} {\bibfnamefont {D.}~\bibnamefont {Lacroix}}, \ and\ \bibinfo {author} {\bibfnamefont {E.~A.~R.}\ \bibnamefont {Guzman}},\ }\href {\doibase 10.1140/epja/s10050-023-01141-1} {\bibfield  {journal} {\bibinfo  {journal} {The European Physical Journal A}\ }\textbf {\bibinfo {volume} {59}},\ \bibinfo {pages} {227} (\bibinfo {year} {2023})}\BibitemShut {NoStop}%
\bibitem [{\citenamefont {Corporation}(2020)}]{Intel2020}%
  \BibitemOpen
  \bibfield  {author} {\bibinfo {author} {\bibfnamefont {I.}~\bibnamefont {Corporation}},\ }\href {https://arxiv.org/abs/2001.10554} {\bibfield  {journal} {\bibinfo  {journal} {arXiv preprint arXiv:2001.10554}\ } (\bibinfo {year} {2020})},\ \bibinfo {note} {accessed: 2025-02-24}\BibitemShut {NoStop}%
\bibitem [{\citenamefont {Lund}\ \emph {et~al.}(2017)\citenamefont {Lund}, \citenamefont {Bremner},\ and\ \citenamefont {Ralph}}]{samplingreview}%
  \BibitemOpen
  \bibfield  {author} {\bibinfo {author} {\bibfnamefont {A.~P.}\ \bibnamefont {Lund}}, \bibinfo {author} {\bibfnamefont {M.~J.}\ \bibnamefont {Bremner}}, \ and\ \bibinfo {author} {\bibfnamefont {T.~C.}\ \bibnamefont {Ralph}},\ }\href {https://www.nature.com/articles/s41534-017-0018-2} {\bibfield  {journal} {\bibinfo  {journal} {njp quantum information}\ }\textbf {\bibinfo {volume} {3}} (\bibinfo {year} {2017})}\BibitemShut {NoStop}%
\bibitem [{\citenamefont {Brakerski}\ \emph {et~al.}(2018)\citenamefont {Brakerski}, \citenamefont {Christiano}, \citenamefont {Mahadev}, \citenamefont {Vazirani},\ and\ \citenamefont {Vidick}}]{BCMVV18}%
  \BibitemOpen
  \bibfield  {author} {\bibinfo {author} {\bibfnamefont {Z.}~\bibnamefont {Brakerski}}, \bibinfo {author} {\bibfnamefont {P.~F.}\ \bibnamefont {Christiano}}, \bibinfo {author} {\bibfnamefont {U.}~\bibnamefont {Mahadev}}, \bibinfo {author} {\bibfnamefont {U.~V.}\ \bibnamefont {Vazirani}}, \ and\ \bibinfo {author} {\bibfnamefont {T.}~\bibnamefont {Vidick}},\ }in\ \href {\doibase 10.1109/FOCS.2018.00038} {\emph {\bibinfo {booktitle} {59th {IEEE} Annual Symposium on Foundations of Computer Science, {FOCS} 2018, Paris, France, October 7-9, 2018}}},\ \bibinfo {editor} {edited by\ \bibinfo {editor} {\bibfnamefont {M.}~\bibnamefont {Thorup}}}\ (\bibinfo  {publisher} {{IEEE} Computer Society},\ \bibinfo {year} {2018})\ pp.\ \bibinfo {pages} {320--331}\BibitemShut {NoStop}%
\bibitem [{\citenamefont {Liu}\ and\ \citenamefont {Cai}(2025)}]{liu2025quantum}%
  \BibitemOpen
  \bibfield  {author} {\bibinfo {author} {\bibfnamefont {K.}~\bibnamefont {Liu}}\ and\ \bibinfo {author} {\bibfnamefont {Z.}~\bibnamefont {Cai}},\ }\href {https://arxiv.org/abs/2502.11285} {\bibfield  {journal} {\bibinfo  {journal} {arXiv preprint arXiv:2502.11285}\ } (\bibinfo {year} {2025})}\BibitemShut {NoStop}%
\bibitem [{\citenamefont {{Qedma Quantum Computing}}(2025{\natexlab{a}})}]{QESEMpaper}%
  \BibitemOpen
  \bibfield  {author} {\bibinfo {author} {\bibnamefont {{Qedma Quantum Computing}}},\ }\href@noop {} {\bibfield  {journal} {\bibinfo  {journal} {To appear}\ } (\bibinfo {year} {2025}{\natexlab{a}})}\BibitemShut {NoStop}%
\bibitem [{\citenamefont {Preskill}(1998)}]{preskillFT}%
  \BibitemOpen
  \bibfield  {author} {\bibinfo {author} {\bibfnamefont {J.}~\bibnamefont {Preskill}},\ }\enquote {\bibinfo {title} {Introduction to quantum computation and information},}\ \ (\bibinfo  {publisher} {World Scientific},\ \bibinfo {year} {1998})\ Chap.\ \bibinfo {chapter} {FAULT-TOLERANT QUANTUM COMPUTATION}, pp.\ \bibinfo {pages} {213--269}\BibitemShut {NoStop}%
\bibitem [{\citenamefont {Ben-Or}\ \emph {et~al.}(2013)\citenamefont {Ben-Or}, \citenamefont {Gottesman},\ and\ \citenamefont {Hassidim}}]{quantumrefrigerator}%
  \BibitemOpen
  \bibfield  {author} {\bibinfo {author} {\bibfnamefont {M.}~\bibnamefont {Ben-Or}}, \bibinfo {author} {\bibfnamefont {D.}~\bibnamefont {Gottesman}}, \ and\ \bibinfo {author} {\bibfnamefont {A.}~\bibnamefont {Hassidim}},\ }\href {https://arxiv.org/abs/1301.1995} {\enquote {\bibinfo {title} {Quantum refrigerator},}\ } (\bibinfo {year} {2013}),\ \Eprint {http://arxiv.org/abs/1301.1995} {arXiv:1301.1995 [quant-ph]} \BibitemShut {NoStop}%
\bibitem [{\citenamefont {Xiong}\ \emph {et~al.}(2020{\natexlab{a}})\citenamefont {Xiong}, \citenamefont {Chandra}, \citenamefont {Ng},\ and\ \citenamefont {Hanzo}}]{xiong2020sampling}%
  \BibitemOpen
  \bibfield  {author} {\bibinfo {author} {\bibfnamefont {Y.}~\bibnamefont {Xiong}}, \bibinfo {author} {\bibfnamefont {D.}~\bibnamefont {Chandra}}, \bibinfo {author} {\bibfnamefont {S.~X.}\ \bibnamefont {Ng}}, \ and\ \bibinfo {author} {\bibfnamefont {L.}~\bibnamefont {Hanzo}},\ }\href {https://ieeexplore.ieee.org/abstract/document/9294106} {\bibfield  {journal} {\bibinfo  {journal} {IEEE Access}\ }\textbf {\bibinfo {volume} {8}},\ \bibinfo {pages} {228967} (\bibinfo {year} {2020}{\natexlab{a}})}\BibitemShut {NoStop}%
\bibitem [{\citenamefont {Ferracin}\ \emph {et~al.}(2024)\citenamefont {Ferracin}, \citenamefont {Hashim}, \citenamefont {Ville}, \citenamefont {Naik}, \citenamefont {Carignan-Dugas}, \citenamefont {Qassim}, \citenamefont {Morvan}, \citenamefont {Santiago}, \citenamefont {Siddiqi},\ and\ \citenamefont {Wallman}}]{ferracin2024efficiently}%
  \BibitemOpen
  \bibfield  {author} {\bibinfo {author} {\bibfnamefont {S.}~\bibnamefont {Ferracin}}, \bibinfo {author} {\bibfnamefont {A.}~\bibnamefont {Hashim}}, \bibinfo {author} {\bibfnamefont {J.-L.}\ \bibnamefont {Ville}}, \bibinfo {author} {\bibfnamefont {R.}~\bibnamefont {Naik}}, \bibinfo {author} {\bibfnamefont {A.}~\bibnamefont {Carignan-Dugas}}, \bibinfo {author} {\bibfnamefont {H.}~\bibnamefont {Qassim}}, \bibinfo {author} {\bibfnamefont {A.}~\bibnamefont {Morvan}}, \bibinfo {author} {\bibfnamefont {D.~I.}\ \bibnamefont {Santiago}}, \bibinfo {author} {\bibfnamefont {I.}~\bibnamefont {Siddiqi}}, \ and\ \bibinfo {author} {\bibfnamefont {J.~J.}\ \bibnamefont {Wallman}},\ }\href {https://quantum-journal.org/papers/q-2024-07-15-1410/} {\bibfield  {journal} {\bibinfo  {journal} {Quantum}\ }\textbf {\bibinfo {volume} {8}},\ \bibinfo {pages} {1410} (\bibinfo {year} {2024})}\BibitemShut {NoStop}%
\bibitem [{\citenamefont {Van Den~Berg}\ \emph {et~al.}(2023)\citenamefont {Van Den~Berg}, \citenamefont {Minev}, \citenamefont {Kandala},\ and\ \citenamefont {Temme}}]{van2023probabilistic}%
  \BibitemOpen
  \bibfield  {author} {\bibinfo {author} {\bibfnamefont {E.}~\bibnamefont {Van Den~Berg}}, \bibinfo {author} {\bibfnamefont {Z.~K.}\ \bibnamefont {Minev}}, \bibinfo {author} {\bibfnamefont {A.}~\bibnamefont {Kandala}}, \ and\ \bibinfo {author} {\bibfnamefont {K.}~\bibnamefont {Temme}},\ }\href {https://www.nature.com/articles/s41567-023-02042-2} {\bibfield  {journal} {\bibinfo  {journal} {Nature physics}\ }\textbf {\bibinfo {volume} {19}},\ \bibinfo {pages} {1116} (\bibinfo {year} {2023})}\BibitemShut {NoStop}%
\bibitem [{\citenamefont {Granet}\ and\ \citenamefont {Dreyer}(2024)}]{granet2024dilution}%
  \BibitemOpen
  \bibfield  {author} {\bibinfo {author} {\bibfnamefont {E.}~\bibnamefont {Granet}}\ and\ \bibinfo {author} {\bibfnamefont {H.}~\bibnamefont {Dreyer}},\ }\href {https://arxiv.org/abs/2409.04254} {\bibfield  {journal} {\bibinfo  {journal} {arXiv preprint arXiv:2409.04254}\ } (\bibinfo {year} {2024})}\BibitemShut {NoStop}%
\bibitem [{\citenamefont {Tran}\ \emph {et~al.}(2023)\citenamefont {Tran}, \citenamefont {Sharma},\ and\ \citenamefont {Temme}}]{tran2023locality}%
  \BibitemOpen
  \bibfield  {author} {\bibinfo {author} {\bibfnamefont {M.~C.}\ \bibnamefont {Tran}}, \bibinfo {author} {\bibfnamefont {K.}~\bibnamefont {Sharma}}, \ and\ \bibinfo {author} {\bibfnamefont {K.}~\bibnamefont {Temme}},\ }\href {https://arxiv.org/abs/2303.06496} {\bibfield  {journal} {\bibinfo  {journal} {arXiv preprint arXiv:2303.06496}\ } (\bibinfo {year} {2023})}\BibitemShut {NoStop}%
\bibitem [{\citenamefont {Eddins}\ \emph {et~al.}(2024)\citenamefont {Eddins}, \citenamefont {Tran},\ and\ \citenamefont {Rall}}]{eddins2024lightcone}%
  \BibitemOpen
  \bibfield  {author} {\bibinfo {author} {\bibfnamefont {A.}~\bibnamefont {Eddins}}, \bibinfo {author} {\bibfnamefont {M.~C.}\ \bibnamefont {Tran}}, \ and\ \bibinfo {author} {\bibfnamefont {P.}~\bibnamefont {Rall}},\ }\href {https://arxiv.org/abs/2409.04401} {\bibfield  {journal} {\bibinfo  {journal} {arXiv preprint arXiv:2409.04401}\ } (\bibinfo {year} {2024})}\BibitemShut {NoStop}%
\bibitem [{\citenamefont {Kliesch}\ and\ \citenamefont {Roth}(2021)}]{PRXQuantum.2.010201}%
  \BibitemOpen
  \bibfield  {author} {\bibinfo {author} {\bibfnamefont {M.}~\bibnamefont {Kliesch}}\ and\ \bibinfo {author} {\bibfnamefont {I.}~\bibnamefont {Roth}},\ }\href {\doibase 10.1103/PRXQuantum.2.010201} {\bibfield  {journal} {\bibinfo  {journal} {PRX Quantum}\ }\textbf {\bibinfo {volume} {2}},\ \bibinfo {pages} {010201} (\bibinfo {year} {2021})}\BibitemShut {NoStop}%
\bibitem [{\citenamefont {Aharonov}\ \emph {et~al.}(1996)\citenamefont {Aharonov}, \citenamefont {Ben-Or}, \citenamefont {Impagliazzo},\ and\ \citenamefont {Nisan}}]{ABIN96}%
  \BibitemOpen
  \bibfield  {author} {\bibinfo {author} {\bibfnamefont {D.}~\bibnamefont {Aharonov}}, \bibinfo {author} {\bibfnamefont {M.}~\bibnamefont {Ben-Or}}, \bibinfo {author} {\bibfnamefont {R.}~\bibnamefont {Impagliazzo}}, \ and\ \bibinfo {author} {\bibfnamefont {N.}~\bibnamefont {Nisan}},\ }\href {https://arxiv.org/abs/quant-ph/9611028} {\enquote {\bibinfo {title} {Limitations of noisy reversible computation},}\ } (\bibinfo {year} {1996}),\ \Eprint {http://arxiv.org/abs/quant-ph/9611028} {arXiv:quant-ph/9611028 [quant-ph]} \BibitemShut {NoStop}%
\bibitem [{\citenamefont {Gao}\ and\ \citenamefont {Duan}(2018)}]{GaoDuan2018}%
  \BibitemOpen
  \bibfield  {author} {\bibinfo {author} {\bibfnamefont {X.}~\bibnamefont {Gao}}\ and\ \bibinfo {author} {\bibfnamefont {L.}~\bibnamefont {Duan}},\ }\href {https://arxiv.org/abs/1810.03176} {\enquote {\bibinfo {title} {Efficient classical simulation of noisy quantum computation},}\ } (\bibinfo {year} {2018}),\ \Eprint {http://arxiv.org/abs/1810.03176} {arXiv:1810.03176 [quant-ph]} \BibitemShut {NoStop}%
\bibitem [{\citenamefont {Deshpande}\ \emph {et~al.}(2022)\citenamefont {Deshpande}, \citenamefont {Niroula}, \citenamefont {Shtanko}, \citenamefont {Gorshkov}, \citenamefont {Fefferman},\ and\ \citenamefont {Gullans}}]{Deshpande2022}%
  \BibitemOpen
  \bibfield  {author} {\bibinfo {author} {\bibfnamefont {A.}~\bibnamefont {Deshpande}}, \bibinfo {author} {\bibfnamefont {P.}~\bibnamefont {Niroula}}, \bibinfo {author} {\bibfnamefont {O.}~\bibnamefont {Shtanko}}, \bibinfo {author} {\bibfnamefont {A.~V.}\ \bibnamefont {Gorshkov}}, \bibinfo {author} {\bibfnamefont {B.}~\bibnamefont {Fefferman}}, \ and\ \bibinfo {author} {\bibfnamefont {M.~J.}\ \bibnamefont {Gullans}},\ }\href {\doibase 10.1103/prxquantum.3.040329} {\bibfield  {journal} {\bibinfo  {journal} {PRX Quantum}\ }\textbf {\bibinfo {volume} {3}} (\bibinfo {year} {2022}),\ 10.1103/prxquantum.3.040329}\BibitemShut {NoStop}%
\bibitem [{\citenamefont {Kastoryano}\ and\ \citenamefont {Temme}(2013)}]{10.1063/1.4804995}%
  \BibitemOpen
  \bibfield  {author} {\bibinfo {author} {\bibfnamefont {M.~J.}\ \bibnamefont {Kastoryano}}\ and\ \bibinfo {author} {\bibfnamefont {K.}~\bibnamefont {Temme}},\ }\href {\doibase 10.1063/1.4804995} {\bibfield  {journal} {\bibinfo  {journal} {Journal of Mathematical Physics}\ }\textbf {\bibinfo {volume} {54}},\ \bibinfo {pages} {052202} (\bibinfo {year} {2013})},\ \Eprint {http://arxiv.org/abs/https://pubs.aip.org/aip/jmp/article-pdf/doi/10.1063/1.4804995/13369764/052202\_1\_online.pdf} {https://pubs.aip.org/aip/jmp/article-pdf/doi/10.1063/1.4804995/13369764/052202\_1\_online.pdf} \BibitemShut {NoStop}%
\bibitem [{\citenamefont {McKay}\ \emph {et~al.}(2023)\citenamefont {McKay}, \citenamefont {Hincks}, \citenamefont {Pritchett}, \citenamefont {Carroll}, \citenamefont {Govia},\ and\ \citenamefont {Merkel}}]{mckay2023benchmarking}%
  \BibitemOpen
  \bibfield  {author} {\bibinfo {author} {\bibfnamefont {D.~C.}\ \bibnamefont {McKay}}, \bibinfo {author} {\bibfnamefont {I.}~\bibnamefont {Hincks}}, \bibinfo {author} {\bibfnamefont {E.~J.}\ \bibnamefont {Pritchett}}, \bibinfo {author} {\bibfnamefont {M.}~\bibnamefont {Carroll}}, \bibinfo {author} {\bibfnamefont {L.~C.}\ \bibnamefont {Govia}}, \ and\ \bibinfo {author} {\bibfnamefont {S.~T.}\ \bibnamefont {Merkel}},\ }\href@noop {} {\bibfield  {journal} {\bibinfo  {journal} {arXiv preprint arXiv:2311.05933}\ } (\bibinfo {year} {2023})}\BibitemShut {NoStop}%
\bibitem [{\citenamefont {Shepherd}\ and\ \citenamefont {Bremner}(2009)}]{bremnerIQP1}%
  \BibitemOpen
  \bibfield  {author} {\bibinfo {author} {\bibfnamefont {D.}~\bibnamefont {Shepherd}}\ and\ \bibinfo {author} {\bibfnamefont {M.~J.}\ \bibnamefont {Bremner}},\ }\href@noop {} {\bibfield  {journal} {\bibinfo  {journal} {Proc. R. Soc. A}\ }\textbf {\bibinfo {volume} {465}},\ \bibinfo {pages} {1413} (\bibinfo {year} {2009})}\BibitemShut {NoStop}%
\bibitem [{\citenamefont {Bremner}\ and\ \citenamefont {Shepherd}(2011)}]{bremnerIQP2}%
  \BibitemOpen
  \bibfield  {author} {\bibinfo {author} {\bibfnamefont {M.~J.}\ \bibnamefont {Bremner}}\ and\ \bibinfo {author} {\bibfnamefont {R.~J.~D.}\ \bibnamefont {Shepherd}},\ }\href@noop {} {\bibfield  {journal} {\bibinfo  {journal} {Proc. R. Soc. A}\ }\textbf {\bibinfo {volume} {467}},\ \bibinfo {pages} {459} (\bibinfo {year} {2011})}\BibitemShut {NoStop}%
\bibitem [{\citenamefont {Bremner}\ \emph {et~al.}(2016)\citenamefont {Bremner}, \citenamefont {Shepherd},\ and\ \citenamefont {Montanaro}}]{bremnerIQP3}%
  \BibitemOpen
  \bibfield  {author} {\bibinfo {author} {\bibfnamefont {M.~J.}\ \bibnamefont {Bremner}}, \bibinfo {author} {\bibfnamefont {D.}~\bibnamefont {Shepherd}}, \ and\ \bibinfo {author} {\bibfnamefont {A.}~\bibnamefont {Montanaro}},\ }\href@noop {} {\bibfield  {journal} {\bibinfo  {journal} {Phys. Rev. Lett.}\ }\textbf {\bibinfo {volume} {117}} (\bibinfo {year} {2016})}\BibitemShut {NoStop}%
\bibitem [{\citenamefont {Bremner}\ \emph {et~al.}(2017)\citenamefont {Bremner}, \citenamefont {Montanaro},\ and\ \citenamefont {Shepherd}}]{bremnerIQP4}%
  \BibitemOpen
  \bibfield  {author} {\bibinfo {author} {\bibfnamefont {M.~J.}\ \bibnamefont {Bremner}}, \bibinfo {author} {\bibfnamefont {A.}~\bibnamefont {Montanaro}}, \ and\ \bibinfo {author} {\bibfnamefont {D.}~\bibnamefont {Shepherd}},\ }\href@noop {} {\bibfield  {journal} {\bibinfo  {journal} {Quantum}\ }\textbf {\bibinfo {volume} {1}} (\bibinfo {year} {2017})}\BibitemShut {NoStop}%
\bibitem [{\citenamefont {Bergamaschi}\ \emph {et~al.}(2024)\citenamefont {Bergamaschi}, \citenamefont {Chen},\ and\ \citenamefont {Liu}}]{IQP5}%
  \BibitemOpen
  \bibfield  {author} {\bibinfo {author} {\bibfnamefont {T.}~\bibnamefont {Bergamaschi}}, \bibinfo {author} {\bibfnamefont {C.-F.}\ \bibnamefont {Chen}}, \ and\ \bibinfo {author} {\bibfnamefont {Y.}~\bibnamefont {Liu}},\ }in\ \href@noop {} {\emph {\bibinfo {booktitle} {65th {IEEE} Annual Symposium on Foundations of Computer Science, {FOCS}}}}\ (\bibinfo  {publisher} {{IEEE} Computer Society},\ \bibinfo {year} {2024})\BibitemShut {NoStop}%
\bibitem [{\citenamefont {Aharonov}\ \emph {et~al.}(2022)\citenamefont {Aharonov}, \citenamefont {Gao}, \citenamefont {Landau}, \citenamefont {Liu},\ and\ \citenamefont {Vazirani}}]{Aharonov2022}%
  \BibitemOpen
  \bibfield  {author} {\bibinfo {author} {\bibfnamefont {D.}~\bibnamefont {Aharonov}}, \bibinfo {author} {\bibfnamefont {X.}~\bibnamefont {Gao}}, \bibinfo {author} {\bibfnamefont {Z.}~\bibnamefont {Landau}}, \bibinfo {author} {\bibfnamefont {Y.}~\bibnamefont {Liu}}, \ and\ \bibinfo {author} {\bibfnamefont {U.}~\bibnamefont {Vazirani}},\ }\href {https://arxiv.org/abs/2211.03999} {\bibfield  {journal} {\bibinfo  {journal} {arXiv preprint arXiv:2211.03999}\ } (\bibinfo {year} {2022})}\BibitemShut {NoStop}%
\bibitem [{\citenamefont {L{\"o}schnauer}\ \emph {et~al.}(2024)\citenamefont {L{\"o}schnauer}, \citenamefont {Toba}, \citenamefont {Hughes}, \citenamefont {King}, \citenamefont {Weber}, \citenamefont {Srinivas}, \citenamefont {Matt}, \citenamefont {Nourshargh}, \citenamefont {Allcock}, \citenamefont {Ballance} \emph {et~al.}}]{loschnauer2024scalable}%
  \BibitemOpen
  \bibfield  {author} {\bibinfo {author} {\bibfnamefont {C.}~\bibnamefont {L{\"o}schnauer}}, \bibinfo {author} {\bibfnamefont {J.~M.}\ \bibnamefont {Toba}}, \bibinfo {author} {\bibfnamefont {A.}~\bibnamefont {Hughes}}, \bibinfo {author} {\bibfnamefont {S.}~\bibnamefont {King}}, \bibinfo {author} {\bibfnamefont {M.}~\bibnamefont {Weber}}, \bibinfo {author} {\bibfnamefont {R.}~\bibnamefont {Srinivas}}, \bibinfo {author} {\bibfnamefont {R.}~\bibnamefont {Matt}}, \bibinfo {author} {\bibfnamefont {R.}~\bibnamefont {Nourshargh}}, \bibinfo {author} {\bibfnamefont {D.}~\bibnamefont {Allcock}}, \bibinfo {author} {\bibfnamefont {C.}~\bibnamefont {Ballance}},  \emph {et~al.},\ }\href {https://arxiv.org/abs/2407.07694} {\bibfield  {journal} {\bibinfo  {journal} {arXiv preprint arXiv:2407.07694}\ } (\bibinfo {year} {2024})}\BibitemShut {NoStop}%
\bibitem [{\citenamefont {McArdle}\ \emph {et~al.}(2020)\citenamefont {McArdle}, \citenamefont {Endo}, \citenamefont {Aspuru-Guzik}, \citenamefont {Benjamin},\ and\ \citenamefont {Yuan}}]{McArdle2020}%
  \BibitemOpen
  \bibfield  {author} {\bibinfo {author} {\bibfnamefont {S.}~\bibnamefont {McArdle}}, \bibinfo {author} {\bibfnamefont {S.}~\bibnamefont {Endo}}, \bibinfo {author} {\bibfnamefont {A.}~\bibnamefont {Aspuru-Guzik}}, \bibinfo {author} {\bibfnamefont {S.~C.}\ \bibnamefont {Benjamin}}, \ and\ \bibinfo {author} {\bibfnamefont {X.}~\bibnamefont {Yuan}},\ }\href {\doibase 10.1103/REVMODPHYS.92.015003/FIGURES/13/THUMBNAIL} {\bibfield  {journal} {\bibinfo  {journal} {Reviews of Modern Physics}\ }\textbf {\bibinfo {volume} {92}},\ \bibinfo {pages} {015003} (\bibinfo {year} {2020})},\ \Eprint {http://arxiv.org/abs/1808.10402} {arXiv:1808.10402} \BibitemShut {NoStop}%
\bibitem [{\citenamefont {Filippov}\ \emph {et~al.}(2023)\citenamefont {Filippov}, \citenamefont {Leahy}, \citenamefont {Rossi},\ and\ \citenamefont {Garc{\'\i}a-P{\'e}rez}}]{filippov2023scalable}%
  \BibitemOpen
  \bibfield  {author} {\bibinfo {author} {\bibfnamefont {S.}~\bibnamefont {Filippov}}, \bibinfo {author} {\bibfnamefont {M.}~\bibnamefont {Leahy}}, \bibinfo {author} {\bibfnamefont {M.~A.}\ \bibnamefont {Rossi}}, \ and\ \bibinfo {author} {\bibfnamefont {G.}~\bibnamefont {Garc{\'\i}a-P{\'e}rez}},\ }\href {https://arxiv.org/abs/2307.11740} {\bibfield  {journal} {\bibinfo  {journal} {arXiv preprint arXiv:2307.11740}\ } (\bibinfo {year} {2023})}\BibitemShut {NoStop}%
\bibitem [{\citenamefont {Lindner}\ and\ \citenamefont {Yunoki}(2024)}]{lindner_yunoki_2024}%
  \BibitemOpen
  \bibfield  {author} {\bibinfo {author} {\bibfnamefont {N.}~\bibnamefont {Lindner}}\ and\ \bibinfo {author} {\bibfnamefont {S.}~\bibnamefont {Yunoki}},\ }\href {https://www.youtube.com/watch?v=tQW6FdLc6zo} {\enquote {\bibinfo {title} {Q2b24 tokyo | pushing forward for quantum advantage with error mitigation and hpc},}\ } (\bibinfo {year} {2024}),\ \bibinfo {note} {accessed: 2025-03-10}\BibitemShut {NoStop}%
\bibitem [{\citenamefont {Fuller}\ \emph {et~al.}(2025)\citenamefont {Fuller}, \citenamefont {Tran}, \citenamefont {Lykov}, \citenamefont {Johnson}, \citenamefont {Rossmannek}, \citenamefont {Wei}, \citenamefont {He}, \citenamefont {Kim}, \citenamefont {Vu}, \citenamefont {Sharma} \emph {et~al.}}]{fuller2025improved}%
  \BibitemOpen
  \bibfield  {author} {\bibinfo {author} {\bibfnamefont {B.}~\bibnamefont {Fuller}}, \bibinfo {author} {\bibfnamefont {M.~C.}\ \bibnamefont {Tran}}, \bibinfo {author} {\bibfnamefont {D.}~\bibnamefont {Lykov}}, \bibinfo {author} {\bibfnamefont {C.}~\bibnamefont {Johnson}}, \bibinfo {author} {\bibfnamefont {M.}~\bibnamefont {Rossmannek}}, \bibinfo {author} {\bibfnamefont {K.~X.}\ \bibnamefont {Wei}}, \bibinfo {author} {\bibfnamefont {A.}~\bibnamefont {He}}, \bibinfo {author} {\bibfnamefont {Y.}~\bibnamefont {Kim}}, \bibinfo {author} {\bibfnamefont {D.}~\bibnamefont {Vu}}, \bibinfo {author} {\bibfnamefont {K.}~\bibnamefont {Sharma}},  \emph {et~al.},\ }\href {https://arxiv.org/abs/2502.01897} {\bibfield  {journal} {\bibinfo  {journal} {arXiv preprint arXiv:2502.01897}\ } (\bibinfo {year} {2025})}\BibitemShut {NoStop}%
\bibitem [{\citenamefont {{TOP500 Authors}}(2024)}]{TOP500_Nov2024}%
  \BibitemOpen
  \bibfield  {author} {\bibinfo {author} {\bibnamefont {{TOP500 Authors}}},\ }\href@noop {} {\enquote {\bibinfo {title} {Top500 november 2024 list},}\ }\bibinfo {howpublished} {\url{https://top500.org/lists/top500/2024/11/}} (\bibinfo {year} {2024}),\ \bibinfo {note} {accessed: 2024-12-10}\BibitemShut {NoStop}%
\bibitem [{\citenamefont {Fang}\ \emph {et~al.}(2022)\citenamefont {Fang}, \citenamefont {{\"O}zkaya}, \citenamefont {Li}, \citenamefont {{\c{C}}ataly{\"u}rek},\ and\ \citenamefont {Krishnamoorthy}}]{fang2022efficient}%
  \BibitemOpen
  \bibfield  {author} {\bibinfo {author} {\bibfnamefont {B.}~\bibnamefont {Fang}}, \bibinfo {author} {\bibfnamefont {M.~Y.}\ \bibnamefont {{\"O}zkaya}}, \bibinfo {author} {\bibfnamefont {A.}~\bibnamefont {Li}}, \bibinfo {author} {\bibfnamefont {{\"U}.~V.}\ \bibnamefont {{\c{C}}ataly{\"u}rek}}, \ and\ \bibinfo {author} {\bibfnamefont {S.}~\bibnamefont {Krishnamoorthy}},\ }in\ \href {https://ieeexplore.ieee.org/abstract/document/9912699?casa_token=R1HbCvPDy64AAAAA:cj78nVAUjvqkyno-3jnBWwZN0PKmP2mBvfpPAf9ed1AkbeyU5ip8oC1zTbE_J_aTgyvhI77O-CU} {\emph {\bibinfo {booktitle} {2022 IEEE International Conference on Cluster Computing (CLUSTER)}}}\ (\bibinfo {organization} {IEEE},\ \bibinfo {year} {2022})\ pp.\ \bibinfo {pages} {289--300}\BibitemShut {NoStop}%
\bibitem [{\citenamefont {Nahum}\ \emph {et~al.}(2018)\citenamefont {Nahum}, \citenamefont {Vijay},\ and\ \citenamefont {Haah}}]{Haah1}%
  \BibitemOpen
  \bibfield  {author} {\bibinfo {author} {\bibfnamefont {A.}~\bibnamefont {Nahum}}, \bibinfo {author} {\bibfnamefont {S.}~\bibnamefont {Vijay}}, \ and\ \bibinfo {author} {\bibfnamefont {J.}~\bibnamefont {Haah}},\ }\href {\doibase 10.1103/PhysRevX.8.021014} {\bibfield  {journal} {\bibinfo  {journal} {Phys. Rev. X}\ }\textbf {\bibinfo {volume} {8}},\ \bibinfo {pages} {021014} (\bibinfo {year} {2018})}\BibitemShut {NoStop}%
\bibitem [{\citenamefont {Larocca}\ \emph {et~al.}(2024)\citenamefont {Larocca}, \citenamefont {Thanasilp}, \citenamefont {Wang}, \citenamefont {Sharma}, \citenamefont {Biamonte}, \citenamefont {Coles}, \citenamefont {Cincio}, \citenamefont {McClean}, \citenamefont {Holmes},\ and\ \citenamefont {Cerezo}}]{larocca2024review}%
  \BibitemOpen
  \bibfield  {author} {\bibinfo {author} {\bibfnamefont {M.}~\bibnamefont {Larocca}}, \bibinfo {author} {\bibfnamefont {S.}~\bibnamefont {Thanasilp}}, \bibinfo {author} {\bibfnamefont {S.}~\bibnamefont {Wang}}, \bibinfo {author} {\bibfnamefont {K.}~\bibnamefont {Sharma}}, \bibinfo {author} {\bibfnamefont {J.}~\bibnamefont {Biamonte}}, \bibinfo {author} {\bibfnamefont {P.~J.}\ \bibnamefont {Coles}}, \bibinfo {author} {\bibfnamefont {L.}~\bibnamefont {Cincio}}, \bibinfo {author} {\bibfnamefont {J.~R.}\ \bibnamefont {McClean}}, \bibinfo {author} {\bibfnamefont {Z.}~\bibnamefont {Holmes}}, \ and\ \bibinfo {author} {\bibfnamefont {M.}~\bibnamefont {Cerezo}},\ }\href {https://arxiv.org/abs/2405.00781} {\bibfield  {journal} {\bibinfo  {journal} {arXiv preprint arXiv:2405.00781}\ } (\bibinfo {year} {2024})}\BibitemShut {NoStop}%
\bibitem [{\citenamefont {Kechedzhi}\ \emph {et~al.}(2024)\citenamefont {Kechedzhi}, \citenamefont {Isakov}, \citenamefont {Mandr{\`a}}, \citenamefont {Villalonga}, \citenamefont {Mi}, \citenamefont {Boixo},\ and\ \citenamefont {Smelyanskiy}}]{kechedzhi2024effective}%
  \BibitemOpen
  \bibfield  {author} {\bibinfo {author} {\bibfnamefont {K.}~\bibnamefont {Kechedzhi}}, \bibinfo {author} {\bibfnamefont {S.~V.}\ \bibnamefont {Isakov}}, \bibinfo {author} {\bibfnamefont {S.}~\bibnamefont {Mandr{\`a}}}, \bibinfo {author} {\bibfnamefont {B.}~\bibnamefont {Villalonga}}, \bibinfo {author} {\bibfnamefont {X.}~\bibnamefont {Mi}}, \bibinfo {author} {\bibfnamefont {S.}~\bibnamefont {Boixo}}, \ and\ \bibinfo {author} {\bibfnamefont {V.}~\bibnamefont {Smelyanskiy}},\ }\href {https://www.sciencedirect.com/science/article/pii/S0167739X23004569?casa_token=VOTef10MzeoAAAAA:bkcdBspT4CF6pUdHHOybsr3b7xSbR1vi7x_WmP0hCmKDyeJemqMNxJMdBUczqqF25CrTpnDzICE} {\bibfield  {journal} {\bibinfo  {journal} {Future Generation Computer Systems}\ }\textbf {\bibinfo {volume} {153}},\ \bibinfo {pages} {431} (\bibinfo {year} {2024})}\BibitemShut {NoStop}%
\bibitem [{\citenamefont {Vallero}\ \emph {et~al.}(2024)\citenamefont {Vallero}, \citenamefont {Vella},\ and\ \citenamefont {Rech}}]{vallero2024state}%
  \BibitemOpen
  \bibfield  {author} {\bibinfo {author} {\bibfnamefont {M.}~\bibnamefont {Vallero}}, \bibinfo {author} {\bibfnamefont {F.}~\bibnamefont {Vella}}, \ and\ \bibinfo {author} {\bibfnamefont {P.}~\bibnamefont {Rech}},\ }\href {https://arxiv.org/abs/2401.06188} {\bibfield  {journal} {\bibinfo  {journal} {arXiv preprint arXiv:2401.06188}\ } (\bibinfo {year} {2024})}\BibitemShut {NoStop}%
\bibitem [{\citenamefont {Pan}\ and\ \citenamefont {Zhang}(2022)}]{PhysRevLett.128.030501}%
  \BibitemOpen
  \bibfield  {author} {\bibinfo {author} {\bibfnamefont {F.}~\bibnamefont {Pan}}\ and\ \bibinfo {author} {\bibfnamefont {P.}~\bibnamefont {Zhang}},\ }\href {\doibase 10.1103/PhysRevLett.128.030501} {\bibfield  {journal} {\bibinfo  {journal} {Phys. Rev. Lett.}\ }\textbf {\bibinfo {volume} {128}},\ \bibinfo {pages} {030501} (\bibinfo {year} {2022})}\BibitemShut {NoStop}%
\bibitem [{\citenamefont {Markov}\ and\ \citenamefont {Shi}(2008)}]{markov2008simulating}%
  \BibitemOpen
  \bibfield  {author} {\bibinfo {author} {\bibfnamefont {I.~L.}\ \bibnamefont {Markov}}\ and\ \bibinfo {author} {\bibfnamefont {Y.}~\bibnamefont {Shi}},\ }\href {https://epubs.siam.org/doi/abs/10.1137/050644756?casa_token=R43IX8wqh10AAAAA:xk1i8DIkileWE5pvCukBp4vr6uOTUB_IM1qo4QoQ2RvkSnnVOnXQCYjOmNG7fg5JEjg_DsyRFNk} {\bibfield  {journal} {\bibinfo  {journal} {SIAM Journal on Computing}\ }\textbf {\bibinfo {volume} {38}},\ \bibinfo {pages} {963} (\bibinfo {year} {2008})}\BibitemShut {NoStop}%
\bibitem [{\citenamefont {Gray}\ and\ \citenamefont {Kourtis}(2021)}]{gray2021hyper}%
  \BibitemOpen
  \bibfield  {author} {\bibinfo {author} {\bibfnamefont {J.}~\bibnamefont {Gray}}\ and\ \bibinfo {author} {\bibfnamefont {S.}~\bibnamefont {Kourtis}},\ }\href {https://quantum-journal.org/papers/q-2021-03-15-410/} {\bibfield  {journal} {\bibinfo  {journal} {Quantum}\ }\textbf {\bibinfo {volume} {5}},\ \bibinfo {pages} {410} (\bibinfo {year} {2021})}\BibitemShut {NoStop}%
\bibitem [{\citenamefont {Vidal}(2003)}]{PhysRevLett.91.147902}%
  \BibitemOpen
  \bibfield  {author} {\bibinfo {author} {\bibfnamefont {G.}~\bibnamefont {Vidal}},\ }\href {\doibase 10.1103/PhysRevLett.91.147902} {\bibfield  {journal} {\bibinfo  {journal} {Phys. Rev. Lett.}\ }\textbf {\bibinfo {volume} {91}},\ \bibinfo {pages} {147902} (\bibinfo {year} {2003})}\BibitemShut {NoStop}%
\bibitem [{\citenamefont {Tindall}\ \emph {et~al.}(2024)\citenamefont {Tindall}, \citenamefont {Fishman}, \citenamefont {Stoudenmire},\ and\ \citenamefont {Sels}}]{PRXQuantum.5.010308}%
  \BibitemOpen
  \bibfield  {author} {\bibinfo {author} {\bibfnamefont {J.}~\bibnamefont {Tindall}}, \bibinfo {author} {\bibfnamefont {M.}~\bibnamefont {Fishman}}, \bibinfo {author} {\bibfnamefont {E.~M.}\ \bibnamefont {Stoudenmire}}, \ and\ \bibinfo {author} {\bibfnamefont {D.}~\bibnamefont {Sels}},\ }\href {\doibase 10.1103/PRXQuantum.5.010308} {\bibfield  {journal} {\bibinfo  {journal} {PRX Quantum}\ }\textbf {\bibinfo {volume} {5}},\ \bibinfo {pages} {010308} (\bibinfo {year} {2024})}\BibitemShut {NoStop}%
\bibitem [{\citenamefont {Nahum}\ \emph {et~al.}(2017)\citenamefont {Nahum}, \citenamefont {Ruhman}, \citenamefont {Vijay},\ and\ \citenamefont {Haah}}]{Haah2}%
  \BibitemOpen
  \bibfield  {author} {\bibinfo {author} {\bibfnamefont {A.}~\bibnamefont {Nahum}}, \bibinfo {author} {\bibfnamefont {J.}~\bibnamefont {Ruhman}}, \bibinfo {author} {\bibfnamefont {S.}~\bibnamefont {Vijay}}, \ and\ \bibinfo {author} {\bibfnamefont {J.}~\bibnamefont {Haah}},\ }\href {\doibase 10.1103/PhysRevX.7.031016} {\bibfield  {journal} {\bibinfo  {journal} {Phys. Rev. X}\ }\textbf {\bibinfo {volume} {7}},\ \bibinfo {pages} {031016} (\bibinfo {year} {2017})}\BibitemShut {NoStop}%
\bibitem [{\citenamefont {Bravyi}\ \emph {et~al.}(2019)\citenamefont {Bravyi}, \citenamefont {Browne}, \citenamefont {Calpin}, \citenamefont {Campbell}, \citenamefont {Gosset},\ and\ \citenamefont {Howard}}]{Bravyi2019simulationofquantum}%
  \BibitemOpen
  \bibfield  {author} {\bibinfo {author} {\bibfnamefont {S.}~\bibnamefont {Bravyi}}, \bibinfo {author} {\bibfnamefont {D.}~\bibnamefont {Browne}}, \bibinfo {author} {\bibfnamefont {P.}~\bibnamefont {Calpin}}, \bibinfo {author} {\bibfnamefont {E.}~\bibnamefont {Campbell}}, \bibinfo {author} {\bibfnamefont {D.}~\bibnamefont {Gosset}}, \ and\ \bibinfo {author} {\bibfnamefont {M.}~\bibnamefont {Howard}},\ }\href {\doibase 10.22331/q-2019-09-02-181} {\bibfield  {journal} {\bibinfo  {journal} {{Quantum}}\ }\textbf {\bibinfo {volume} {3}},\ \bibinfo {pages} {181} (\bibinfo {year} {2019})}\BibitemShut {NoStop}%
\bibitem [{\citenamefont {Reardon-Smith}\ \emph {et~al.}(2024)\citenamefont {Reardon-Smith}, \citenamefont {Oszmaniec},\ and\ \citenamefont {Korzekwa}}]{ReardonSmith2024improvedsimulation}%
  \BibitemOpen
  \bibfield  {author} {\bibinfo {author} {\bibfnamefont {O.}~\bibnamefont {Reardon-Smith}}, \bibinfo {author} {\bibfnamefont {M.}~\bibnamefont {Oszmaniec}}, \ and\ \bibinfo {author} {\bibfnamefont {K.}~\bibnamefont {Korzekwa}},\ }\href {\doibase 10.22331/q-2024-12-04-1549} {\bibfield  {journal} {\bibinfo  {journal} {{Quantum}}\ }\textbf {\bibinfo {volume} {8}},\ \bibinfo {pages} {1549} (\bibinfo {year} {2024})}\BibitemShut {NoStop}%
\bibitem [{\citenamefont {Zhou}\ \emph {et~al.}(2020)\citenamefont {Zhou}, \citenamefont {Stoudenmire},\ and\ \citenamefont {Waintal}}]{PhysRevX.10.041038}%
  \BibitemOpen
  \bibfield  {author} {\bibinfo {author} {\bibfnamefont {Y.}~\bibnamefont {Zhou}}, \bibinfo {author} {\bibfnamefont {E.~M.}\ \bibnamefont {Stoudenmire}}, \ and\ \bibinfo {author} {\bibfnamefont {X.}~\bibnamefont {Waintal}},\ }\href {\doibase 10.1103/PhysRevX.10.041038} {\bibfield  {journal} {\bibinfo  {journal} {Phys. Rev. X}\ }\textbf {\bibinfo {volume} {10}},\ \bibinfo {pages} {041038} (\bibinfo {year} {2020})}\BibitemShut {NoStop}%
\bibitem [{\citenamefont {Suzuki}\ \emph {et~al.}(2022)\citenamefont {Suzuki}, \citenamefont {Endo}, \citenamefont {Fujii},\ and\ \citenamefont {Tokunaga}}]{Suzuki2022}%
  \BibitemOpen
  \bibfield  {author} {\bibinfo {author} {\bibfnamefont {Y.}~\bibnamefont {Suzuki}}, \bibinfo {author} {\bibfnamefont {S.}~\bibnamefont {Endo}}, \bibinfo {author} {\bibfnamefont {K.}~\bibnamefont {Fujii}}, \ and\ \bibinfo {author} {\bibfnamefont {Y.}~\bibnamefont {Tokunaga}},\ }\href {\doibase 10.1103/PRXQuantum.3.010345} {\bibfield  {journal} {\bibinfo  {journal} {PRX Quantum}\ }\textbf {\bibinfo {volume} {3}},\ \bibinfo {pages} {010345} (\bibinfo {year} {2022})}\BibitemShut {NoStop}%
\bibitem [{\citenamefont {Piveteau}\ \emph {et~al.}(2021)\citenamefont {Piveteau}, \citenamefont {Sutter}, \citenamefont {Bravyi}, \citenamefont {Gambetta},\ and\ \citenamefont {Temme}}]{Piveteau2021}%
  \BibitemOpen
  \bibfield  {author} {\bibinfo {author} {\bibfnamefont {C.}~\bibnamefont {Piveteau}}, \bibinfo {author} {\bibfnamefont {D.}~\bibnamefont {Sutter}}, \bibinfo {author} {\bibfnamefont {S.}~\bibnamefont {Bravyi}}, \bibinfo {author} {\bibfnamefont {J.~M.}\ \bibnamefont {Gambetta}}, \ and\ \bibinfo {author} {\bibfnamefont {K.}~\bibnamefont {Temme}},\ }\href {\doibase 10.1103/PhysRevLett.127.200505} {\bibfield  {journal} {\bibinfo  {journal} {Phys. Rev. Lett.}\ }\textbf {\bibinfo {volume} {127}},\ \bibinfo {pages} {200505} (\bibinfo {year} {2021})}\BibitemShut {NoStop}%
\bibitem [{\citenamefont {Lostaglio}\ and\ \citenamefont {Ciani}(2021)}]{Lostaglio2021}%
  \BibitemOpen
  \bibfield  {author} {\bibinfo {author} {\bibfnamefont {M.}~\bibnamefont {Lostaglio}}\ and\ \bibinfo {author} {\bibfnamefont {A.}~\bibnamefont {Ciani}},\ }\href {\doibase 10.1103/PhysRevLett.127.200506} {\bibfield  {journal} {\bibinfo  {journal} {Phys. Rev. Lett.}\ }\textbf {\bibinfo {volume} {127}},\ \bibinfo {pages} {200506} (\bibinfo {year} {2021})}\BibitemShut {NoStop}%
\bibitem [{\citenamefont {Xiong}\ \emph {et~al.}(2020{\natexlab{b}})\citenamefont {Xiong}, \citenamefont {Chandra}, \citenamefont {Ng},\ and\ \citenamefont {Hanzo}}]{Xiong2020}%
  \BibitemOpen
  \bibfield  {author} {\bibinfo {author} {\bibfnamefont {Y.}~\bibnamefont {Xiong}}, \bibinfo {author} {\bibfnamefont {D.}~\bibnamefont {Chandra}}, \bibinfo {author} {\bibfnamefont {S.~X.}\ \bibnamefont {Ng}}, \ and\ \bibinfo {author} {\bibfnamefont {L.}~\bibnamefont {Hanzo}},\ }\href {\doibase 10.1109/ACCESS.2020.3045016} {\bibfield  {journal} {\bibinfo  {journal} {IEEE Access}\ }\textbf {\bibinfo {volume} {8}},\ \bibinfo {pages} {228967} (\bibinfo {year} {2020}{\natexlab{b}})}\BibitemShut {NoStop}%
\bibitem [{\citenamefont {Tsubouchi}\ \emph {et~al.}(2024)\citenamefont {Tsubouchi}, \citenamefont {Mitsuhashi}, \citenamefont {Sharma},\ and\ \citenamefont {Yoshioka}}]{Tsubouchi2024}%
  \BibitemOpen
  \bibfield  {author} {\bibinfo {author} {\bibfnamefont {K.}~\bibnamefont {Tsubouchi}}, \bibinfo {author} {\bibfnamefont {Y.}~\bibnamefont {Mitsuhashi}}, \bibinfo {author} {\bibfnamefont {K.}~\bibnamefont {Sharma}}, \ and\ \bibinfo {author} {\bibfnamefont {N.}~\bibnamefont {Yoshioka}},\ }\href {https://arxiv.org/abs/2405.07720} {\bibfield  {journal} {\bibinfo  {journal} {arXiv preprint arXiv:2405.07720}\ } (\bibinfo {year} {2024})}\BibitemShut {NoStop}%
\bibitem [{\citenamefont {Zhang}\ \emph {et~al.}(2025)\citenamefont {Zhang}, \citenamefont {Xie}, \citenamefont {Gao}, \citenamefont {Yang}, \citenamefont {Bao}, \citenamefont {Zhu}, \citenamefont {Chen}, \citenamefont {Wang}, \citenamefont {Zhang}, \citenamefont {Zhong} \emph {et~al.}}]{zhang2025demonstrating}%
  \BibitemOpen
  \bibfield  {author} {\bibinfo {author} {\bibfnamefont {A.}~\bibnamefont {Zhang}}, \bibinfo {author} {\bibfnamefont {H.}~\bibnamefont {Xie}}, \bibinfo {author} {\bibfnamefont {Y.}~\bibnamefont {Gao}}, \bibinfo {author} {\bibfnamefont {J.-N.}\ \bibnamefont {Yang}}, \bibinfo {author} {\bibfnamefont {Z.}~\bibnamefont {Bao}}, \bibinfo {author} {\bibfnamefont {Z.}~\bibnamefont {Zhu}}, \bibinfo {author} {\bibfnamefont {J.}~\bibnamefont {Chen}}, \bibinfo {author} {\bibfnamefont {N.}~\bibnamefont {Wang}}, \bibinfo {author} {\bibfnamefont {C.}~\bibnamefont {Zhang}}, \bibinfo {author} {\bibfnamefont {J.}~\bibnamefont {Zhong}},  \emph {et~al.},\ }\href {https://arxiv.org/abs/2501.09079} {\bibfield  {journal} {\bibinfo  {journal} {arXiv preprint arXiv:2501.09079}\ } (\bibinfo {year} {2025})}\BibitemShut {NoStop}%
\bibitem [{\citenamefont {{Qedma Quantum Computing}}(2025{\natexlab{b}})}]{SALEMpaper}%
  \BibitemOpen
  \bibfield  {author} {\bibinfo {author} {\bibnamefont {{Qedma Quantum Computing}}},\ }\href@noop {} {\bibfield  {journal} {\bibinfo  {journal} {To appear}\ } (\bibinfo {year} {2025}{\natexlab{b}})}\BibitemShut {NoStop}%
\bibitem [{\citenamefont {Reichardt}(2020)}]{Reichardt_2021}%
  \BibitemOpen
  \bibfield  {author} {\bibinfo {author} {\bibfnamefont {B.~W.}\ \bibnamefont {Reichardt}},\ }\href {\doibase 10.1088/2058-9565/abc6f4} {\bibfield  {journal} {\bibinfo  {journal} {Quantum Science and Technology}\ }\textbf {\bibinfo {volume} {6}},\ \bibinfo {pages} {015007} (\bibinfo {year} {2020})}\BibitemShut {NoStop}%
\bibitem [{\citenamefont {Smith}\ \emph {et~al.}(2024)\citenamefont {Smith}, \citenamefont {Brown},\ and\ \citenamefont {Bartlett}}]{smith2024mitigating}%
  \BibitemOpen
  \bibfield  {author} {\bibinfo {author} {\bibfnamefont {S.~C.}\ \bibnamefont {Smith}}, \bibinfo {author} {\bibfnamefont {B.~J.}\ \bibnamefont {Brown}}, \ and\ \bibinfo {author} {\bibfnamefont {S.~D.}\ \bibnamefont {Bartlett}},\ }\href@noop {} {\bibfield  {journal} {\bibinfo  {journal} {Communications Physics}\ }\textbf {\bibinfo {volume} {7}},\ \bibinfo {pages} {386} (\bibinfo {year} {2024})}\BibitemShut {NoStop}%
\bibitem [{\citenamefont {Prabhu}\ and\ \citenamefont {Reichardt}(2024)}]{PrabhuReichardt2024}%
  \BibitemOpen
  \bibfield  {author} {\bibinfo {author} {\bibfnamefont {P.}~\bibnamefont {Prabhu}}\ and\ \bibinfo {author} {\bibfnamefont {B.~W.}\ \bibnamefont {Reichardt}},\ }\href {\doibase 10.1103/PhysRevA.110.012419} {\bibfield  {journal} {\bibinfo  {journal} {Phys. Rev. A}\ }\textbf {\bibinfo {volume} {110}},\ \bibinfo {pages} {012419} (\bibinfo {year} {2024})}\BibitemShut {NoStop}%
\bibitem [{\citenamefont {Clinton}\ \emph {et~al.}(2024)\citenamefont {Clinton}, \citenamefont {Cubitt}, \citenamefont {Flynn}, \citenamefont {Gambetta}, \citenamefont {Klassen}, \citenamefont {Montanaro}, \citenamefont {Piddock}, \citenamefont {Santos},\ and\ \citenamefont {Sheridan}}]{clinton2024towards}%
  \BibitemOpen
  \bibfield  {author} {\bibinfo {author} {\bibfnamefont {L.}~\bibnamefont {Clinton}}, \bibinfo {author} {\bibfnamefont {T.}~\bibnamefont {Cubitt}}, \bibinfo {author} {\bibfnamefont {B.}~\bibnamefont {Flynn}}, \bibinfo {author} {\bibfnamefont {F.~M.}\ \bibnamefont {Gambetta}}, \bibinfo {author} {\bibfnamefont {J.}~\bibnamefont {Klassen}}, \bibinfo {author} {\bibfnamefont {A.}~\bibnamefont {Montanaro}}, \bibinfo {author} {\bibfnamefont {S.}~\bibnamefont {Piddock}}, \bibinfo {author} {\bibfnamefont {R.~A.}\ \bibnamefont {Santos}}, \ and\ \bibinfo {author} {\bibfnamefont {E.}~\bibnamefont {Sheridan}},\ }\href@noop {} {\bibfield  {journal} {\bibinfo  {journal} {Nature Communications}\ }\textbf {\bibinfo {volume} {15}},\ \bibinfo {pages} {211} (\bibinfo {year} {2024})}\BibitemShut {NoStop}%
\bibitem [{\citenamefont {Zimbor{\'a}s}\ \emph {et~al.}(2025)\citenamefont {Zimbor{\'a}s}, \citenamefont {Koczor}, \citenamefont {Holmes}, \citenamefont {Borrelli}, \citenamefont {Gily{\'e}n}, \citenamefont {Huang}, \citenamefont {Cai}, \citenamefont {Ac{\'\i}n}, \citenamefont {Aolita}, \citenamefont {Banchi} \emph {et~al.}}]{zimboras2025myths}%
  \BibitemOpen
  \bibfield  {author} {\bibinfo {author} {\bibfnamefont {Z.}~\bibnamefont {Zimbor{\'a}s}}, \bibinfo {author} {\bibfnamefont {B.}~\bibnamefont {Koczor}}, \bibinfo {author} {\bibfnamefont {Z.}~\bibnamefont {Holmes}}, \bibinfo {author} {\bibfnamefont {E.-M.}\ \bibnamefont {Borrelli}}, \bibinfo {author} {\bibfnamefont {A.}~\bibnamefont {Gily{\'e}n}}, \bibinfo {author} {\bibfnamefont {H.-Y.}\ \bibnamefont {Huang}}, \bibinfo {author} {\bibfnamefont {Z.}~\bibnamefont {Cai}}, \bibinfo {author} {\bibfnamefont {A.}~\bibnamefont {Ac{\'\i}n}}, \bibinfo {author} {\bibfnamefont {L.}~\bibnamefont {Aolita}}, \bibinfo {author} {\bibfnamefont {L.}~\bibnamefont {Banchi}},  \emph {et~al.},\ }\href {https://arxiv.org/abs/2501.05694} {\bibfield  {journal} {\bibinfo  {journal} {arXiv preprint arXiv:2501.05694}\ } (\bibinfo {year} {2025})}\BibitemShut {NoStop}%
\end{thebibliography}


%


\appendix

\stoptoc
\section{Appendix}

\subsection{Known lower bounds on sampling overhead of general EM methods}
\label{app:lowerbounds}

References \cite{Tsubouchi2023universal, Quek_Eisert} demonstrated the existence of circuits for which the sample complexity of any EM protocol must be exponential in the total circuit volume $nD/2$, and not only in $D$. The construction is based on random circuits made of layers sampled from an $n$-qubit unitary 2-design.  Assuming a fixed local error channel after each layer, and averaging over the above ensemble of circuits, leads to a \textit{twirling} of the local error channel into an $n$-qubit depolarizing channel. The latter leads to a decay of the output state towards the maximally mixed state, with a decay constant that is proportional to the total circuit volume (this prototypical example is discussed in more detail in Sec.~\ref{Sec: evidence for exp}). Accordingly, there exists an individual circuit in the ensemble which exhibits the same decay. Reference \cite{Quek_Eisert} showed that the same conclusion holds when constructing the above 2-designs from 2-qubit Clifford gates followed by a local error channel. By transpiling the above construction to QPUs with the connectivity of a $d$-dimensional lattice, Ref.~\cite{Quek_Eisert} further showed compelling evidence that, for the above circuit ensemble, it is the active volume (or `light cone') that controls the sampling overhead of EM, as opposed to the total circuit volume. Technically, the bounds in Ref.~\cite{Quek_Eisert} are phrased in terms of the \textit{total volume}, but are only valid for $D$ large enough such that all $n$ qubits can be entangled by the circuit, given the restricted circuit connectivity.

\subsection{Deducing exponential lower bounds on number of shots, from classical simulation of noisy circuits} \label{app:samplingLB}

Here we turn back to the first question discussed in Section \ref{Sec: EM alone cannot}, and ask whether the results of Ref.~\cite{Schuster_Yao} can be used to deduce the lower bound of Eq.~\eqref{Eq: T_EM} for generic EM protocols.
As mentioned above, there are specific counter examples  in which this bound does not hold \cite{granet2024dilution}; however, one might expect the lower bound to still hold generically. 

We focus on classically-hard circuit families $I_n$ (see Definition~\ref{def:ome}) and assume that classical simulation algorithms for the given class of \textit{ideal} circuits and observables $I_n$ must scale as $\Omega(2^{n})$. 
We consider any EM protocol, which for any circuit and observable  $(C,O)\in I_n$, achieves an accuracy $\epsilon$ on average over computational input states.
Using Eq.~\eqref{Eq: Yao}, we can then reverse the logic of Corollary \ref{cor}, yielding a lower bound $M=\Omega(e^{\tilde{\Omega}(\gamma^2 n)})$ for the shot overhead. This may be compared with the known result $M=\Omega(e^{\Omega(\gamma D)})$ for any EM protocol, applied to any $(C,O)$, that achieves accuracy $\epsilon$ on a fixed input state \cite{takagi2022fundamental, Tsubouchi2023universal, Takagi2023universal, Quek_Eisert}. In cases where $n\notin O(D)$ the bound obtained from Ref.~\cite{Schuster_Yao} is stronger (asymptotically, though with a weaker constant $\gamma^2$ as opposed to $\gamma$). 
For example, in 2d circuits where $n\sim D^2$, the sampling overhead scales exponentially in $D^2$ rather than in $D$. Note that this still does not guarantee an overhead exponential in $V$ as in Eq.~\eqref{eq:emscaling} (which is, e.g.,  $\sim D^3$ in the $2d$ circuit case). Interestingly, this still leaves open the following possibility (suggested in  Ref.~\cite{granet2024dilution}): there may be a family with  fine tuned input states, which are on one hand classically hard (namely their classical simulation must scale as $\Omega(2^{n})$) for which there exists an EM protocol that exhibits a sampling overhead which is exponential only in $D$ (and not in $n$) when restricted to those fine tuned input states. Such cases, if they exist, would lead to a quasi-polynomial QA (in between polynomial and exponential), such that, e.g., for 2d circuits, $T_c=\Omega(2^n)$, but $T_{EM}=O(e^{\gamma \sqrt{n}})$.

\subsection{Calculating $V=V(n,v)$ for the HPC run time plot in Fig.~\ref{fig: QESEM runtime}\label{app:velocity}}
The plots of the classical run time estimations for Fig.~\ref{fig: QESEM runtime} require calculating the active volume $V$ as a function of $n$ and the operator spread velocity $v$. We detail here how these calculations are done to derive Eq.~\eqref{eq:Tc-of-V}. 

Let us start with the simple case: $v=1$. Here, the active volume corresponds to the ``na\"ive'' light-cone that is determined by the circuit connectivity. In such cases the qubit support of this lightcone (which is the number of active qubits), is 
$n=\left(\frac{2}{d}\cdot D\right)^d$. The $2/d$ factor accounts
for the fact that each layer extends the support only in one of the
$d$ dimensions, but in both the positive and negative directions
along that dimension. Similarly, for a general $v\in [0,1]$, the
qubit support is given by
\begin{equation}\label{eq:nten}
  n=\left(\frac{2v}{d} D\right)^d .
\end{equation}

Once $n$ is fixed, the active volume is given by
the volume of a $(d+1)$-dimensional pyramid with base area $n$ and
height $D$. We additionally divide by a factor of 2, to
take into account that a dense layer on $m$ qubits involves $m/2$
2-qubit gates. We thus derive:
\begin{align}
  V=\frac{nD}{2(d+1)}=\frac{c_d}{4v}n^{1/c_d},
\label{Eq: V(n) exmple}
\end{align}
where $c_d=d/(d+1)$. This relation can be easily inverted to give $n$ as a function of $V$:
\begin{align}
  n = \left(4vV/c_d\right)^{c_d}.
\end{align}

\end{document}